\documentclass[review]{elsarticle}

\usepackage{hyperref}
\usepackage{amsmath}
\usepackage{pgfplots}

\journal{arxiv}

\begin{document}

\begin{frontmatter}

\title{A reduced order Kalman Filter model for sequential Data Assimilation of turbulent flows}


\author[Poitiers]{M. Meldi*}
\cortext[Poitiers]{Corresponding author, \textit{marcello.meldi@ensma.fr}}
\author[Poitiers]{A. Poux**}

\address[Poitiers]{Institut PPRIME, Department of Fluid Flow, Heat Transfer and Combustion, CNRS -
ENSMA - Universit\'{e} de Poitiers, UPR 3346, SP2MI - T´el´eport, 211 Bd. Marie et Pierre Curie,
B.P. 30179 F86962 Futuroscope Chasseneuil Cedex, France}

\cortext[Poitiers]{Current address : CNRS UMR 6614 CORIA - Universit\'{e} de Rouen - Site Universitaire du Madrillet, 675, avenue de l’Université, BP 12, 76801 Saint-Etienne du Rouvray Cedex, France}

\begin{abstract}
A Kalman filter based sequential estimator is presented in the present work.
The estimator is integrated in the structure of segregated solvers for the analysis of incompressible flows.
This technique provides an \textit{augmented} flow state integrating available observation in the CFD model, naturally preserving a zero-divergence condition for the velocity field.
Because of the prohibitive costs associated with a complete Kalman Filter application, two model reduction strategies have been proposed and assessed.
These strategies dramatically reduce the increase in computational costs of the model, which can be quantified in an increase of $10\% - 15\%$ with respect to the classical numerical simulation.
In addition, an extended analysis of the behavior of the numerical model covariance $Q$ has been performed.
The results have shown that optimized values are strongly linked to the truncation error of the discretization procedure.
The estimator has been applied to the analysis of a number of test cases exhibiting increasing complexity, including turbulent flow configurations.
The results show that the \textit{augmented} flow successfully improves the prediction of the physical quantities investigated, even when the observation is provided in a limited region of the physical domain.
In addition, the present work indicates that these Data Assimilation techniques, which are at an embryonic stage of development in CFD, can be pushed even further using the \textit{augmented} prediction as a powerful tool for the optimization of the free parameters in the numerical simulation.
\end{abstract}


\end{frontmatter}


\section{Introduction}

The accurate prediction of turbulent flow configurations is one of the ultimate open challenges in fluid mechanics studies.
Most of industrial / environmental applications aim to provide accurate and robust estimation of aspects which are governed by turbulence statistical moments such as aerodynamic forces, transport of particles and heat exchange.
Traditional investigative tools, such as experiments and numerical simulation, are not completely successful in producing robust descriptions of turbulent configurations, because of important fundamental drawbacks.

Measurements obtained via Experimental Fluid Mechanics (EFD), such as those sampled by surface sensors, provide a local description of flow dynamics.
Because of the non-linear, strongly inertial behavior of the flow, the determination of a complete map of flow behavior is problematic.
Information can be reconstructed by the use of reduced-order models, such as POD \cite{Lumley_AP1970}.
However, these approximated models usually provide an incomplete reconstruction of turbulent flows.
One of the reasons is that strong non-linear interactions occur between the modeling error and the bias associated to the measurement, which is tied to epistemic uncertainties in the experimental sampling.
This aspect, which is amplified by the multi-scale nature of turbulence, usually results in poor characteristics of robustness and precision of the reduced-order model.

Under this perspective, the use of Computational Fluid Dynamics (CFD) can provide more complete maps of flow characteristics, including regions of the physical domain were experimental sampling is problematic because of structural difficulties.
However, numerical simulation is affected as well by errors / epistemic uncertainties.
In this case, the parametric set-up of the simulation (physical characterization of the flow, boundary conditions...) can not exactly reproduce the subtle perturbations and in-homogeneity of the real flow, which are unknown a priori.
This is particularly true for turbulence, where small perturbations present in the environment are amplified and they ultimately drive the evolution of the flow.
In addition, owing to computational resources constraints, analyses of very high Reynolds flows are presently limited to reduced order simulations via RANS / LES approaches \cite{Pope2000_cambridge,Wilcox2006_DCW,Sagaut2006_springer}.
Turbulence / subgrid scale modeling is usually a main source of error in CFD, in particular because of its non-linear interaction observed with the numerical / boundary condition error.
Thus, while both EFD and CFD are affected by bias, the confidence level of the results is affected by uncertainties of a completely different nature.
This is the reason why the comparison of experiments and numerical results is a complex task even for the classical case of grid turbulence decay \cite{Davidson2004_oxford}.

In the last decade, new methodological approaches coming from Estimation Theory (ET) have been employed by the fluid mechanics community to obtain an optimized prediction of flow configurations.
ET is a branch of statistics dealing with the estimation of optimal parametric description, using data which is affected by a level of uncertainty / stochasticity \cite{Simon2006_wiley}.
In particular, Data Assimilation (DA) includes a wide spectrum of tools which aim at the estimation of an optimal state integrating a model and observations which are affected by uncertainties.
They are usually referred to as \textit{estimators}.
Some studies in EFD have been proposed, where the DA tool combines experimental sampling with reduced order numerical solvers in order to provide the zero-divergence condition of incompressible flows \cite{Artana2012_jcp,Leroux2014_ep}.
Early CFD applications mainly deal with variational approaches based on the adjoint method, which have been a classical choices for meteorological studies since the 1970s \cite{Daley1991_cambridge}.
These methods are defined as an optimization problem where a given measure is minimized under the constraint of the governing equation.
The resolution of this problem determines the distribution of a basis of parameters (typical choices are boundary / initial conditions) which optimizes the flow configuration.
Recent applications deal both with fundamental studies \cite{Foures2014_jfm,Mons2016_jcp} and industrial oriented analyses \cite{Othmer2014_jmi,Onder2016_cf}.
While these approaches are very precise and they allow for sensitivity analyses considering a very large number of variables, their application to turbulent flow investigation over a long observation window is problematic \cite{Onder2016_cf,Sirkes1997_mwr}.
A much less commonly investigated path for CFD is represented by the use of sequential methods, which are based on Bayesian inference and they occasionally require the resolution of Riccati-type equations.
Examples based on techniques such as the Kalman filter \cite{Kalman1960_jbe} or the ensemble Kalman filter \cite{Evensen2009_IEEE} have been reported in the literature \cite{Chevalier2006_jfm,Suzuki2010_jfm,Rochoux2014_nhess,Suzuki2015_fdr,Kato2015_jcp}.

These techniques are showing enormous potential, because the coupling between experimental / numerical data can potentially exclude the bias which can not be identified in the two methods alone.
However, reliable tools for the optimal prediction of complex flow configurations are still far out of reach, because of the level of maturity of application of these techniques in fluid mechanics analyses.
In fact, these methods are still largely unexplored, in particular for the analysis of turbulent flow configurations.
In the present work we propose a methodological approach (estimator) for sequential Data Assimilation, which efficiently integrates information (usually experimental data) in CFD solvers.
The approach is based on a reduced order Kalman filter \cite{Kalman1960_jbe}, which exploits structural characteristics of the segregated solvers commonly implemented in commercial CFD software.
While similar approaches have been rigorously derived for coupled solvers for incompressible flows \cite{Suzuki2012_jfm}, the present model proposes practical, computational inexpensive solutions for the analysis of three dimensional turbulent flow configurations.

The paper is structured as follows.
In Section \ref{sec:numerics} the state of the art is presented introducing all the background elements which synergically interact in the estimator.
These elements include a description of the numerical solver as well as the Kalman Filter.
In Section \ref{sec:model} the Kalman Filter based estimator is introduced and discussed.
In Section \ref{subsec:Manufactured} an extensive analysis of the structure of the model covariance matrix $Q$ is performed.
In section \ref{sec:experimentations} the estimator is applied to the analysis of laminar flows.
In particular, the property of synchronization of the model with available observation is investigated.
In Section \ref{sec:turbFlows} the analysis is extended to turbulent flow configurations and their statistical behavior.
In Section \ref{sec:future} future development are discussed, including parametric optimization of the numerical model.
Finally, in Section \ref{sec:conclusions} conclusions are drawn.

\section{Numerical \& Methodological Ingredients: State of the Art}
\label{sec:numerics}

The basic elements for the development of the reduced-order Kalman filter estimator are now introduced.
This includes a description of the segregated numerical CFD solver used for the resolution of the Navier-Stokes equations, as well as fundamental elements for Kalman Filter application. 

\subsection{The Pressure Implicit with Spitting of Operator (PISO) algorithm}
\label{subsec:PISO}

Most of the CFD solvers available in commercial and open-source codes are based on numerical discretization of the Navier--Stokes equations.
For incompressible flows of Newtonian fluids, the evolution of the velocity $\mathbf{u}$ and the normalized pressure $p = p^{\prime} / \rho$ is described as:

\begin{eqnarray}
\label{eq:NavStokes}
\frac{\partial \mathbf{u}}{\partial t} + \nabla \cdot \left( \mathbf{u} \, \mathbf{u} \right) - \nabla \cdot \left( \nu \, \nabla \mathbf{u} \right)&=& - \nabla p \\
\label{eq:MassBudget}
\nabla \cdot \mathbf{u} &=& 0 
\end{eqnarray}

$\nabla \cdot$ and $\nabla$ are the divergence operator and the gradient operator, respectively.
In the case of reduced-order simulation, simplified variables are considered (average velocity $\overline{\mathbf{u}}$ in the case of RANS, filtered velocity $\mathbf{u^*}$ in the case of LES, ...) and a term $T$ representing the turbulence / subgridscale modeling must be included in the vectorial equation \ref{eq:NavStokes}.

Several different discretization strategies can be proposed for the numerical resolution of this system.
In the following, the Pressure Implicit with Splitting of Operator (PISO) \cite{Issa1986_jcp} as implemented in the open-source code \textit{OpenFOAM} \cite{Jasak1996_phd} is presented.
This code, which elaborates a finite volume discretization of the Navier--Stokes equations, has been identified as the best demonstrator for this research activity thanks to the flexibility of its algorithmic structure and the ease in implementation.

At the beginning of every time step a \textit{prediction} of the flow field is performed discretizing the momentum equation \ref{eq:NavStokes}.
In particular, an implicit discretization of the time derivative is applied.
The PISO algorithm operates on the discretization at this level, simplifying the non-linear term in equation \ref{eq:NavStokes}.
This approximation, which relies on the hypothesis of small values of the time step $\Delta t$, resolves the non-linearity of the system via a linearization of the new state around the source state (previous time step).
The approximation error of this operation is proportional to $\Delta t^2$\cite{Ferziger2002_springer}.
The resulting discretized system, known as \textit{predictor}, is:

\begin{equation}
\label{eq:predictor}
a_P \mathbf{u}_P = - \sum_N a_N \mathbf{u_N} + \Phi_0(\mathbf{u}_0) -\nabla p = \Phi (\mathbf{u}) -\nabla{p}
\end{equation}  

Here, the term with subscript $0$ indicates a source velocity from previous time steps.
The subscript $P$ represents the discretized velocity field in the considered mesh element, while the subscript $N$ indicates its neighbors.
The coefficients $a_P$ and $a_N$ represent the result of the discretization process of the velocity budget terms in equation \ref{eq:NavStokes}.
As a first guess, the term $\nabla{p}$ is calculated using the pressure field at the previous time step.
Because of the underlying hypothesis of the PISO algorithm over $\Delta t$, a Courant-Friedrick limit (CFL) for numerical stability must be satisfied for the time advancement.
However, the resulting discretized system can be investigated by the resolution of a linear system.
The coefficient $a_P$ and the operator $\Phi$ capture the essential non-linearity of equation \ref{eq:NavStokes}.
Instead of performing a computationally expensive matrix inversion, the predictor is usually solved via iterative resolution techniques.
In addition, the pressure-velocity coupling is avoided performing an iterative loop.
First, it is observed that the velocity field predicted by equation \ref{eq:predictor} does not usually comply with the zero-divergence condition:

\begin{eqnarray}
\label{eq:divdiscr}
\nabla \cdot \mathbf{u} &=& \sum_f S \times \mathbf{u}_f = 0 \\
\label{eq:Ufaces}
\mathbf{u}_f &=& \left(\frac{\Phi(\mathbf{u})}{a_P} \right)_f - \left( \frac{\nabla p}{a_P} \right)_f
\end{eqnarray}

where the subscript $f$ stands for an interpolation on the face centers of the mesh element and $S$ is the corresponding surface area.
If equation \ref{eq:divdiscr} and \ref{eq:Ufaces} are combined, a Poisson equation for the pressure field is obtained:

\begin{equation}
\label{eq:PoissonDisc}
\nabla \cdot \left( \frac{\nabla p}{a_P} \right) = \nabla \cdot \left( \frac{\Phi(\mathbf{u})}{a_P}\right) = \sum_f S \times \left( \frac{\Phi(\mathbf{u})}{a_P}\right)_f
\end{equation}

The pressure field calculated in equation \ref{eq:PoissonDisc} does not necessarily satisfy equation \ref{eq:predictor}.
This value is used to correct the velocity field obtained in the \textit{predictor} step.
This operation is referred to as \textit{corrector} step.
The loop updating the pressure / velocity fields continues until a prescribed convergence criterion is satisfied.
Summarizing, the loop can be structured in the following phases:

\begin{enumerate}
\item{A first prediction of the velocity field is obtained by the discretized momentum equation \ref{eq:predictor}.
This velocity field is usually not divergence-free.}
\item{A Poisson equation is resolved to update the pressure field $p$ as a function of the new predicted velocity field.
The pressure gradient does not usually satisfy equation \ref{eq:predictor}, so that the velocity field can be updated accounting for the new information.}
\end{enumerate}

In practice, the physical system bounces from a momentum-predicted state to a zero-divergence condition until the convergence is reached.
An essential feature is that, while the Navier--Stokes equation are strongly non-linear, the discretization proposed includes the non-linearity in a linear system resolution, with an error scaling as $\Delta t^2$.
This last aspect, which is shared by a number of CFD numerical strategies, is essential for the integration of the proposed estimator, as it will be shown in Section \ref{sec:model}.

\subsection{The Kalman Filter}
\label{sec:KalmanFilter}

The Kalman Filter \cite{Kalman1960_jbe} is a powerful sequential tool used in Data Assimilation studies.
It allows for robust prediction of an optimized physical state, accounting for the level of uncertainty in the prediction provided by multiple investigative tools.
Let us considered a physical phenomenon $\mathbf{u}$ described by the following discrete system at the instant $k$:

\begin{eqnarray}
\label{eq:KFmodel}
\mathbf{u}_k &=& \Phi_k \mathbf{u}_{k-1} + B_k \, \mathbf{c}_k + \mathbf{w}_k \\
\label{eq:KFobservation}
\mathbf{z}_k &=& H_k \mathbf{u}_k +\mathbf{v}_k
\end{eqnarray}

Equation \ref{eq:KFmodel} represents a \textit{model} for the time evolution of $\mathbf{u}$.
A similar structure with respect to the \textit{predictor} equation \ref{eq:predictor} is recognizable.
On the other hand, the term $\mathbf{c}$ represents a source term that can be manipulated by the user.
Equation \ref{eq:KFobservation} represents available \textit{observation} $\mathbf{z}$ of the same physical phenomenon at the instant $k$.
Both the equations are affected by a degree of epistemic uncertainty, represented by the terms $\mathbf{w}_k=\mathcal{N}(0,Q_k)$ and $\mathbf{v}_k=\mathcal{N}(0,R_k)$.
This uncertainty can represent both bias associated with the prediction / observation as well as measurement noise.
The terms $w$ and $v$ are supposed to be zero-mean and they are characterized by a time-dependent variance $Q_k$ and $R_k$, respectively.
The smaller the value of the variance, the higher the confidence in the model / observation.

The Kalman filter provides an optimized prediction of $\mathbf{u}$, which will be referred to as $\mathbf{\hat{u}}$, accounting for the level of confidence in the model and in the available observation.
More precisely, the optimized state provides a minimization of the error covariance of the physical system $P=E((\mathbf{u}-\mathbf{\hat{u}}) \times (\mathbf{u}-\mathbf{\hat{u}})^T)$, which is a quadratic function describing the total level of confidence of the system.
The discrete version of the Kalman Filer operates through two steps:

\begin{enumerate}
\item{A \textit{predictor} step, where $\mathbf{u}$ and $P$ are advanced in time using information derived via model only:
\begin{eqnarray}
\label{eq:KFpred_u}
\mathbf{\hat{u}}_{k|k-1} &=& \Psi_k \mathbf{\hat{u}}_{k-1} + B_k \, \mathbf{c}_k \\
\label{eq:KFpred_P}
P_{k|k-1} &=& \Psi_k \,P_{k-1|k-1} \, \Psi_k^T + Q_k
\end{eqnarray}
}
\item{An \textit{update} step, when the observation is integrated through a weighted measure provided by the \textit{Kalman Gain} operator $K_k$:
\begin{eqnarray}
\label{eq:KalmanGain}    
S_k &=& H_k \, P_{k|k-1} \, H_k^T + R_k \, , \qquad K_k = P_{k|k-1} \, H_k^T \, S_k^{-1} \\
\label{eq:KFupdate_u}
\mathbf{\hat{u}}_{k|k} &=& \mathbf{\hat{u}}_{k|k-1} + K_f \, \left( z_k - H_k \, \mathbf{\hat{u}}_{k|k-1} \right) \\
\label{eq:KFupdate_P}
P_{k|k} &=& \left(I -K_k \, H_k \right) \,P_{k-1|k-1}
\end{eqnarray}  
}
\end{enumerate}

The two-step structure of the discrete Kalman filter is reminiscent of the predictor / corrector strategy presented for the PISO algorithm in section \ref{subsec:PISO}.
This feature will be exploited to derive an estimator in section \ref{sec:model}.

\section{Integrating CFD solver \& Kalman filter: a DA estimator}
\label{sec:model}

The present estimator combines the CFD PISO scheme with the discrete Kalman Filter in order to produce a divergence free \textit{augmented} prediction.
This is obtained considering the momentum equation \ref{eq:predictor} as the \textit{model} for the Kalman Filter, while equation \ref{eq:PoissonDisc} is used to regularize a target velocity field.
The model articulates in three macro-steps, which perform different operations whether observation is provided at the current time step of not:

\begin{enumerate}
\item{\textbf{Predictor step}.
At the beginning of the time step $k$, the momentum equation \ref{eq:predictor} and the error covariance matrix $P$ time advancement in equation \ref{eq:KFpred_P} are performed, whether or not observation is provided.
This procedure is equivalent to the \textit{update} step of the Kalman Filter.
The matrix $\Psi_k$ can be exactly obtained by manipulation of the operator $\Phi$, which includes a matrix inversion.
This problematic aspect will be discussed in detail when the reduced-order strategies will be presented.
If observation is available, the Kalman gain $K$ is calculated using equation \ref{eq:KalmanGain}.}
\item{\textbf{Corrector step} If observation is not available, the corrector step exactly replicates the PISO algorithm:
\begin{eqnarray}
\label{eq:Poisson1}
\nabla \cdot \left( \frac{\nabla p}{a_P} \right)_f & =&  \sum_f S \times \left( \frac{\Phi(\mathbf{u})}{a_P}\right)_f \\
\label{eq:velCorr1}
\mathbf{u} &=& \frac{\Phi (\mathbf{u})}{a_P} - \frac{\nabla{p}}{a_P}
\end{eqnarray}
These two equations are solved iteratively, until the model solution converges towards a zero-divergence condition.
If observation is available, the Poisson equation is instead used to impose the zero-divergence condition for the \textit{augmented} prediction:
\begin{eqnarray}
\label{eq:AugmDA}
\mathbf{\hat{u}} &=& \mathbf{u} + K \, \left(\mathbf{z} - H \, \mathbf{u} \right) = \frac{\Phi (\mathbf{u})}{a_P} - \frac{\nabla{p}}{a_P} + F \\
\label{eq:PoissonDA}
\nabla \cdot \left( \frac{\nabla p}{a_P} \right)_f & =&  \sum_f S \times \left( \frac{\Phi(\mathbf{u})}{a_P} + F \right)_f \\
\label{eq:velCorrDA}
\mathbf{u} &=& \frac{\Phi (\mathbf{u})}{a_P} - \frac{\nabla{p}}{a_P}
\end{eqnarray}

Similarly to the classical version of the corrector, the loop of equations \ref{eq:AugmDA}, \ref{eq:PoissonDA} and \ref{eq:velCorrDA} is performed until convergence.
The resulting algorithm assures that the augmented velocity $\mathbf{\hat{u}}$ always respect the zero-divergence constraint, while this condition is observed for the velocity field $\mathbf{u}$ derived by model only if $\nabla \cdot F =0$.
However, re-normalization of the observation in order to force it complying with the zero-divergence condition \cite{Suzuki2010_jfm} is not needed here.}
\item{\textbf{Regularization} step.
This last procedure is triggered only when observation is available.
First, the error covariance matrix $P$ is updated via equation \ref{eq:KFupdate_P}.
In addition, the augmented state $\mathbf{\hat{u}}$ is used to improve the performance of the numerical model.
Within the framework of this research activity, the most simple solution of imposing $\mathbf{u}=\mathbf{\hat{u}}$ is chosen.
In this way, the model will use the augmented velocity as a source term $\mathbf{u_0}$ at the beginning of the next time step.}  
\end{enumerate}  

The analysis of the proposed algorithm reveals that the classical PISO algorithm is modified only if observation is available, otherwise a classical numerical resolution is obtained.
However, the resolution of the full algorithm presented is problematic when considering large dynamic systems such as those needed for the analysis of turbulent flow configurations.
The principal reasons are:

\begin{enumerate}
\item{The derivation of the matrix $\Psi$ from the operator $\Phi$ is expensive, as it demands a matrix inversion each time observation is provided.
Considering a discretization in $N$ mesh elements, the size of the matrix should be $3N \, \times \, 3N$ because of the interactions of the three velocity components.
Thanks to the linearization of the PISO algorithm, this problem is here reduced to three $N \, \times \, N$ matrices.
This operation can be performed only once if the matrix $\Phi$ in constant in time, such as for laminar flows with an explicit treatment of the inertia.
However, this is clearly not feasible for numerical simulation of turbulent flows, which is the target of the present research work.}
\item{Similar critical issues must be solved for the resolution of the error covariance matrix $P$.
Taking benefit from the linearization of the PISO algorithm, the procedure consists of eight $N \, \times \, N$ matrix product and a matrix inversion for each velocity component.
This operation demands computational resources and RAM availability which are orders of magnitude larger than the costs associated with the numerical simulation.}
\item{The structure of the matrix $Q$ is difficult to predict.
Suzuki \cite{Suzuki2012_jfm} proposed to estimate its value using reference experimental data.
While his pioneering development is remarkable, this approach excludes the bias associated with the experiments and, more in general, with the observation, which is not always true.
In addition, the quantification of the level of confidence of the numerical simulation should be ideally derived accounting for model results only.
Studies in the optimal control field \cite{Kim2007_arfm} indicate that the most suitable structure for the matrix $Q$ is diagonal.
This implies that the uncertainty generated by the model on a discrete element is local and it is not initially tied to the other elements.
Similar considerations are usually provided for the observation covariance matrix $R$.
This hypothesis is reasonable if the considered model is a numerical solver, and it will be used in the present framework.}
\end{enumerate}

In the following, approximated solutions for the three problematic aspects introduced are proposed.
The goal is to produce an efficient reduced method, which provides an optimized flow prediction with moderate increase of computational resources when compared to classical numerical simulation.

\subsection{Sequential observer}
\label{sec:Observer}

The first reduced model presented is based on very strong hypothesis based on the behavior of the model matrix $\Psi$.
During the update step, the Kalman Filter propagates the model uncertainty introduced by the matrix $Q$ using the correlation between state variables provided by $\Psi$.
This information is stored in the error covariance matrix $P$.
The general formulation of the Kalman Filter is efficient for every structure of the matrix $\Psi$.
However, applications in fluid mechanics exhibit diagonal-dominant state matrices.
This is particularly true for the analysis of turbulent configurations, where a very small time step $\Delta t$ has to be imposed to capture the dynamics of the flow.
In addition, small time steps are required for application of the PISO algorithm, as previously discussed.
Thus, while the matrix $P$ is technically a full matrix, it shows a large number of elements $\approx 0$ corresponding to the zeros of the matrix $\Psi + \Psi^T$ in the applications currently investigated.

So, the matrix $\Psi$ is here approximated as:

\begin{equation}
\label{eq:ObserverMat}
\Psi_{OB} = \frac{1}{a_P} \Phi_0 (\mathbf{u_0}) = \frac{1}{a_P \, \Delta t} \, I  
\end{equation}

The second equality is valid for the first order Euler time discretization, and it is reported for sake of clarity.
This approximation implies that, if $P$ is initially set to be diagonal and $Q$ and $R$ are diagonal during the simulation, the matrix $P$ will be diagonal as well.
It also implies that the uncertainty produced by the model at each time step is not propagated to other elements of the mesh.
This family of estimators is usually referred to as \textit{observers} \cite{Luenberger1960_wiley,Luenberger1964_ieee}.
The resulting Kalman gain $K_{OB}$ (which is as well a diagonal matrix) is sub-optimal, as the information is not shared to neighbors cells by equation \ref{eq:AugmDA}.
However, this drawback is mitigated by the resolution of the Poisson equation \ref{eq:PoissonDA}, which instantaneously propagates the information through all the physical domain.
In addition, as all the matrices describing the Kalman procedure are diagonal, all the matrix products in equations \ref{eq:KFpred_P}-\ref{eq:KFupdate_P} can be replaced by local algebraic relations for every mesh element.
This implies that the cost associated with the Kalman filter becomes negligible with respect to the resources required by the solver for the resolution of the momentum equation and the Poisson equation.
However, integrating observation in the PISO loop usually demands a higher number of iterations to reach the prescribed level of converge of the solution.
An augmentation of the 10\% - 15\% of the computational cost with respect to classical CFD has been observed in the practical application to the test cases investigated.

\subsection{Sequential estimator based on filtering of the matrix $P$}
\label{sec:filteredModel}

A second reduced order model based on the Kalman Filter is here proposed, which similarly exploits characteristics of the dynamic system.
It will be referred to as \textit{filtering} estimator.
In this case, a filtered error covariance matrix $\tilde{P}$ is used for the computations.
The only elements stored are the non-zero elements of the matrix $\Psi+\Psi^T$.
In the case of three-dimensional simulations using hexahedral elements, $\tilde{P}$ is epta-diagonal, as each mesh cell communicates with a maximum of six neighbors.
Within this framework, we propose approximated formulae for matrix /matrix product and matrix inversion, which dramatically reduce the computational demands for these expensive operations.
For sake of clarity the operation is now detailed for a penta-diagonal matrix, which may correspond to a two-dimensional simulation.
Extension to epta-diagonal matrix is direct.   
Let us consider two penta-diagonal matrices $A$ and $B$, which are both diagonal-dominant.
The matrix $\widetilde{C}$, which is the resulting penta-diagonal matrix obtained filtering the result of the product $C = A \times B^T$, can be approximated such that:

\begin{eqnarray*}
\label{eq:approximated_product}  
\tilde{c}_1^i &=& a_1^i b_3^i + a_3^i b_1^i\\
\tilde{c}_2^i &=& a_2^i b_3^i + a_3^i b_2^i\\
\tilde{c}_3^i &=& \sum_{j=1}^5 a_j^i b_j^i\\
\tilde{c}_4^i &=& a_4^i b_3^i + a_3^i b_4^i\\
\tilde{c}_5^i &=& a_5^i b_3^i + a_3^i b_5^i
\end{eqnarray*}

Where the subscripts $1,\,2,\, \cdots,\, 5$ represent the five diagonals stored for the filtered matrix. $a_j^i$ represents the element of the matrix $A$ located in line $i$ and in the $j^{th}$ diagonal.
In this product, each line is independent of the others, making it appropriate for parallelization and reducing the total number of operations from $N \times N$ to $5 \times N$.
If we now consider a matrix $C = I$, it is also possible to define an approximation of the right inverse of a matrix $A$ via the resolution of a simplified system for each line ($X^T = A^{-1}$):

\begin{eqnarray*}
\label{eq:approximated_inverse}  
a_1^i x_3^i + a_3^i x_1^i &=& 0\\
a_2^i x_3^i + a_3^i x_2^i &=& 0\\
\sum_{j=1}^5 a_j^i x_j^i &=& 1\\
a_4^i x_3^i + a_3^i x_4^i &=& 0\\
a_5^i x_3^i + a_3^i x_5^i &=& 0
\end{eqnarray*}

This approximation allows for very fast calculations of large matrices with an approximation error that decreases the more the matrix is diagonal-dominant.
In order to practically illustrate the procedure, two penta-diagonal random matrix (inversible and strictly diagonal dominant) $A$ and $B$ are generated.
Their exact product $A \times B=C$ is as well computed:

\begin{eqnarray*}
\small
A &= \left(\begin{array}{lllll}
  8.02e-01   &9.85e-02  & 0.00e+00  & 1.17e-01  & 0.00e+00 \\
  1.02e-01 &  7.38e-01  & 8.91e-02  & 0.00e+00 &  1.01e-01 \\
  0.00e+00  & 9.80e-02 &  9.43e-01  & 1.00e-01 &  0.00e+00 \\
  1.04e-01   &0.00e+00 &  9.55e-02 &  1.02e+00 &  9.61e-02 \\
  0.00e+00 &  8.54e-02 &  0.00e+00 &  1.11e-01 &  9.10e-01 \end{array} \right)\\
B &= \left(\begin{array}{lllll}
  1.04e+00 &  1.04e-01 &  0.00e+00 &  1.03e-01&   0.00e+00 \\
  9.90e-02 &  1.14e+00 &  1.04e-01 &  0.00e+00 &  9.93e-02 \\
  0.00e+00 &  9.45e-02 & 8.57e-01  & 1.19e-01  & 0.00e+00 \\
  1.06e-01   &0.00e+00 &  9.11e-02  & 1.01e+00  & 9.54e-02 \\
  0.00e+00 &  8.96e-02 &  0.00e+00 &  1.10e-01 &  9.56e-01  \end{array} \right)\\
A \times B =C &= \left(\begin{array}{lllll}
  8.57e-01  & 1.96e-01 &  2.09e-02  & 2.01e-01  &2.10e-02 \\
  1.79e-01  & 8.71e-01 &  1.53e-01&   3.21e-02 &  1.69e-01 \\
  2.03e-02  & 2.01e-01 &  8.28e-01  & 2.14e-01  & 1.93e-02 \\
  2.16e-01  & 2.84e-02  & 1.75e-01  & 1.06e+00&   1.89e-01 \\
  2.02e-02  & 1.79e-01 &  1.90e-02  & 2.12e-01  & 8.89e-01   \end{array} \right)
\end{eqnarray*}

The filtered product of A and B is:

\begin{eqnarray*}
\small
\widetilde{C} &= \left(\begin{array}{lllll}
  8.34e-01 &  1.03e-01 &  0.00e+00  & 1.22e-01 &  0.00e+00 \\
  1.86e-01  & 8.62e-01  & 1.78e-01 &  0.00e+00  & 1.15e-01 \\
  0.00e+00  & 1.70e-01  & 8.28e-01  & 1.84e-01  & 0.00e+00 \\
  1.05e-01  & 0.00e+00  & 2.09e-01  & 1.05e+00  & 2.19e-01 \\
  0.00e+00  & 8.16e-02 &  0.00e+00 &  1.06e-01 &  8.70e-01   \end{array} \right)
\end{eqnarray*}

And the filtered $\widetilde{B}$ obtained as $\widetilde{B}=A^{-1} \widetilde{\times} \widetilde{C}$:

\begin{eqnarray*}
\small
\tilde{B} &= \left(\begin{array}{lllll}
  1.04e+00  & 8.38e-02  & 0.00e+00 &  1.08e-01  & 0.00e+00 \\
  1.17e-01 &  1.15e+00 &  1.25e-01  & 0.00e+00  & 1.07e-01 \\
  0.00e+00 &  6.85e-02 &  8.50e-01 &  7.61e-02 &  0.00e+00 \\
  9.86e-02 &  0.00e+00 &  1.36e-01 &  1.02e+00 &  1.17e-01 \\
  0.00e+00  & 7.28e-02 &  0.00e+00 &  8.96e-02 &  9.52e-01    \end{array} \right)
\end{eqnarray*}

Observation of the previous results proves that the filtered matrices $\tilde{B}$ and $\tilde{C}$ are reasonable approximations of the exact matrices $B$ and $C$. The analysis has been extended to larger matrices and the relative error has been quantified via Frobenius norm $||.||_F$. The results reported in table \ref{table:norm_reduc_matmul} indicate that the approximation error committed is not sensitive to the size of the matrix, but as expected it decreases with increasing dominance of the diagonal terms. In the case of numerical simulation of turbulent flows, the parameter $A_{ii}/A_{ij}, i\neq j$ measuring the dominance of the diagonals terms is usually of the order of $5-10$. 

 \begin{table}
\centering
\small
\begin{tabular}{|l|r|r|}
\hline
Size $N$	&	$\frac{||C-\tilde C||_F}{||C||_F}$	&	$\frac{||B-\tilde B||_F}{||B||_F}$	 \\
\hline
5 &   1.03e-01  &   1.93e-02  \\
10 &   9.23e-02  &   2.27e-02  \\
20 &   7.97e-02  &   2.33e-02  \\
40 &   7.99e-02  &   2.91e-02  \\
80 &   7.39e-02  &   2.82e-02  \\
160 &   6.68e-02  &   2.69e-02  \\
320 &   6.71e-02  &   2.89e-02  \\
640 &   6.77e-02  &   2.84e-02  \\
1280 &   6.62e-02  &   2.90e-02 \\
\hline
\end{tabular}\hfill
\begin{tabular}{|l|r|r|}
\hline
$A_{ii}/A_{ij}, i\neq j$	& $\frac{||C-\tilde C||_F}{||C||_F}$	&	$\frac{||B-\tilde B||_F}{||B||_F}$	 \\
\hline
5 &   1.73e-01  &   6.97e-02  \\
10 &   6.68e-02  &   2.77e-02  \\
100 &   4.92e-03  &   2.81e-03  \\
1000 &   4.97e-04  &   2.78e-04  \\
\hline
\end{tabular}
\caption{Frobenius norm of the relative error committed with the approximated matrix product.
On the left, the sensitivity of the error to the size of the matrix is investigated.
In this case the parameter $A_{ii}/A_{ij}, i\neq j$ measuring the dominance of the diagonal terms is set to 10.
On the right, the sensitivity of the approximation error to the parameter $A_{ii}/A_{ij}, i\neq j$ is studied, fixing a matrix size of $640 \times 640$.
\label{table:norm_reduc_matmul}  }
\end{table}

In summary, the \textit{filtering} estimator resolves the full scheme presented in Section \ref{sec:model} via the use of the approximated matrix products / inversions here discussed.
While the resulting Kalman gain is sub-optimal, the level of precision is supposedly higher than the diagonal approximation of the \textit{estimator} observer presented in Section \ref{sec:Observer}, because information transmitted between neighbors is here conserved.
The Authors would like to stress two important points:
\begin{enumerate}
\item{Even if differences of the order of $10\%$ can be measured by the comparison of $\tilde{K}$ and $K$, this does not absolutely imply that a similar level of magnitude of difference is observed in the flow field predictions.
The Kalman gain is actually a gain, and it just approximately regulates the amount of observation that is included in the model prediction.
Additionally, $K$ is an optimal gain with respect to a prescribed quadratic norm.
While this criterion is robust and reliable, it can not be considered as an exact quantification of the quality of the resulting flow prediction.
Tests on simplified test cases in Section \ref{subsec:Manufactured} have shown that negligible differences are observed when applying a full Kalman Filter, an observer estimator or a filtering estimator.}
\item{The structure of the estimator presented in Section \ref{sec:model} is reminiscent to implementation of porous sources or discrete Immersed Boundary Method (IBM) forcing \cite{Uhlmann2005_jcp,Pinelli2010_jcp}.
The term $K \, \left(\mathbf{z} - H \, \mathbf{u} \right)$ actually forces the predicted flow towards the asymptotic limit $\mathbf{z}$ and is regulated in intensity by the Kalman gain $K$.
In classical discrete IBM application, we would simply have $K=1$.
In addition,  the information is integrated locally and then propagated by the pressure field in the \textit{observer}.
On the other hand, the observation is spread to the cell neighbors through interpolation (governed by $K$) in the \textit{filtering} estimator, before propagation via the Poisson equation.}
\end{enumerate}

In the following, it will be shown that the \textit{augmented prediction} of turbulent flows can be performed with present computational capabilities, and it promises groundbreaking advance in CFD.

\section{Manufactured tests cases}
\label{subsec:Manufactured}

The analysis of two classical test cases via estimator is here proposed.
The dynamics are not represented by the Navier--Stokes equations, but simplified equation of diffusion and advection are considered.
Because of this aspect, the computational resources required to analyze these cases are very small and they allow for a complete investigation using the full Kalman filter and the two reduced order model proposed.
In addition, an exact solution is known, which will be perturbed via controlled noise to produce suitable observation and to provide an exact estimation of the covariance matrix $R$. 

\begin{figure}
\begin{tabular}{cc}  
\includegraphics[trim = 110pt 40pt 70pt 70pt, clip=true,width=0.5\textwidth]{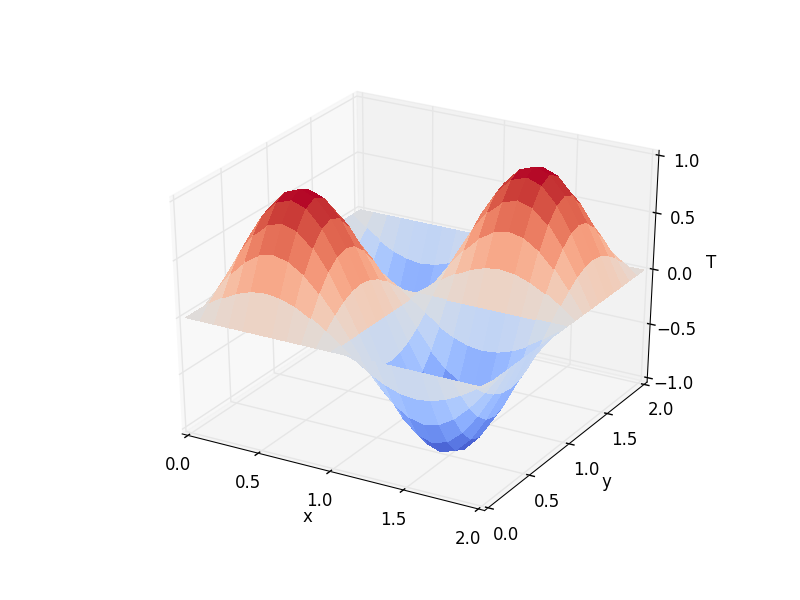} &
\includegraphics[trim = 110pt 40pt 70pt 70pt, clip=true,width=0.5\textwidth]{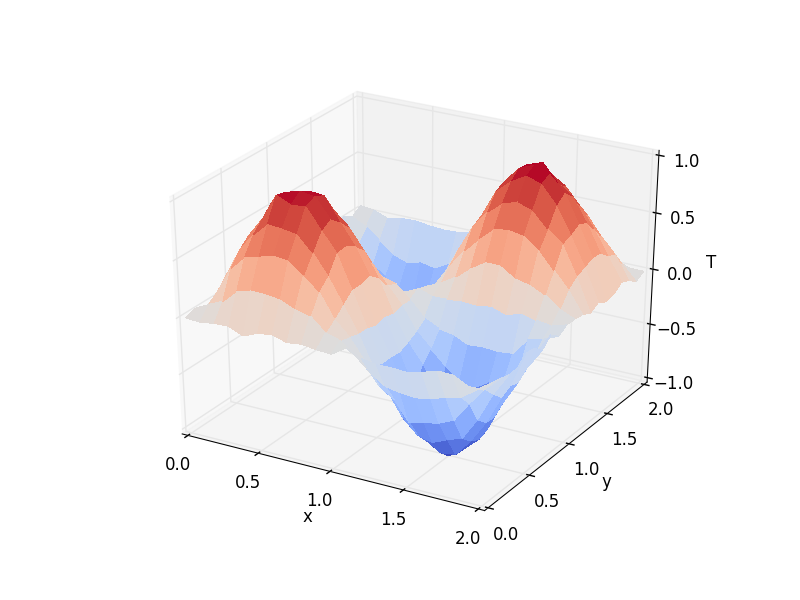} \\
(a) & (b)
\end{tabular}
\caption{Diffusion of a scalar $T$: (a) manufactured solution and (b) \textit{observation} for $t=1$.\label{diffusion}}
\end{figure}

The first test case investigated is a pure diffusion problem of a scalar $T$ in the two-dimensional domain $[0; 2]\times [0;2]$, which is illustrated in figure \ref{diffusion}:
\begin{eqnarray}
\label{eq:Diffusion}    
\partial_t T -\Delta T&=& g(t)
\end{eqnarray}  
$\Delta T$ is here the laplacian of $T$ and Homogeneous Dirichlet boundary conditions are imposed.
A suitable source term $g(t)$ serves a manufactured solution which is also used to set a non-zero initial value:
\begin{eqnarray}
\label{eq:ManufacturedSol1}    
T_{ref}(x,y,t)&=& sin(\pi x)sin(\pi y)cos(2\pi t)
\end{eqnarray}  
The numerical discretization of this case is performed via a finite volume scheme of second order precision in space and explicit Euler (first order) in time.

\begin{figure}
\begin{tabular}{cc} 
\includegraphics[trim = 110pt 40pt 70pt 70pt, clip=true,width=0.5\textwidth]{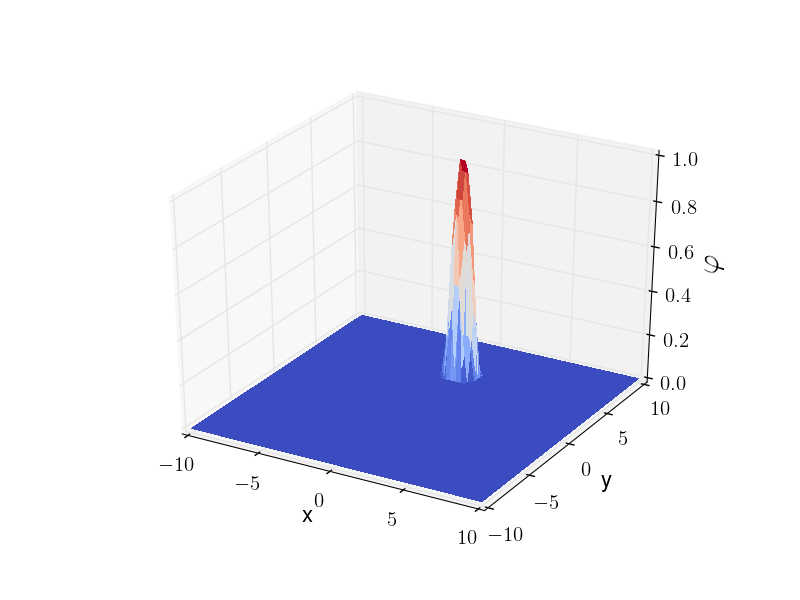} &
\includegraphics[trim = 110pt 40pt 70pt 70pt, clip=true,width=0.5\textwidth]{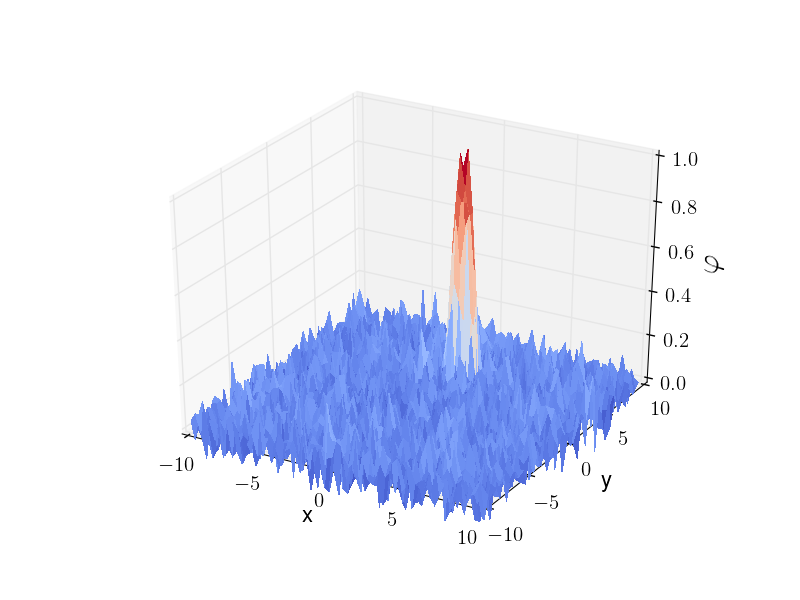} \\
(a) & (b)
\end{tabular}
\caption{Transport of a scalar $\varphi$: (a) manufactured solution and (b) \textit{observation} for $t=5$.\label{advective}}
\end{figure}

The second test case is a pure advection problem of a scalar $\phi$ in the two-dimensional domain $[-10; 10]\times [-10;10]$, which is shown in figure \ref{advective}:
\begin{eqnarray}
\label{eq:advection}    
\partial_t \varphi +\mathbf{u}.\nabla\varphi&=& 0
\end{eqnarray}  
with homogeneous Dirichlet boundary conditions.
A non-zero initial value for $\phi$ is imposed:
\begin{eqnarray}
\label{eq:ManufacturedSol2}    
\varphi_{ref}(x,y,t)&=& max\left[0, 1 - \left(x-5cos(\frac{\pi t}{10})\right)^2 - \left(y-5sin(\frac{\pi t}{10})\right)^2 \right]
\end{eqnarray}  
The advection velocity is defined as $\mathbf{u} = \frac{\pi}{10}(x \vec{e}_y - y \vec{e}_x)= r \frac{\pi}{10}\vec{e}_\theta$, where $(\vec{e}_r,\vec{e}_\theta)$ define a cylindrical coordinate system.
The numerical discretization used for this case is a first order upwind scheme for the spatial derivatives and explicit Euler in time.
This discretization scheme is known to be diffusive.
This feature will be exploited in order to magnify the quality improvement prediction obtained using the Kalman Filter.

\begin{figure}
\begin{tabular}{cc}  
\includegraphics[trim = 110pt 40pt 70pt 70pt, clip=true,width=0.5\textwidth]{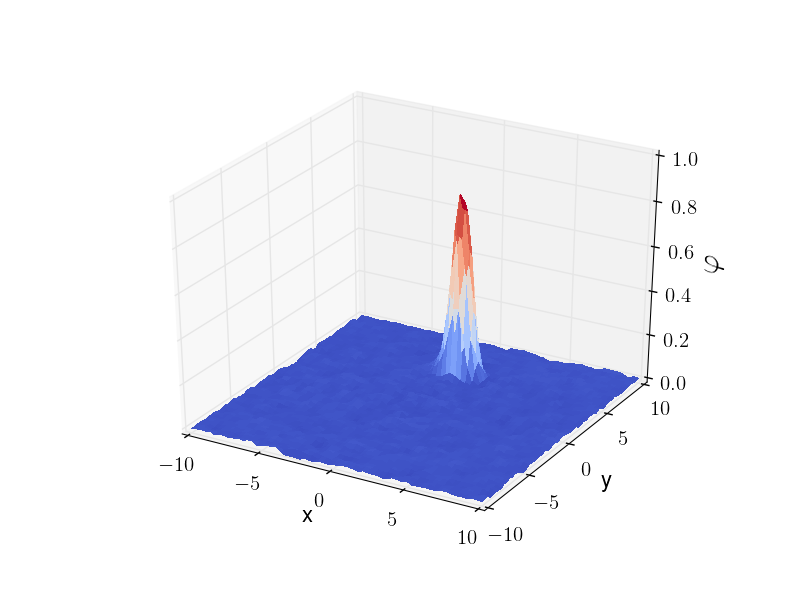} &
\includegraphics[trim = 20pt 0pt 40pt 40pt, clip=true,width=0.5\textwidth]{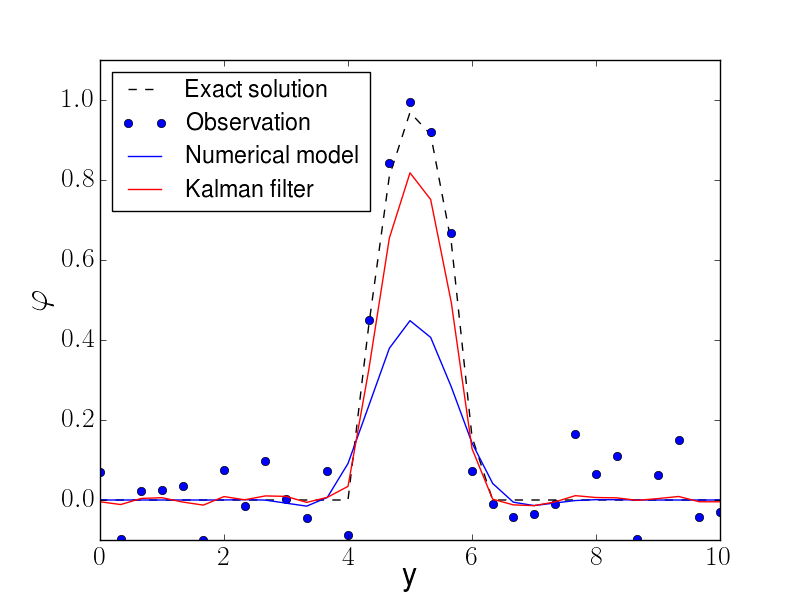} \\
(a) & (b)
\end{tabular}
\caption{Pure advection of a scalar $\varphi$.
(a) Numerical solution provided by application of Kalman filter for $t=1$.
(b) Solution for $x=0$ and $t=5$.
  \label{filter_illustration}}
\end{figure}

In both cases, the \textit{observation} is obtained via perturbation of the manufactured solutions imposing a Gaussian noise of variance $\sigma_r$.
As a consequence, the observation covariance matrix is $R = \sigma_r \, I$.
The model is starting using the manufactured solution as initial condition, so that the error covariance matrix is $P=0$.
However, information about the structure of the model covariance matrix $Q$ is not determined yet.
For this reason, it is initially hypothesized that $Q = \sigma_q \, I$.
A number of test has been performed, choosing different values of the mesh resolution and of the parameters $\sigma_r$ and $\sigma_q$.
Results are reported in figure \ref{filter_illustration} for the advection test case and in table \ref{L2-norms} for both test cases.
In the table, L2 norms are calculated comparing the exact solution with the observation, the model and the prediction by Kalman filter, respectively.
The error for the observation is clearly driven by the parameter $\sigma_r$, while the mesh resolution affect the accuracy of the model prediction.
The configuration obtained via Kalman Filter, which is sensitive to all of the three parameters, is potentially the most accurate.
This is particularly true when the level of confidence in the observation and in the model is similar.
In addition, the analysis of figure \ref{filter_illustration} indicates that Kalman Filter application can reduce drawback and biases associated with the model, such as the numerical diffusion of the upwind schemes using for numerical discretization.

 \begin{table}
\centering
\small
\begin{tabular}{|r|r|r|c|c|c|}
\hline
	&	&		&	\multicolumn{3}{|c|}{Error}	\\
Mesh size 	&	$\sigma_r$	&	$\sigma_q$	&	Observation	&	Model	&	Kalman Filter	\\
\hline
1.05e-01 & 5.91e-05 & 9.77e-07 & 2.17e-02 & 1.09e-02 & 4.56e-03 \\
1.05e-01 & 1.48e-05 & 9.77e-07 & 1.09e-02 & 1.09e-02 & 2.53e-03 \\
1.05e-01 & 3.69e-06 & 9.77e-07 & 5.43e-03 & 1.09e-02 & 2.08e-03 \\
6.90e-02 & 7.00e-06 & 5.41e-07 & 7.48e-03 & 7.47e-03 & 1.63e-03 \\
\hline
2.00e-01 & 1.69e-03 & 6.36e-05 & 1.49e+00 & 7.46e-01 & 3.48e-01 \\
2.00e-01 & 4.22e-04 & 6.36e-05 & 7.45e-01 & 7.46e-01 & 2.57e-01 \\
2.00e-01 & 1.05e-04 & 6.36e-05 & 3.73e-01 & 7.46e-01 & 1.99e-01 \\
1.50e-01 & 2.20e-04 & 2.15e-05 & 7.00e-01 & 7.02e-01 & 2.21e-01 \\
\hline
\end{tabular}
\caption{L2-norm of the error obtained for the observation, the model and the prediction via Kalman Filter when compared to the exact solution.
Parametric analysis of the error with respect to the observation covariance $\sigma_r$, the model covariance $\sigma_q$ and the mesh size is shown.
(top) The diffusion of a scalar $T$ and (bottom) the advection of a scalar $\phi$ are reported, respectively.
Due to the randomness of the Gaussian noise applied to the observation, the error has been averaged over a significantly large number of independent runs.\label{L2-norms}}
\end{table}

\begin{figure}
\centering
\includegraphics[trim = 0pt 0pt 0pt 0pt, clip=true,width=0.8\textwidth]{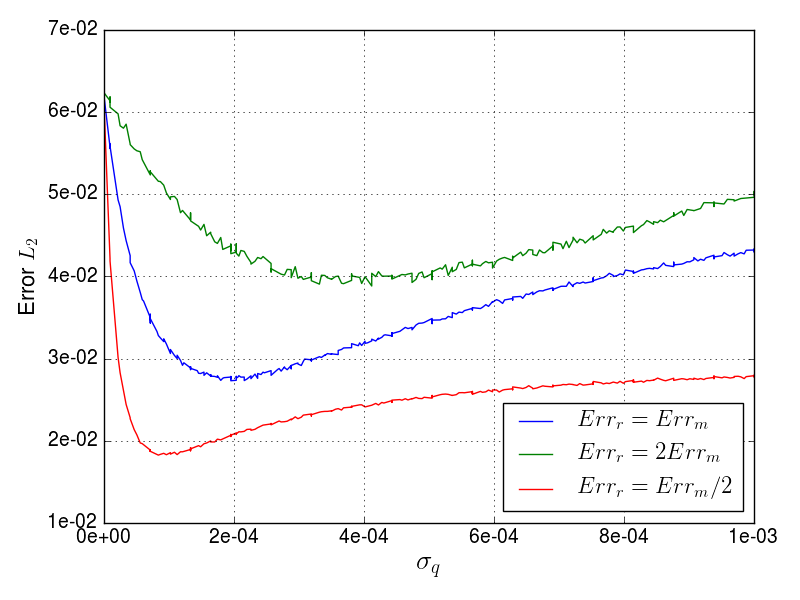}
\caption{Sensitivity of L2 norm of the error of the Kalman Filter prediction as a function of $\sigma_q$.
$Err_r = Err_m$ correspond to a choice of $\sigma_r$ such as the L2 norm the error of the observation is roughly the same as the L2 norm of the pure simulation.
The two other cases correspond respectively to a $\sigma_r$ twice as bigger or smaller.
\label{kalnoise_evolution}}
\end{figure}

\begin{figure}
\centering
\includegraphics[trim = 0pt 0pt 0pt 0pt, clip=true,height=0.65\textwidth]{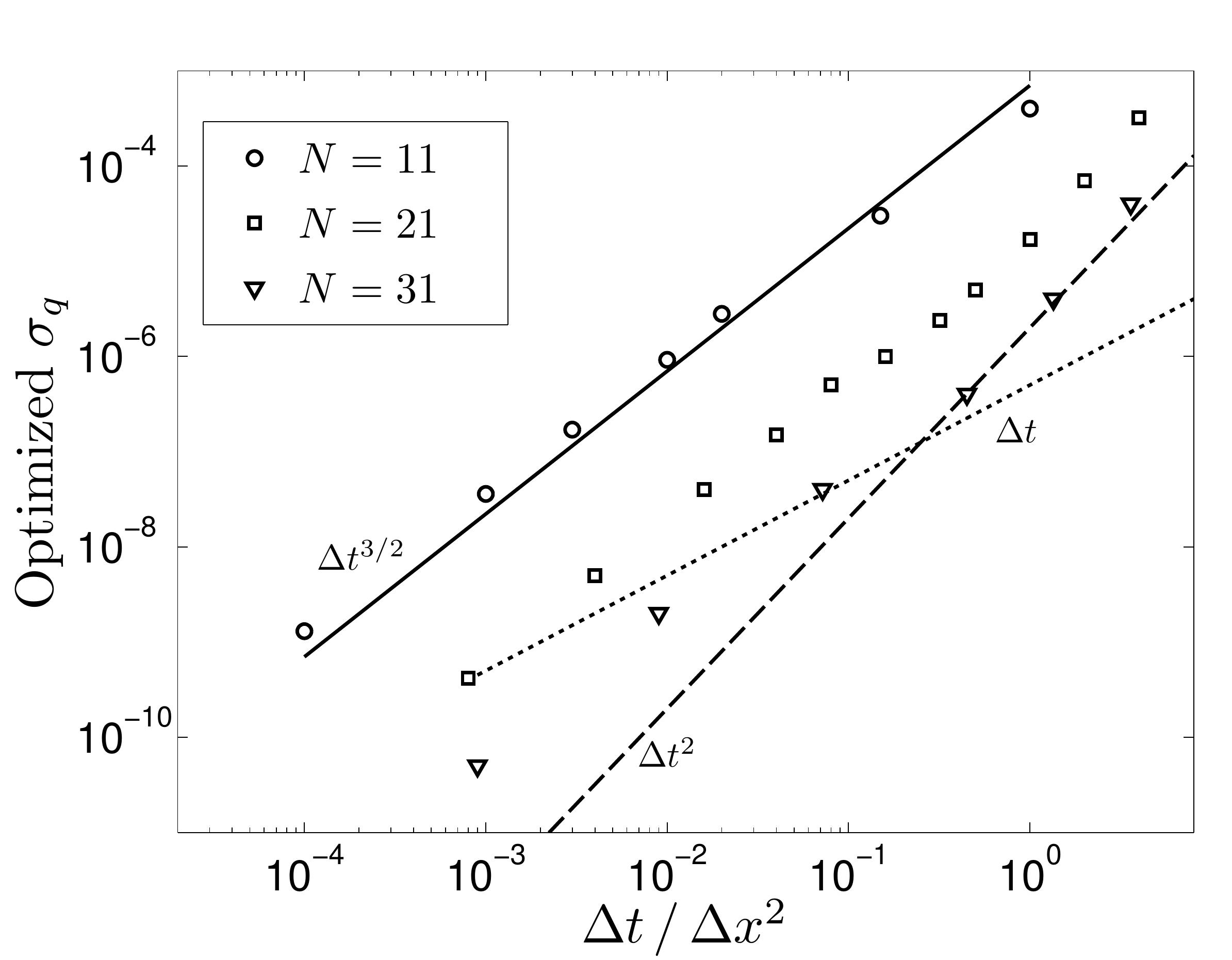}

\caption{Sensitivity of the optimized value of the model variance $\sigma_q$ to $\Delta t, \; \Delta x$. The test case of 2D diffusion is investigated.\label{truncature}}
\end{figure}

 While the definition of the covariance matrix $R$ is straightforward, the determination of the matrix $Q$ is problematic.
The covariance coefficient $\sigma_q$ is now optimized targeting the minimization of the difference between the prediction via Kalman Filter and the exact solution.
This procedure provides insights about the structure of the model covariance matrix $Q$, which is still largely unexplored in CFD applications.
The optimization algorithm is the gradient descent and three different test have been performed: i)confidence in the observation and the model roughly the same, ii) more confidence in the observation and ii) more confidence in the model.
Results are shown in figure \ref{kalnoise_evolution}.
The most important conclusion that can be drawn is that there is a linear correlation between the optimized value of $\sigma_q$ and the variance coefficient $\sigma_r$.
This observation origin from the behavior of the Kalman Filter, which is optimized when the level of confidence in the observation and model are roughly the same. 
The sensitivity of this algorithm has been extensively tested changing the mesh size and the time step $\Delta t$, as shown in figure \ref{truncature} for the pure diffusion test case.
One notable property is that the optimized value of $\sigma_q$ is directly linked with the troncature error of the numerical scheme multiplied by $\Delta t$.
In the case of the diffusion equation, this is equal to:

\begin{equation}
\label{eq:troncErrorDiff}
\Delta t \, E_T = \Delta t^2 \, f_t \left( \frac{\partial^2 T}{\partial t^2}|_{P} \right) + \, \Delta t \,  \Delta x^2 \, f_s \left( \frac{\partial^4 T}{\partial x^4}|_P \, , \, \frac{\partial^4 T}{\partial y^4}|_P \right)
\end{equation}  

where $f_t$, $f_s$ are linear functions of higher order derivatives of the scalar $T$, calculated in the mesh element $P$.
Three cases are analyzed in figure \ref{truncature}. For the first one, a coarse mesh of $N \times N = 11 \times 11$ elements has been used.
In this case, the optimized value of $\sigma_q$ scales as $\Delta t^{3/2} \cdot \Delta x^2$, which is a hybrid behavior of the two sources of error in equation \ref{eq:troncErrorDiff}.
Arguably, boundary effects govern the numerical prediction, considering that $33\%$ of the mesh is composed by boundary elements.
However, the early emergence of two distinct regimes can be observed for the case $N \times N = 31 \times 31$ ($12.4\%$ of boundary elements).
For high $\Delta t$ values, the optimized value of $\sigma_q$ scales as $\Delta t^2$ i.e. error due to the discretization of the time derivative.
As the time step is refined, the overall contribution of the time discretization becomes less and less dominant.
It appears that the limit $\sigma_q \propto \Delta t \cdot \Delta x^2$ is approached increasing the space resolution.
Thus, a transition towards a state where $Q$ is linked with the space discretization error is observed.

In conclusion, the results presented in figure \ref{truncature} actually indicate that the level of confidence in the numerical model is related with the discretization error.
This trend of correlation is logical, because the lack of precision in the numerical results, which represents the uncertainty of the model, is locally produced by the discretization error.
These findings are important because they indicate that the model uncertainty can be at least estimated using model information only.
The functions $f_t$ and $f_s$ can be locally calculated for each mesh element, when a suitable solution is provided.
In section \ref{sec:turbFlows} the determination of the structure of $Q$ for the analysis will be simplified, envisioning direct application without any preliminary information about the flow field.   

 \begin{table}
\centering
\small
\begin{tabular}{|r|r|r|c|c|c|}
\hline
	&	&		&	\multicolumn{3}{|c|}{Error}	\\
Mesh size 	&	$\sigma_r$	&	$\sigma_q$	&	full Kalman	&	filtering estimator	&	observer 	\\
\hline
1.05e-01 & 4.86e-04 & 2.04e-04 & 2.51e-02 & 2.74e-02 & \textbf{2.76e-02} \\
1.05e-01 & 4.86e-04 & 1.57e-04 & 2.38e-02 & \textbf{2.67e-02} & 2.86e-02 \\
1.05e-01 & 4.86e-04 & 1.18e-04 & \textbf{2.28e-02} & 2.88e-02 & 3.04e-02 \\
6.90e-02 & 4.69e-04 & 2.10e-04 & 2.37e-02 & 2.64e-02 & \textbf{2.66e-02}  \\
\hline
3.00e-01 &9.49e-04 &1.78e-04 &  \textbf{3.14e-01} &\textbf{3.10e-01} &\textbf{3.21e-01} \\
2.00e-01 &4.22e-04 &5.12e-05 &\textbf{2.59e-01} &2.97e-01 &3.10e-01\\
2.00e-01 &4.22e-04 &5.63e-05 &2.93e-01 &\textbf{2.56e-01} &3.01e-01\\
2.00e-01 &4.22e-04 &5.74e-05 &2.91e-01 &2.88e-01 &\textbf{2.68e-01}\\
\hline
\end{tabular}
\caption{L2-norms of the error observed for the diffusion case comparing full Kalman Filter, filtering estimator and observer to the exact solution.
Parametric analysis of the error with respect to the observation covariance $\sigma_r$, the model covariance $\sigma_q$ and the mesh size is shown.
Values in bold text correspond to cases where $\sigma_q$ is optimal.
 Due to the randomness of the Gaussian noise applied to the observation, the error has been averaged over a significantly large number of independent runs.\label{L2-norms2}}
 
\end{table}

 At last, the performances of the full Kalman filter, the filtering estimator and the observer have been assessed via comparison for the test cases presented in this section.
 The analysis has encompassed a large spectrum of configurations, including different mesh resolutions and values of $\sigma_r$ and $\sigma_q$.
 For the latter, both optimized and random values have been considered.
 The main results, which are reported in table \ref{L2-norms2}, indicate that three approaches are almost equivalent.
 In addition, optimized values of $\sigma_q$ for the three approaches are very similar as well, reinforcing the idea that the reduced-order methods can provide similar qualitative performance of the full Kalman filter. 
 
\section{Application to laminar test cases: flow around a circular cylinder}
\label{sec:experimentations}

Applications of the reduced order estimator to the analysis of laminar flow configurations are investigated in the presents Section.
The flow around a circular cylinder for $Re=100$ is considered.
This simple geometric case allows for the observation of an unstationary Von-Karman vortex street.
In addition, it is very well documented in the literature (see \cite{Henderson1996,Henderson1999,Henderson1995,Williamson1996,Norberg1993}).
The analysis is here focused on the application of the \textit{observer} introduced in Section \ref{sec:Observer}.

\subsection{Two-dimensional test case}
\label{subsec:2D_Cylinder}

This flow configuration exhibits a two-dimensional unstationary behavior at $Re=100$, so that the numerical domain investigated is initially 2D.
The physical quantities of reference are normalized, so that the asymptotic streamwise velocity in the direction $x$ is $\mathbf{u_{\infty}}=1$ and the diameter of the cylinder is $D=1$.
The resulting normalized kinematic viscosity is derived as $\nu=\mathbf{u_{\infty}} \, D / \, Re = 10^{-2}$.

\begin{figure}
\centering
\includegraphics[trim = 0pt 0pt 0pt 0pt, clip=true,height=0.4\textwidth]{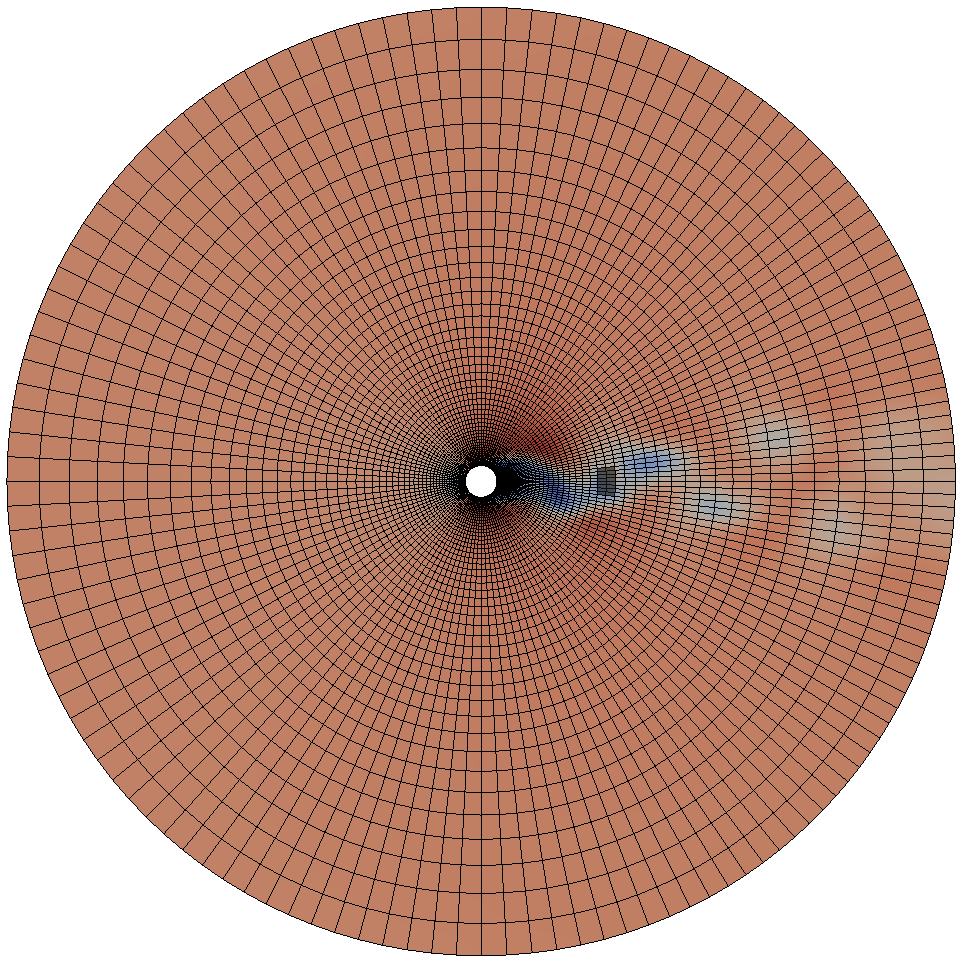}
\includegraphics[trim = 0pt 0pt 950pt 0pt, clip=true,height=0.4\textwidth]{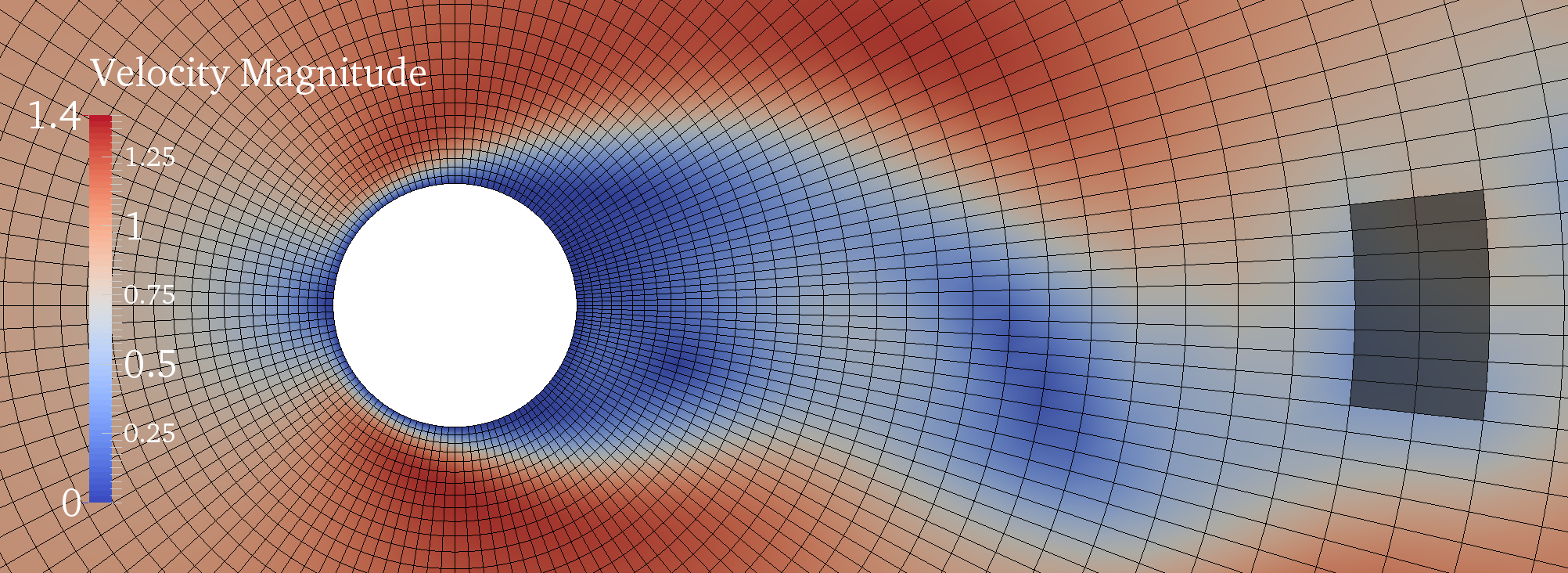}
\caption{Computational domain investigated for the flow around a circular cylinder, $Re=100$.\label{grid_cylindric}}
\end{figure}

The physical domain investigated is circular and defined by a diameter of $D_c=30$.
The inflow boundary conditions imposed is a Dirichlet condition for velocity $\mathbf{u_{}}=\mathbf{u_{\infty}}$.
A mass-conserving advective condition is imposed at the outlet.
A cylindrical coordinate system is chosen to perform the mesh discretization.
The physical domain is discretized in ${140}$ cells in the azimuthal direction and ${50}$ mesh elements in the radial direction, for a total of $7 \times 10^3$ elements (see figure \ref{grid_cylindric}).
A geometric distribution of the mesh elements has been chosen in the radial direction, in order to obtain a finer resolution at the wall.
A constant time step has been chosen for time advancement with ${\Delta t=0.03 t_A = 0.03 \, D / \mathbf{u_{\infty}}}$.
This corresponds to a maximum Courant number of $C_O=1.22$.

The aerodynamic coefficients investigated are defined as:
\begin{itemize}
 \itemsep=0pt
\item The lift coefficient ${C_L = \frac{2F_{y}}{\rho u_{\infty} H_D}}$
\item The drag coefficient ${C_D = \frac{2F_{x}}{\rho u_{\infty} H_D}}$
\end{itemize}

where ${F_{y}}$ and ${F_{x}}$ are the aerodynamic forces in the normal direction ($y$) and in the streamwise direction ($x$).
They include both pressure and viscous contributions.
The surface $H_D$ represents the area of the cylinder normal to the streamwise direction.
In addition $\rho=1$ in the normalized system.
A first test numerical simulation is performed, in order to get rid of the initial transient.
A fully developed configuration is then used as initial condition for the following simulations.

\begin{figure}
\centering
\includegraphics[trim = 100pt 70pt 0pt 70pt, clip=true,width=0.9\textwidth]{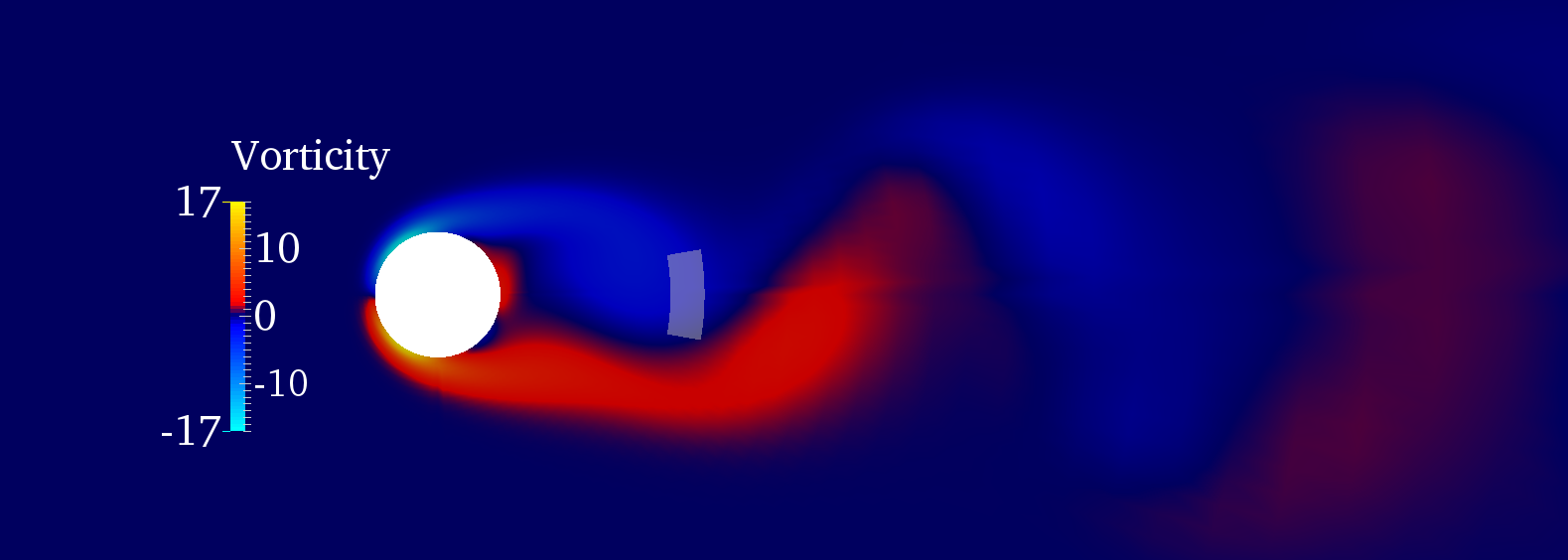}
\caption{Isocontours of the vorticity magnitude. The shaded area corresponds to the observation window chosen for integration of data in the estimator. \label{vortex_street}}
\end{figure}

\begin{figure}
\centering
\includegraphics[trim = 10pt 10pt 10pt 10pt, clip=true,width=\textwidth]{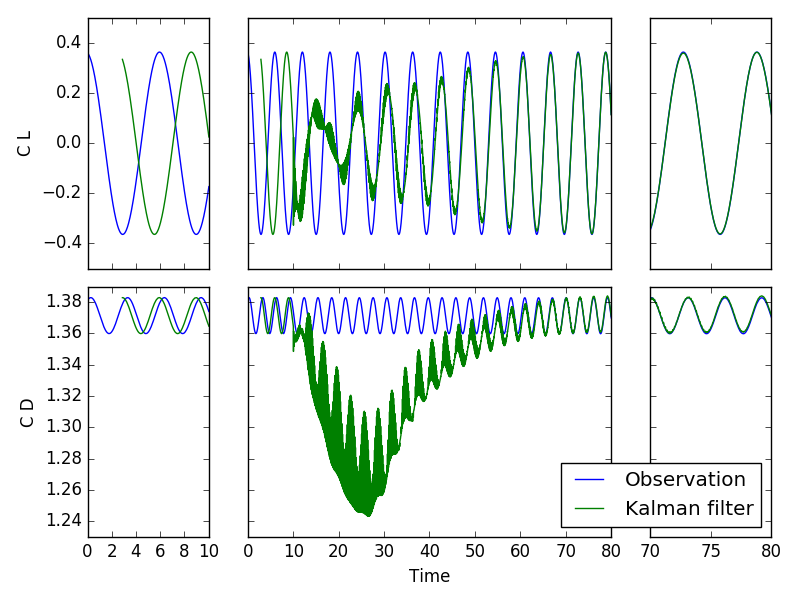}
\caption{Time evolution of the coefficients (top) $C_L$ (bottom) $C_D$ during Data Assimilation for the circular cylinder in 2D.\label{CL_vortex_street_2D}}
\end{figure}

\begin{figure}
\centering
\includegraphics[trim =100pt 70pt 0pt 70pt, clip=true,width=0.9\textwidth]{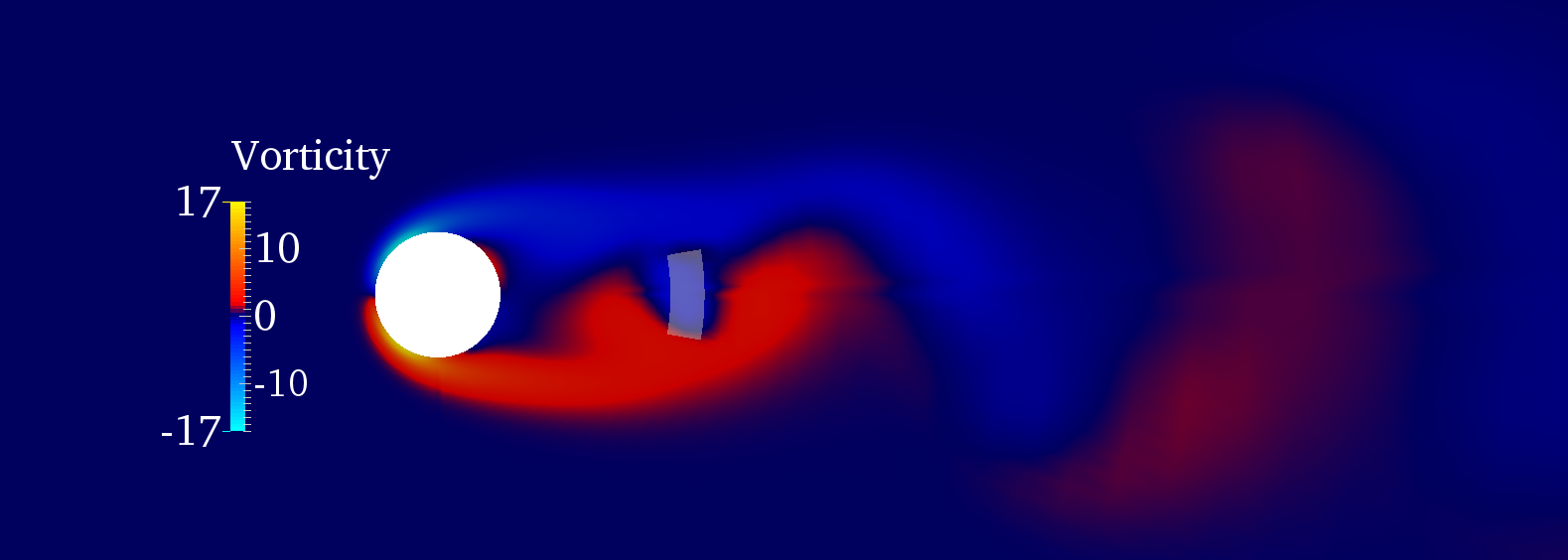}
\caption{Instantaneous vorticity isocontours during the first phases of Data Assimilation via estimator. The model acts as a forcing while synchronizing the numerical model to the observation. \label{async_vortex_street}}
\end{figure}

The strategy adopted for Data Assimilation completely relies on CFD for this case, using the same solver to produce observation and as a model.
The main goals of this analysis are to test the capabilities of the model to i) produce a smooth augmented solution and ii) influence the physical emergence of unsteady phenomena.
Traditionally, sequential method operate efficiently when observation is provided upstream.
On the contrary, they under-perform when information is provided downstream, with respect to variational approaches.
Thus, the observation window for this analysis consists of 24 mesh elements only, which are located in the wake area at approximately one full shedding cycle downstream (see figure \ref{vortex_street}).
Observation is performed via sampling of $u_x$ and $u_y$ produced by a first numerical simulation.
Data is collected every five time steps.
The error covariance matrix $P$ is initially set to zero, while the covariance matrices are chosen as $R= \sigma_r \, I$, $Q= \sigma_q \, I$.
In this case, $\sigma_r$ and $\sigma_q$ are not calculated, but we impose $\sigma_r = \sigma_q$ i.e. the level of confidence in the observation is assumed to be the same as the simulation.
Considering that the Data Assimilation is performed every five time steps, the diagonal values of the Kalman Gain will be slightly larger than $0.5$.
When the Data Assimilation is performed, the CFD model starts using the same initial configuration and solver as for the previous sampling, but the observation is introduced approximately in phase opposition starting from time $t=10$, as shown in figure \ref{CL_vortex_street_2D} for the aerodynamic coefficients.
After a suitable transient, the numerical model completely adjusts the phase of its shedding cycle with the observation by global flow modification.
An instantaneous velocity profile during this transition is shown in figure \ref{async_vortex_street}.
The estimator operates like a forcing in the flow field and the Poisson equation diffuses this result in all the physical domain.
This last result indicates that the sequential estimator is able to efficiently govern upstream flow dynamics when data is provided downstream in the wake.
In addition, data has been voluntarily included in a small section of the physical domain, in order to reproduce most of the occurrences where observation can be only locally provided.
These properties are essential for integration of experimental samples in numerical simulation.     

\subsubsection{Three-dimensional test case}

The flow around a circular cylinder for $Re=100$ is now studied over a three-dimensional domain.
The set-up of the simulation is very similar to the two case, with the following exceptions.
The length of the cylinder in the spanwise direction $z$ is $l=30$ and it is discretized in $20$ mesh elements.
Periodic boundary conditions are imposed on the lateral surfaces.

\begin{figure}
\centering
\includegraphics[trim = 10pt 10pt 10pt 10pt, clip=true,width=\textwidth]{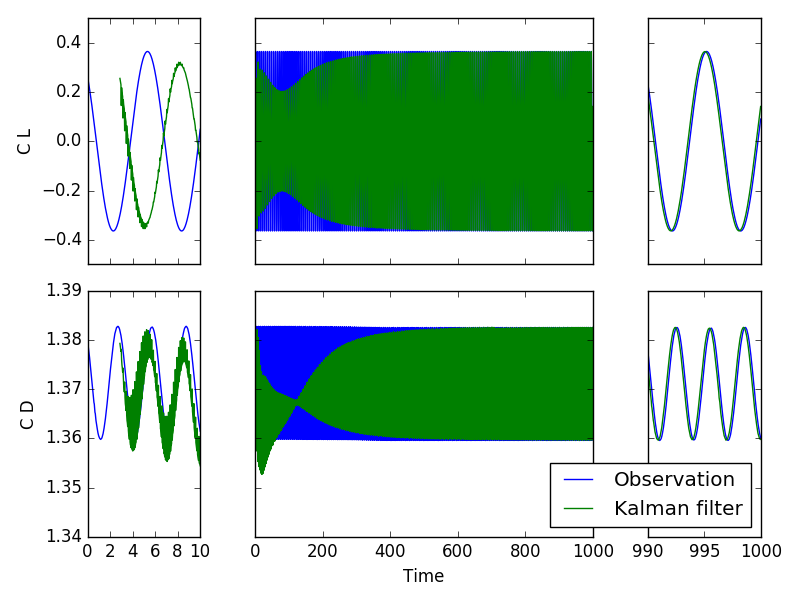}
\caption{Time evolution of the coefficients (top) $C_L$ (bottom) $C_D$ during Data Assimilation for the circular cylinder in 3D.\label{CL_vortex_street_3D}}
\end{figure}

\begin{figure}
\centering
\includegraphics[trim = 30pt 50pt 50pt 10pt, clip=true,width=0.9\textwidth]{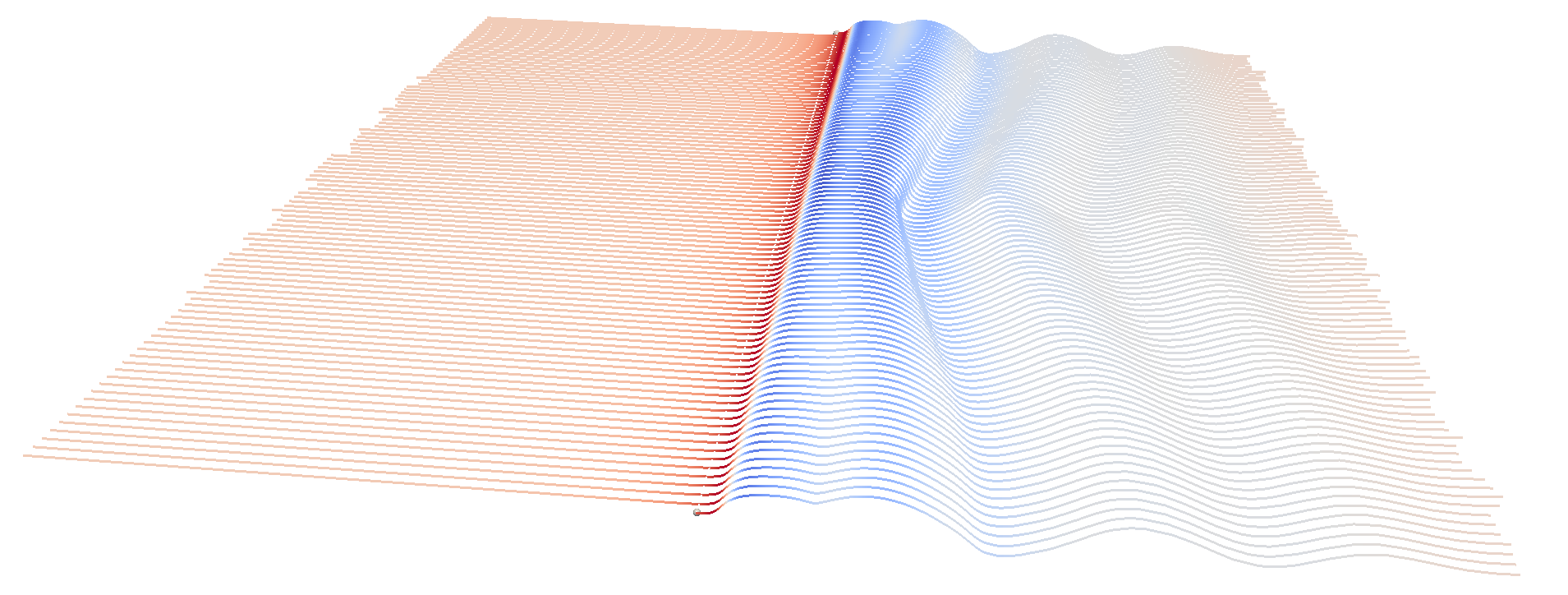}
\caption{Streamlines just above the cylinder showing the flow becomes three-dimensional as the information is propagated by the estimator.\label{async_vortex_street_3D}}
\end{figure}

The definition of the matrices $P$, $R$ and $Q$ is identical to the two-dimensional case, and the development of the analysis is exactly the same. 
However, a two-dimensional observation region has been conserved, which is located at the center of the domain with respect to the spanwise direction.
It also is wider in the streamwise direction (around 100 cells) in order to produce a faster synchronization.
The Data Assimilation process shows similar features to the correspondent two-dimensional case, as shown in figure \ref{CL_vortex_street_3D} for the coefficients $C_D$ and $C_L$.
However, the analysis of the streamlines sampled during the transient in figure \ref{async_vortex_street_3D} indicates that the information is successfully propagated in the spanwise direction, before a complete return to a two-dimensional flow is observed after the synchronization. 

\section{Augmented flow prediction for turbulent flow configurations}
\label{sec:turbFlows}

The reduced order estimator presented in section \ref{sec:model} is now applied to the study of turbulent flow configurations.
This goal is a fundamental research objective for a large spectrum of industrial applications, including transport engineering and environmental sciences.
Two different applications are presented, which deal with simplified flows commonly observed in complex applications.
The focus here has been on the prediction of statistical moments, which is a classical research objective in turbulence studies.
In both cases, the \textit{observer} version of the model has been used, in order to reduce the computational resources demanded.
The estimator required $\approx 10 \%$ resources more than the CFD model.
This result indicate that real-time application for the analysis of turbulent flows are going to be possible in the next decade, in particular for application of esatechnologies resources expected in the next years.
In addition, simplified formulae have been introduced for the estimation of the diagonal elements of the matrix $Q$.
These proposals envision a simple application for an external user, without the need to provide an accurate estimation of high order gradients of the physical quantities as in equation \ref{eq:troncErrorDiff}.
First of all, because of the need to use small time steps $\Delta t$ with respect to the average mesh size $\Delta l = \sqrt[3]{\Delta V}$, the error associated with time discretization is supposed to be negligible with respect to the space discretization.
For this reason, the elements of the matrix $Q$ are locally calculated for the mesh element $i$ as

\begin{equation}
Q_i = C \, \Delta t^{ot} \, \Delta l^{os}
\label{eq:Qapprox}
\end{equation}  
where $C$ is a constant chosen by the user and $ot$, $os$ are related with the order of the numerical schemes used for time / space discretization, respectively.
Clearly, the lower the value of $C$ and the higher the values of $ot$ and $os$, the higher the confidence level in the numerical tool used.

A second aspect that is here considered in the interaction between numerical error and turbulence modeling, which is one of the most problematic issues in particular for Large Eddy Simulation \cite{Ghosal1996_jcp}.
These non linear interactions are responsible for a very high level of uncertainty in the results of the numerical simulation and this challenging aspect is currently subject of investigation by numerous research teams.
In the spirit of a simple approximation, equation \ref{eq:Qapprox} is adapted for the application to reduced-order numerical simulation based on \textit{eddy viscosity} models \cite{Wilcox2006_DCW,Sagaut2006_springer}:

\begin{equation}
Q_i = C \, \left( 1 + \frac{\nu_T}{\nu} \right) \, \Delta t^{ot} \, \Delta l^{os}
\label{eq:QapproxTurb}
\end{equation}
where $\nu_T$ represents the turbulence / subgrid scale viscosity introduced by the model.
While equation \ref{eq:QapproxTurb} does not encompass application for all the turbulence / subgridscale models reported in the literature, it can be applied to most of the models implemented in commercial / open-source codes available.
The mechanics of equation \ref{eq:QapproxTurb} is simple and clear.
If the volume of the mesh element $i$ is large and the value of $\nu_T$ is high, the confidence level in the local performance of the numerical solver is severely penalized.

These techniques are now applied to the analysis of two flow configurations of industrial interest, namely the spatially evolving mixing layer and the flow around a thick plate.

\subsection{Spatially evolving mixing layer, $Re_{\Delta}=100$}
\label{subsec:MixingLayer}

\begin{figure}
 \includegraphics[width=0.8\linewidth]{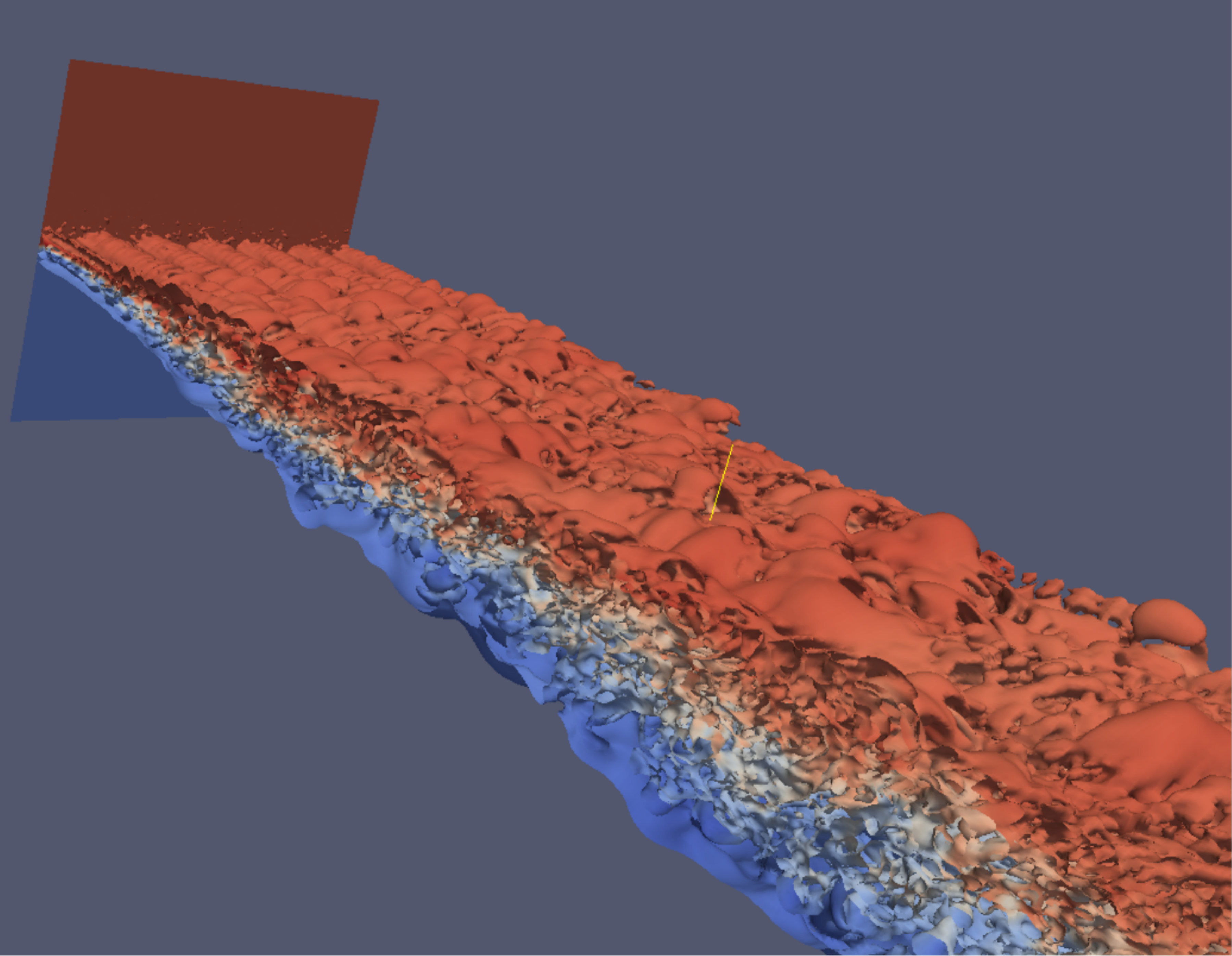}
\caption{Spatially evolving mixing layer flow configuration, $Re_{\Delta}=100$.
Isocontours of the $Q$ criterion colored by the velocity field, obtained by DNS.
The flow exhibit a progressive transition towards turbulence moving downstream.}
\label{fig:SpmixLayA}
\end{figure}

The spatial evolution of a mixing layer \cite{Colonius1997_jfm,Wang2007_jhf,McMullan2009_jhf,Meldi2012_pof} is here investigated.
This classical flow configuration exhibits Kelvin--Helmotz instability structures that are observed in many industrial flows, which emergence from the interaction of the two asymptotic flows described by the velocities $U_1$ and $U_2$, respectively.
A visualization of the case investigated and isocontours of the $Q$ criterion are reported in figure \ref{fig:SpmixLayA}.
If the Reynolds number is sufficiently high, the flow undergoes a turbulent transition sufficiently downstream form the inlet.
The location at which the emergence of turbulent features is observed can be controlled via features of the velocity profile imposed at the inlet.

\begin{figure}
\includegraphics[width=0.8\linewidth]{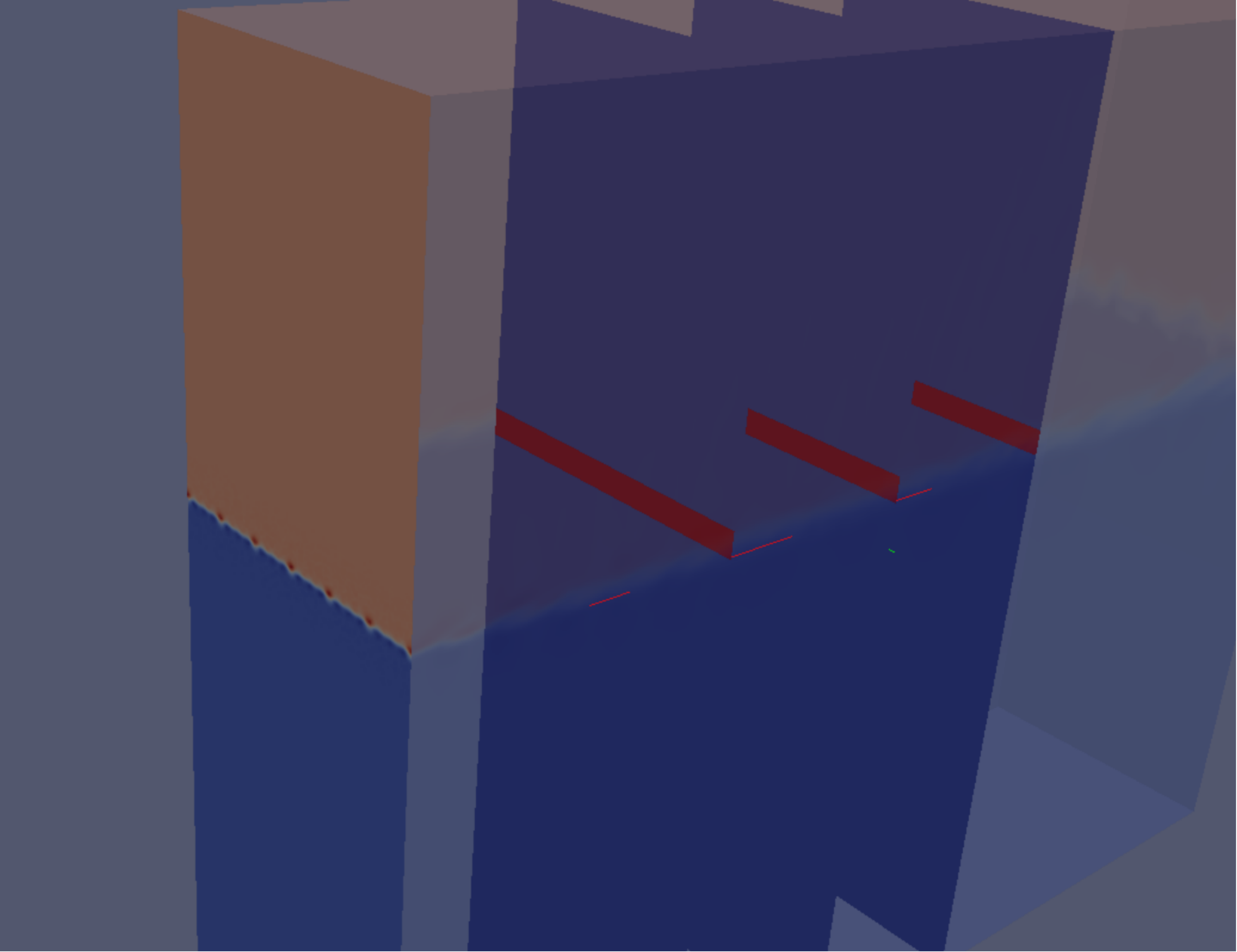}
\caption{Spatially evolving mixing layer flow configuration.
Instantaneous velocity profile imposed at the inlet, as well as the planes where observed data are provided (the exact location is highlighted in red).}
\label{fig:SpmixLayB}
\end{figure}

The flow configuration investigated is described by the following parameters:
\begin{itemize}
\item{$Re_{\Delta} = \frac{\delta_0 (U_1 - U_2)}{2 \nu} = 100$, where $\delta_0 = 5 \times 10^{-4}$ is the vorticity thickness at the inlet and $\nu = 2.5 \times 10^{-5}$ is the molecular viscosity.}
\item{$\alpha = \frac{U_1 - U_2}{U_1 + U_2} = 0.27$.
This parameter, which represents the ratio of shear vs advection effects, is investigated in the literature in the range $\alpha \in [0.2, \, 0.7]$. A value of $\alpha$ closer to the lower end has been chosen here in order to limit to a minimum the interaction of the wake with the boundary conditions in the normal direction.}
\end{itemize}  

The application of the estimator is completely numerical in this case.
In fact, the tools used to derive the augmented prediction are:
\begin{itemize}
\item{An LES solver as a \textit{model}.
Subgrid scale modeling is performed via the Smagorinsky model \cite{Smagorinsky1963_mwr}.}
\item{DNS samples have been used as \textit{observation}.}
\end{itemize}  

These tools are integrated in the \textit{observer} estimator, which will be referred to as DA-LES. The research activity is justified by the analysis of the performance of the Smagorinsky model, which is over-dissipative at the large scales of the flow \cite{Lilly1967_pIBM,Meldi2011_pof}.
In the case of the spatially evolving mixing layer, this features results in an over-prediction of the momentum thickness
\begin{equation}
\overline{\Theta}= \int_{y_m}^{y_M} \, (U_1 - \overline{u_x}) (\overline{u_x} - U_2) dy
\label{eq:momThick}
\end{equation}
where $\overline{u_x}$ is the time averaged streamwise velocity and $y_m$ / $y_M$ represent the lower and upper limit of the physical domain in the normal direction, respectively.
The over-prediction of $\overline{\Theta}$ for the Smagorinsky model is associated with an over-dissipation acting in the transition region of the flow, which is progressively carried downstream in the turbulent wake.
The underlying idea is to assess the capability of the estimator to correct this unfavorable characteristic of the Smagorinsky model via DNS observation in the transition region.
  
For every numerical simulation, the set-up is the following:
\begin{itemize}
\item{The physical domain in the streamwise, normal and spanwise directions $[x,\, y,\, z]$ is $[0, \, 24] \times [-9, \, 6] \times [-3, \, 3]$ in $\Lambda=15.4 \delta_0$ units.
This characteristic length is the principal instability length calculated via linear stability theory.}  
\item{The numerical schemes used for discretization of spatial derivatives are centered second-order schemes.
For the time derivative, a second-order implicit backward differencing scheme has been adopted.}
\item{Periodic conditions have been imposed on the lateral boundaries (planes $x-y$), while traction-free conditions have been imposed on the normal boundaries (planes $x-z$).
An advective mass conserving condition has been imposed at the outlet.
An hyperbolic tangent profile for the velocity field has been imposed at the inlet.
This profile can be written as the sum of three different components $u_{INLET} = u_{th} + u_{sp} + u_{tp}$ where:
\begin{enumerate}
\item{$u_{th} = 0.5 \, (U_1 + U_2) + 0.5 \, (U_1 - U_2) \, tanh (2y / \delta_0)$ is an hyperbolic tangent velocity profile.
This component is imposed for $u_x$ only.}
\item{$u_{sp}= 0.005 \, (U_1 + U_2) \sum\limits_{g=1}^3 cos(2 \, \pi \, g \, z \, / \Lambda) exp(-0.5 \, (y / \delta_0)^2)$ is a spatial modal perturbation, which is exponentially damped moving away from the center-line.
This perturbation is applied to all the velocity components.}
\item{$u_{tp}= 0.02 \, (2 \, rand(t)[-0.5, \, 0.5]) \, (u_{th} + u_{sp}) \, exp(-0.5 \, (y / \delta_0)^2)$ is a white noise time perturbation of maximum intensity equal to $2\%$, which is exponentially damped moving away from the center-line.}  
\end{enumerate}
The velocity profile is represented in figure \ref{fig:SpmixLayB}.
Using this inlet condition, the transition to turbulence is observed at a streamwise distance of $x \approx 10 \Lambda$ \cite{Meldi2012_pof}.
}
\item{The DNS mesh is structured in $384 \times 163 \times 160$ elements.
The distribution in the streamwise $x$ direction and in the spanwise $z$ direction is uniform.
In the normal direction, 64 uniform mesh elements are used to represent the range $y \in [-\Lambda, \, \Lambda]$.
Outside of this range, a local stretching ratio of $\approx 1.1$ is used as the elements become coarser reaching the boundaries in the normal direction.
The resulting computational mesh is composed by $10^7$ elements.
At the center-line, the resolution is estimated to be $[5.86 \eta, \, 2.93 \eta, \, 3.5 \eta]$, where $\eta$ is the Kolmogorov scale calculated using the asymptotic turbulent Reynolds number.
For the LES simulations the mesh is made of $ 96 \times 89 \times 40$ elements.
The distribution pattern is very similar to the DNS strategy, but the resolution in the streamwise and spanwise direction has been reduced by a factor of $4$.
In the normal direction for $y \in [-\Lambda, \, \Lambda]$ the coarsening ratio imposed is equal to $2$.} 
\item{The time step for each numerical simulation has been set to $\Delta t = \frac{t_A}{100}= \frac{\Lambda}{50 \, (U_1+U_2)}$, where $t_A$ is the characteristic advection time.
The maximum Courant number observed is $ Co \approx 0.35$ for the DNS calculation and $Co \approx 0.08$ for the LES.
Each numerical simulation has been run for $20000$ times steps, for a total simulation time $T = 200 t_A$.}
\item{The initial condition for every simulation is a fully developed configuration calculated by a preliminary DNS.
This velocity field has been interpolated on the LES grid.} 

\end{itemize}

\begin{figure}
\begin{tabular}{cc}
 \includegraphics[width=0.48\linewidth]{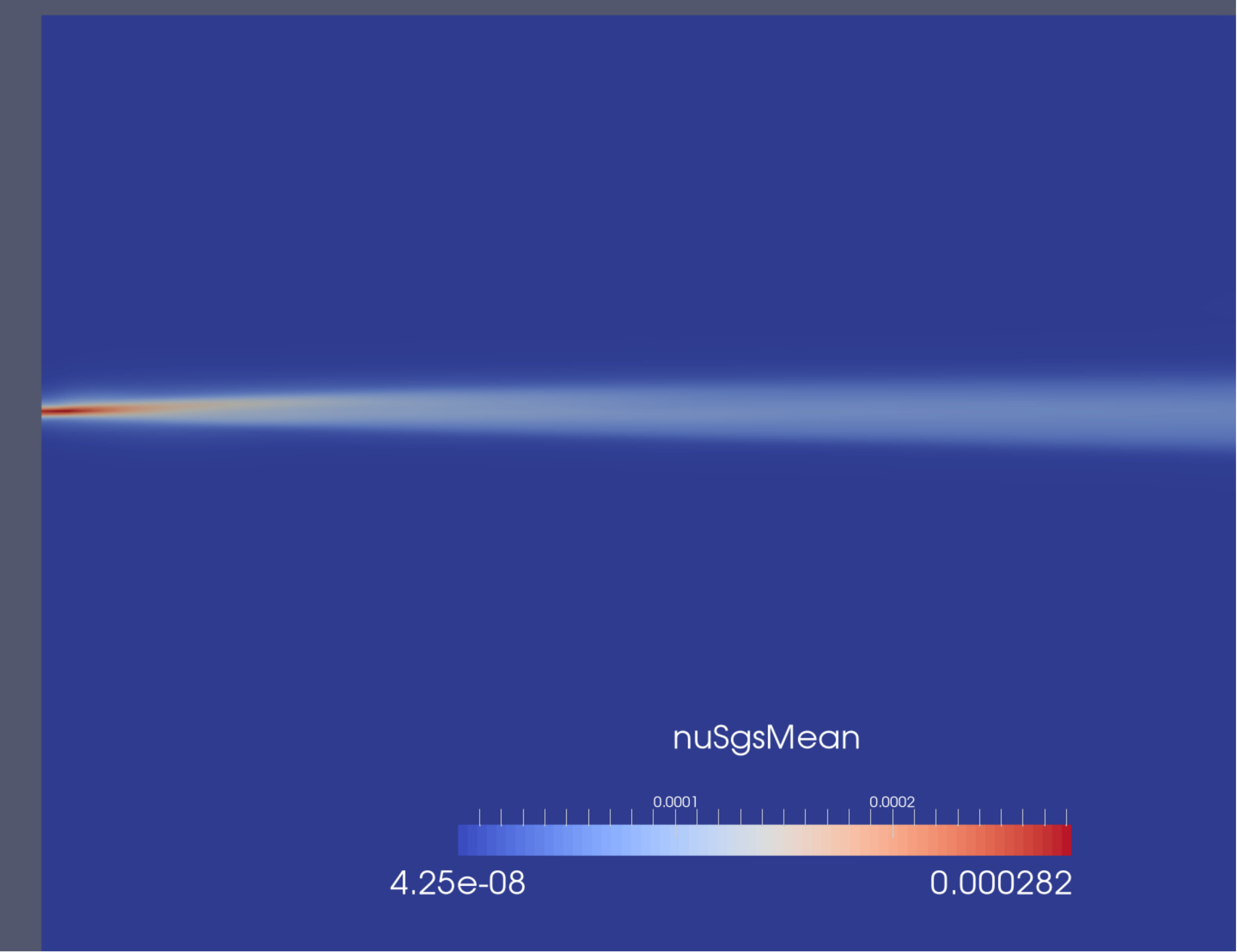} & \includegraphics[width=0.48\linewidth]{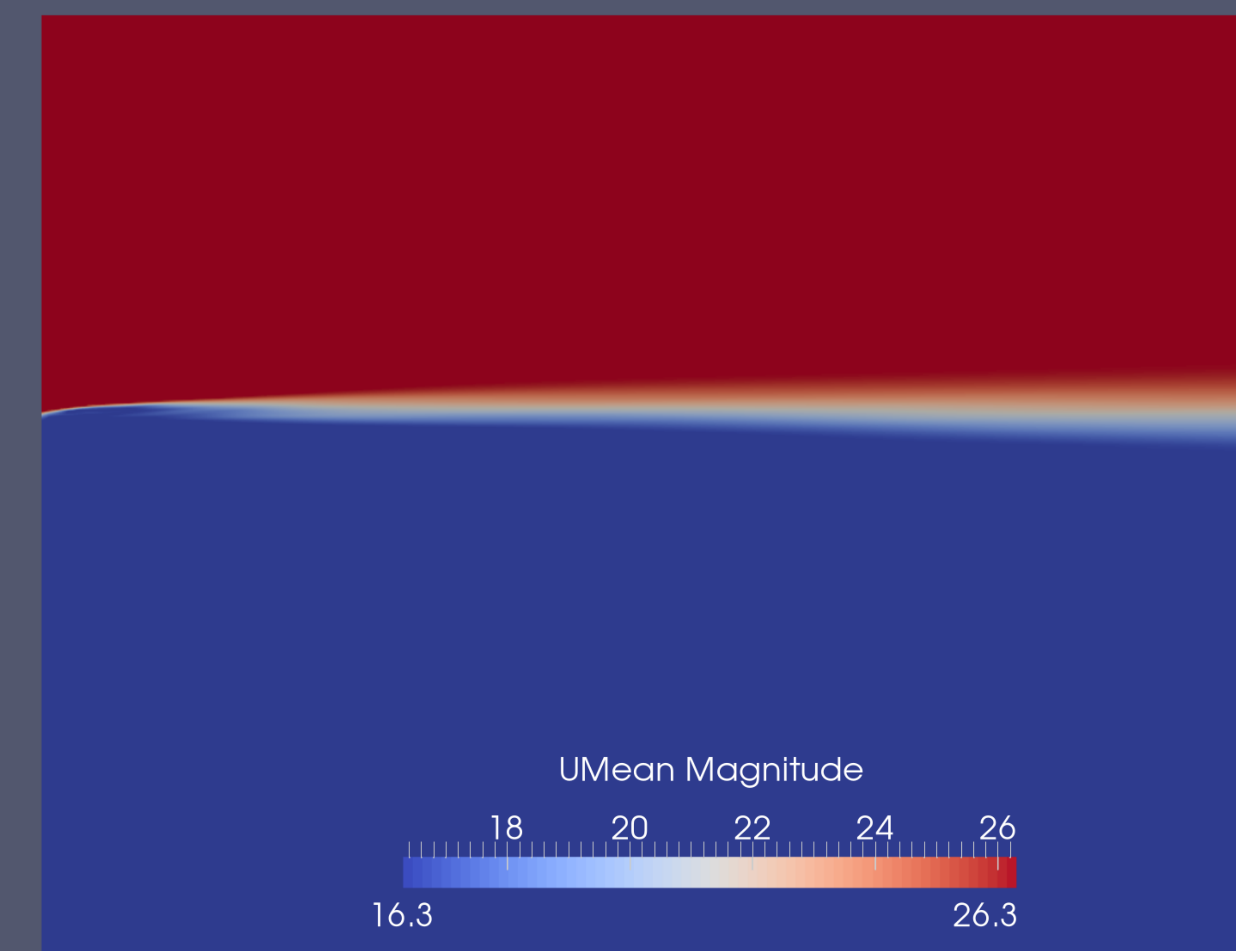} \\
 (a) & (b) \\
 \includegraphics[width=0.48\linewidth]{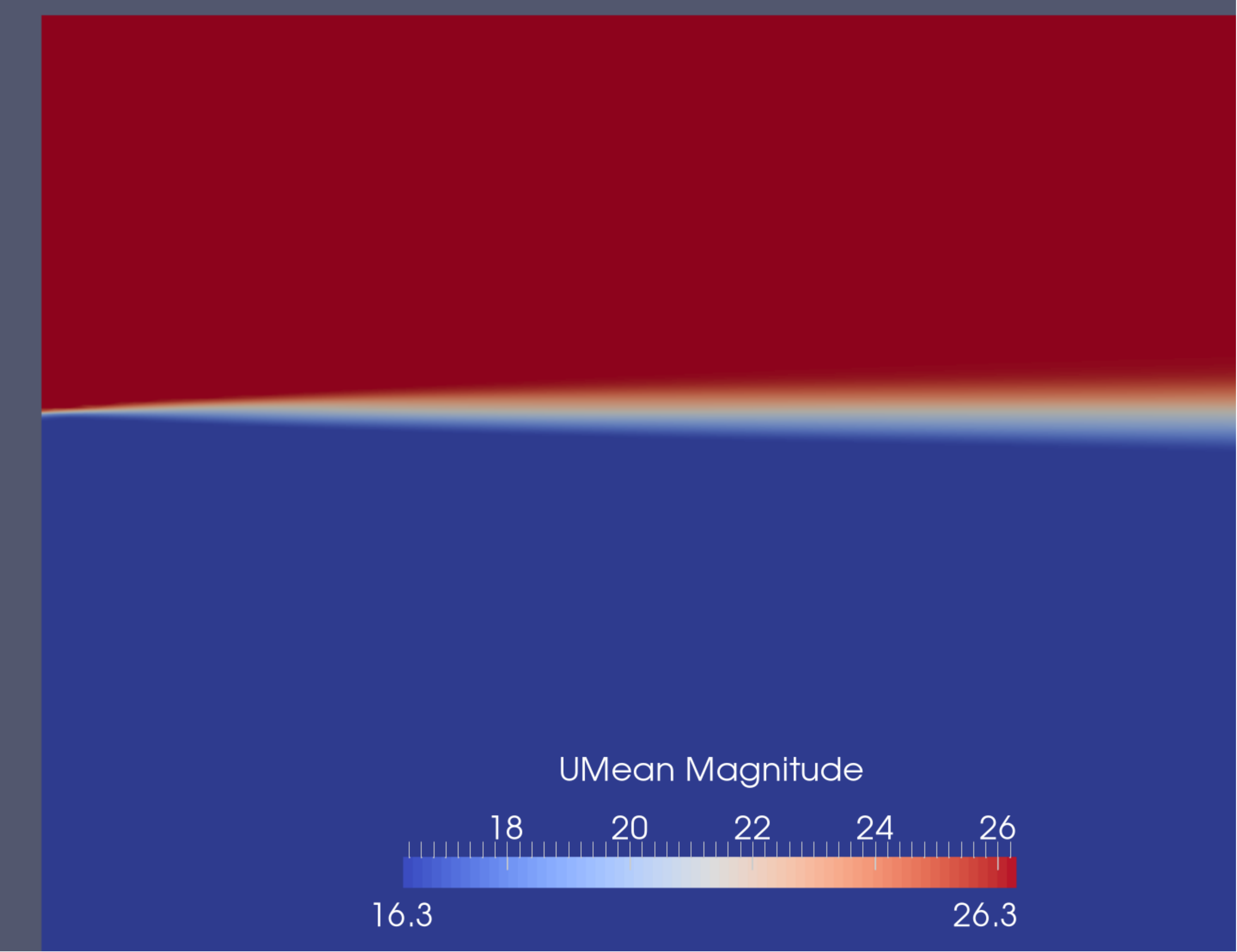} & \includegraphics[width=0.48\linewidth]{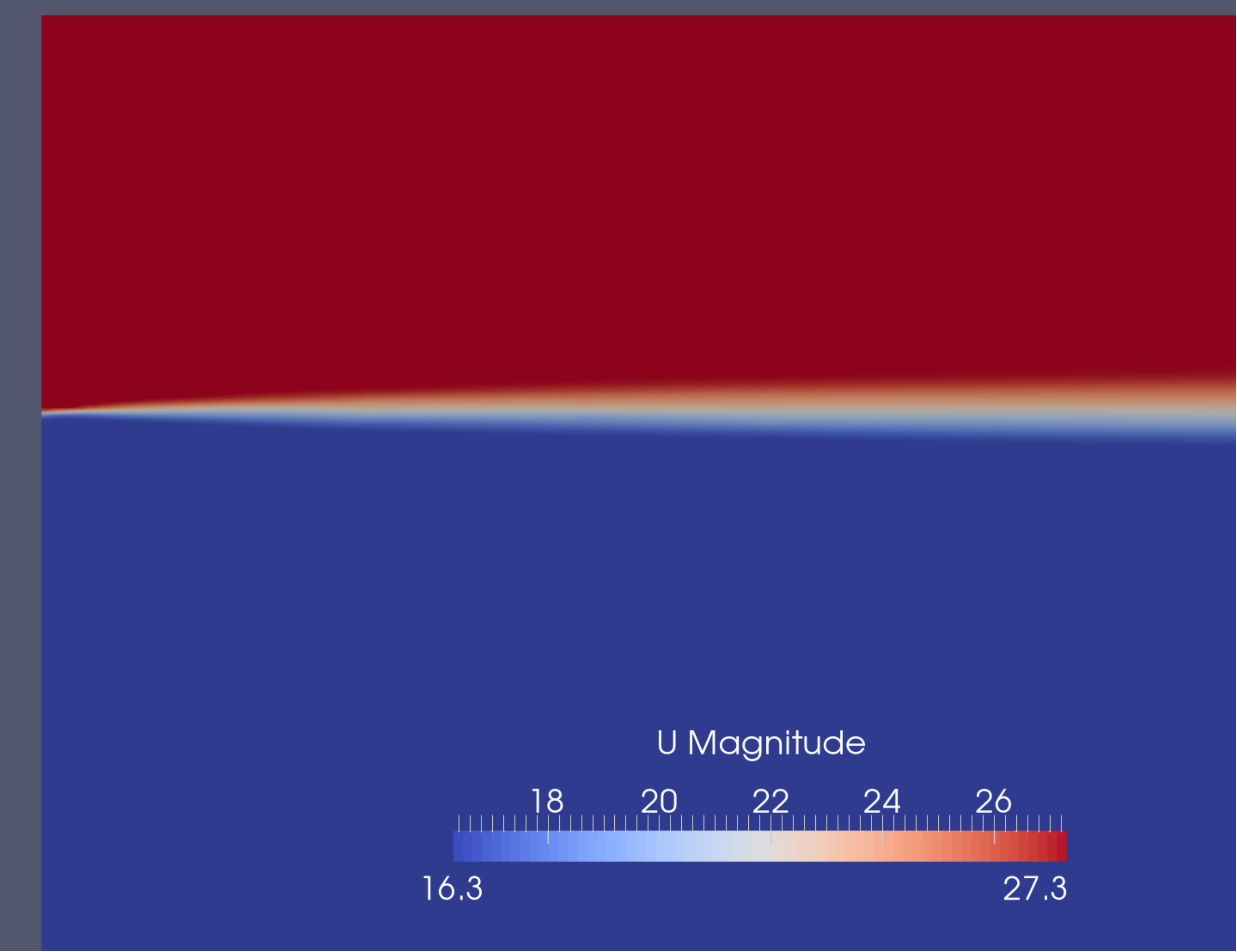} \\
	(c) & (d)
\end{tabular}
\caption{(a) Visualization of the averaged subgrid velocity $\nu_T$ calculated by the Smagorinsky model in the LES simulation.
(b-d) Isocontours of the time average velocity $\|\overline{\mathbf{u}} \|$ on a plane normal to the spanwise direction $z$.
The results are shown for (b) the DNS calculation, (c) the LES simulation and (d) the observer.}
\label{fig:SpmixLayIso}
\end{figure}

\begin{figure}
\begin{tabular}{cc}
 \includegraphics[width=0.48\linewidth]{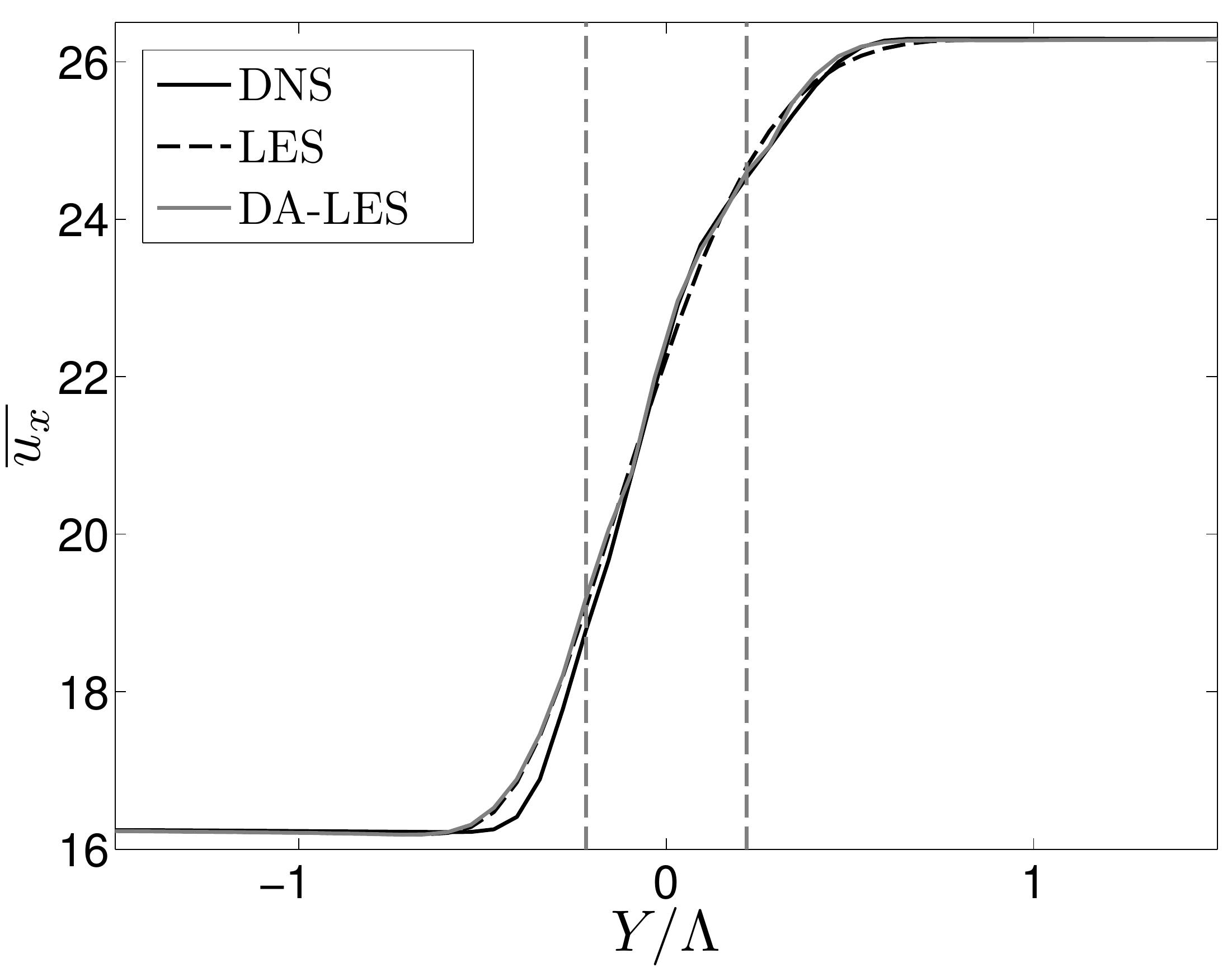} & \includegraphics[width=0.48\linewidth]{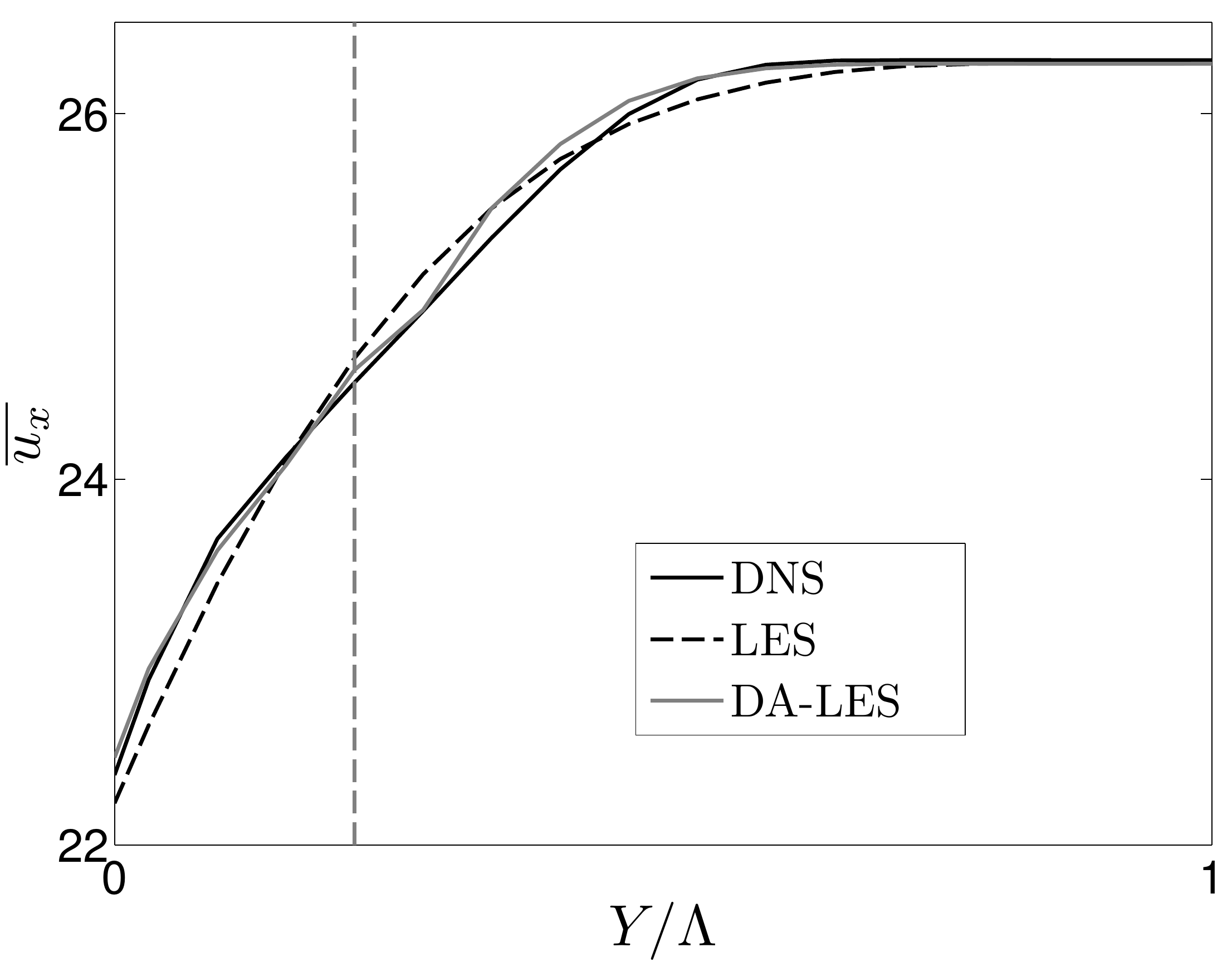} \\
 (a) & (b) \\
 \includegraphics[width=0.48\linewidth]{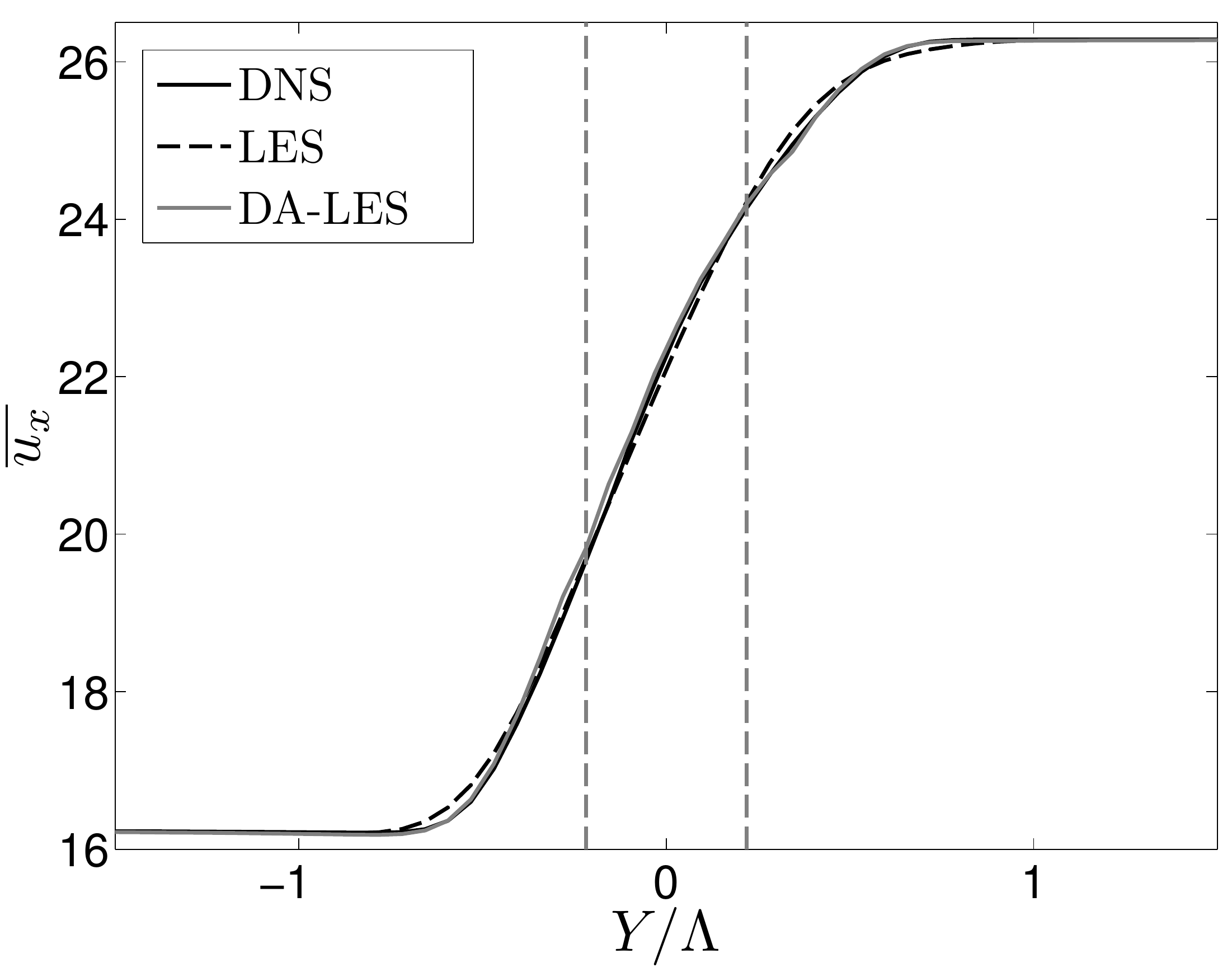} & \includegraphics[width=0.48\linewidth]{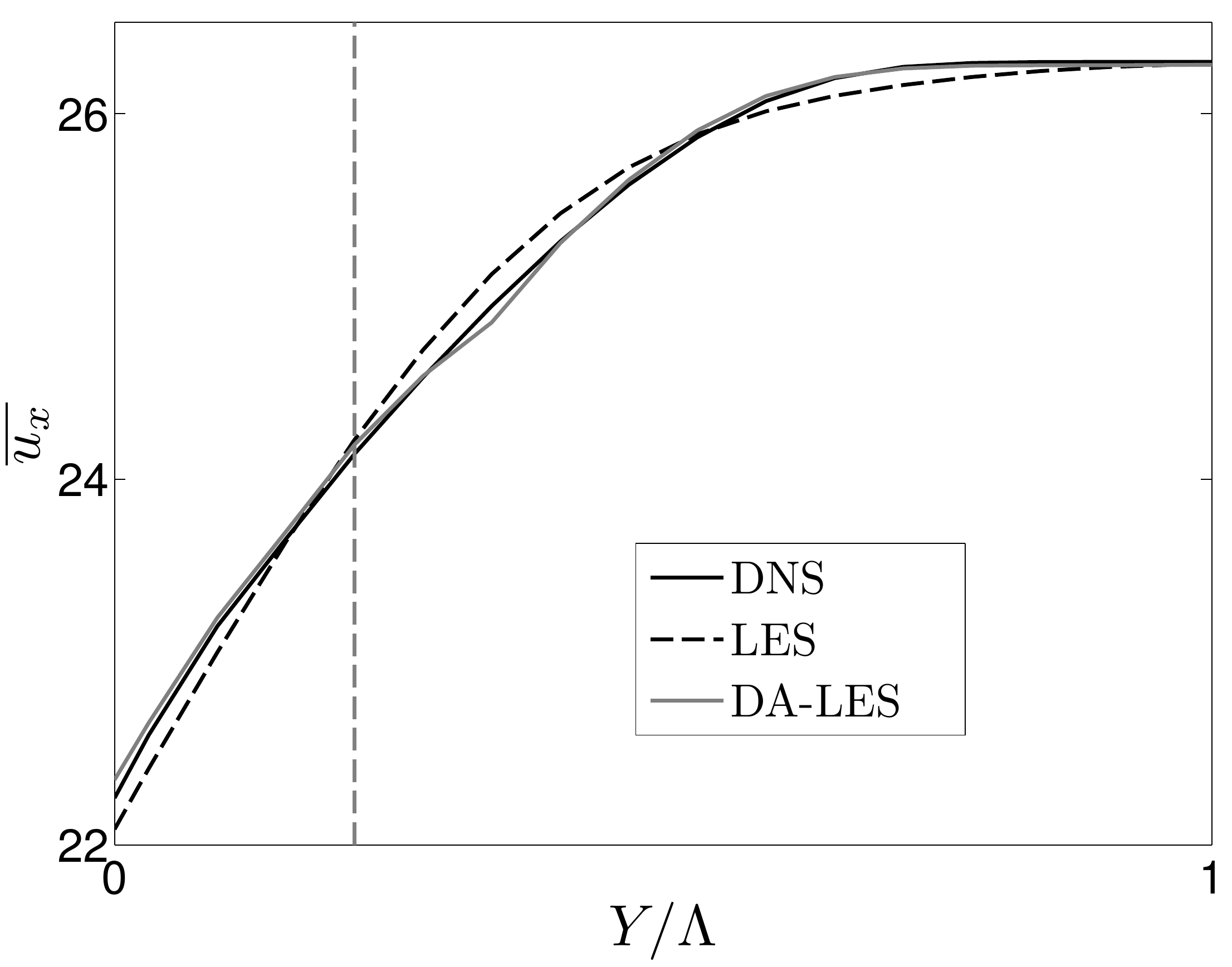} \\
	(c) & (d)
\end{tabular}
\caption{Averaged streamwise velocity $\overline{u_x}$ profiles.
Averages are performed in the spanwise direction and over $T = 200 \, t_A$ times.
The results are shown for (a-b) the streamwise section for $x = 4 \, \Lambda$ i.e.
in the transition region and for (c-d) the streamwise section for $x = 16 \, \Lambda$ where the flow exhibits turbulent features.}
\label{fig:SpmixLayVelProf}
\end{figure}

The choice of the structure of the covariance matrices $Q$ and $R$ is now discussed.
For the LES model, equation \ref{eq:QapproxTurb} is used to determine the value of the elements of the matrix $Q$:  

\begin{equation}
\label{eq:Qvalue_ML}
Q_i = C_{ML} \, \left( 1 + \frac{\nu_T}{\nu} \right) \, \sqrt[3]{{\Delta V}^2} \, {\Delta t}  
\end{equation}
in this case, the confidence level in LES results is clearly lower in the wake region, where the Smagorinsky model introduces the eddy viscosity effect (see figure \ref{fig:SpmixLayVelProf} (a) for time averaged isocontours).
This feature will be exploited by the estimator, in order to limit the unwanted effects of the model in the transition region.
A very similar structure is chosen for the elements $R_i$ of the covariance matrix of the observation by DNS using equation \ref{eq:Qapprox}:

\begin{equation}
\label{eq:Rvalue_ML}
R_i = C_{ML} \, \sqrt[3]{{\Delta V}^2} \, {\Delta t}  
\end{equation}

The DNS observation is integrated in the LES estimators in three different planes normal to the streamwise direction $x$ for $x = 6 \Lambda, \, 10 \Lambda, \, 14\Lambda$.
A visual representation is reported in figure \ref{fig:SpmixLayB}.
 For each plane, the assimilation is limited to $y \in [-1.2 \times 10^{-3} \Lambda, 1.2 \times 10^{-3} \Lambda], \, z \in [-3 \Lambda, \, 3 \Lambda]$, which includes the eight elements closer to the center-line in the normal direction and all the elements in the spanwise direction.
Thus, the observation is provided in $960$ elements of the LES mesh, which corresponds to $\approx 0.3\%$ of the total number of elements.
The choice of the observation points is performed accordingly to previous considerations about the transition of the flow to a fully turbulent state.
In this scenario, the DNS data is expected to improve the performance of the LES solver in the transition region, where the LES performance is supposedly sub-optimal.
It is important to remind that in this case the subgrid scale model is not modified, but that the velocity field is updated comparing the LES and the DNS fields.
The observation has been sampled every 10 time steps, which implies that DNS data are assimilated 2000 times during the simulation.
DNS data are interpolated on the centers of the elements of the LES grid for easier application, and Data Assimilation is performed for all the three components of the instantaneous velocity field.
As the initial physical field introduced in the estimator is interpolated from DNS results, the initial value of the error covariance matrix $P$ is set to zero.
Considering equations \ref{eq:Qvalue_ML} and \ref{eq:Rvalue_ML} and that the time step for the assimilation is 10 times larger than $\Delta t$, one can easily deduce that the diagonal elements of the optimal gain $K$ will be very close to the unity for each velocity component.
This is technically not the optimal application scenario for the estimators, which has proven to perform best when the confidence in the model and in the observation are similar.
Thus, this is a good benchmark to test critical issues of the DA model when applied to the simulation of turbulent flows.

Isocontours of the time-averaged velocity magnitude are reported in figure \ref{fig:SpmixLayIso} (b-d), for the three numerical simulation.
The velocity profiles are very similar, but it can be observed that the transition between the two asymptotic regimes seems to be somehow faster for the LES results in figure \ref{fig:SpmixLayIso} (c).
This actually corresponds to the prediction of a thicker $\overline{\Theta}$, as expected.
This observation is confirmed by the analysis of results in figure \ref{fig:SpmixLayVelProf} for $x= 4 \Lambda$ and $x = 16 \Lambda$.
Here, the average streamwise velocity profiles $\overline{u_x}$ are shown.
Averages are performed in time and in the spanwise direction.
These two sections have been chosen as they are sufficiently upstream / downstream with respect of the first / last assimilation plane, respectively.
The velocity profiles indicate that the estimator is closer on average to the DNS results, even far from the center-line where observation is available for $x= 6, \, 10, \, 14 \, \Lambda$ (gray dashed vertical lines).
This trend is observed for $x = 4 \, \Lambda$ and for $x = 16 \, \Lambda$, downstream with respect to the last assimilation window.
It appears as well that the convergence of the averaged results for the estimator is slower if compared with the pure LES case, but it is more in line with DNS.
Thus, the present results seem to indicate that consistent average features of turbulent flows can be obtained via this sequential estimator, even including a limited amount of observation data.   

\begin{figure}
\begin{tabular}{ccc}
\includegraphics[width=0.33\linewidth]{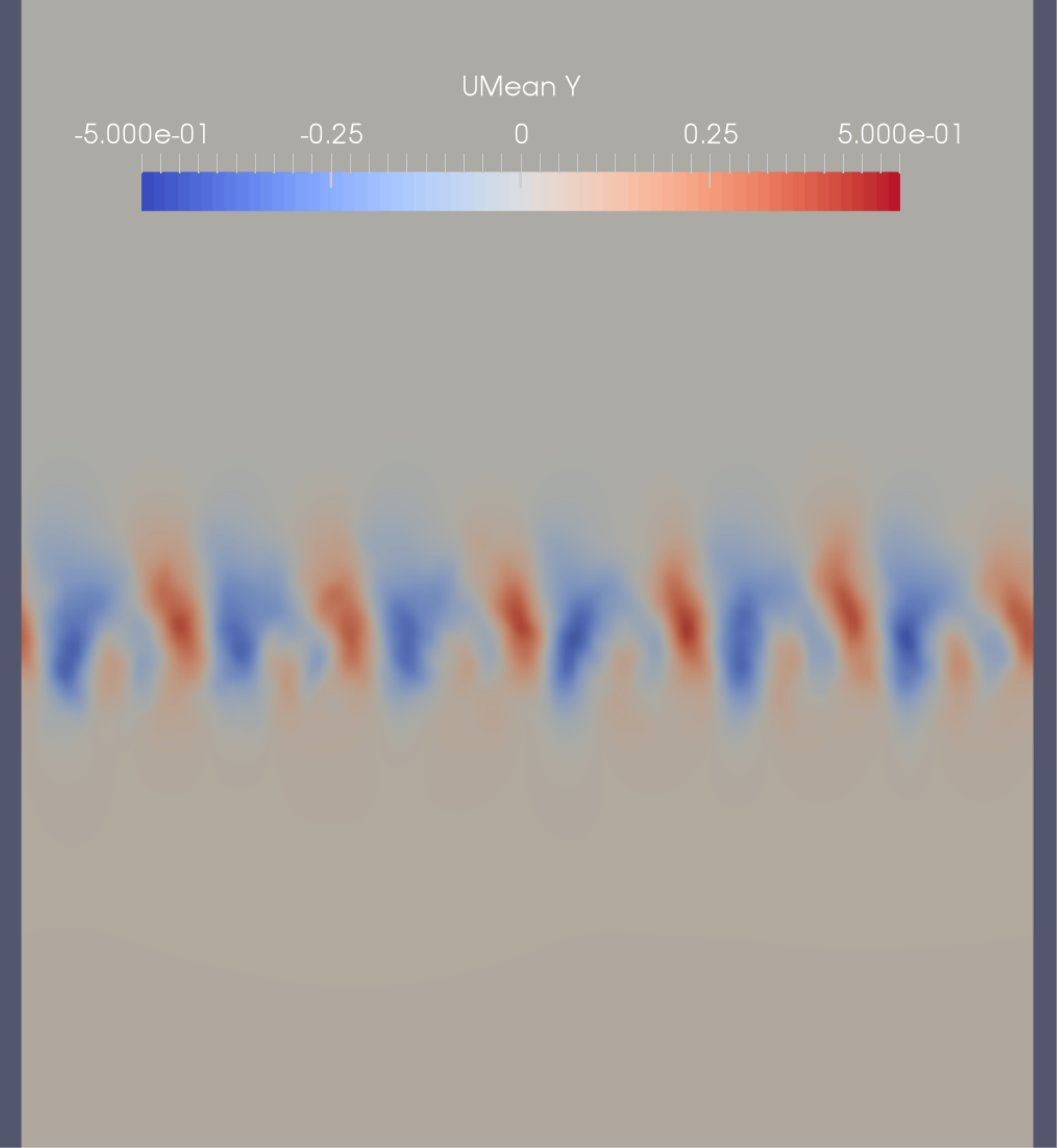} & \includegraphics[width=0.33\linewidth]{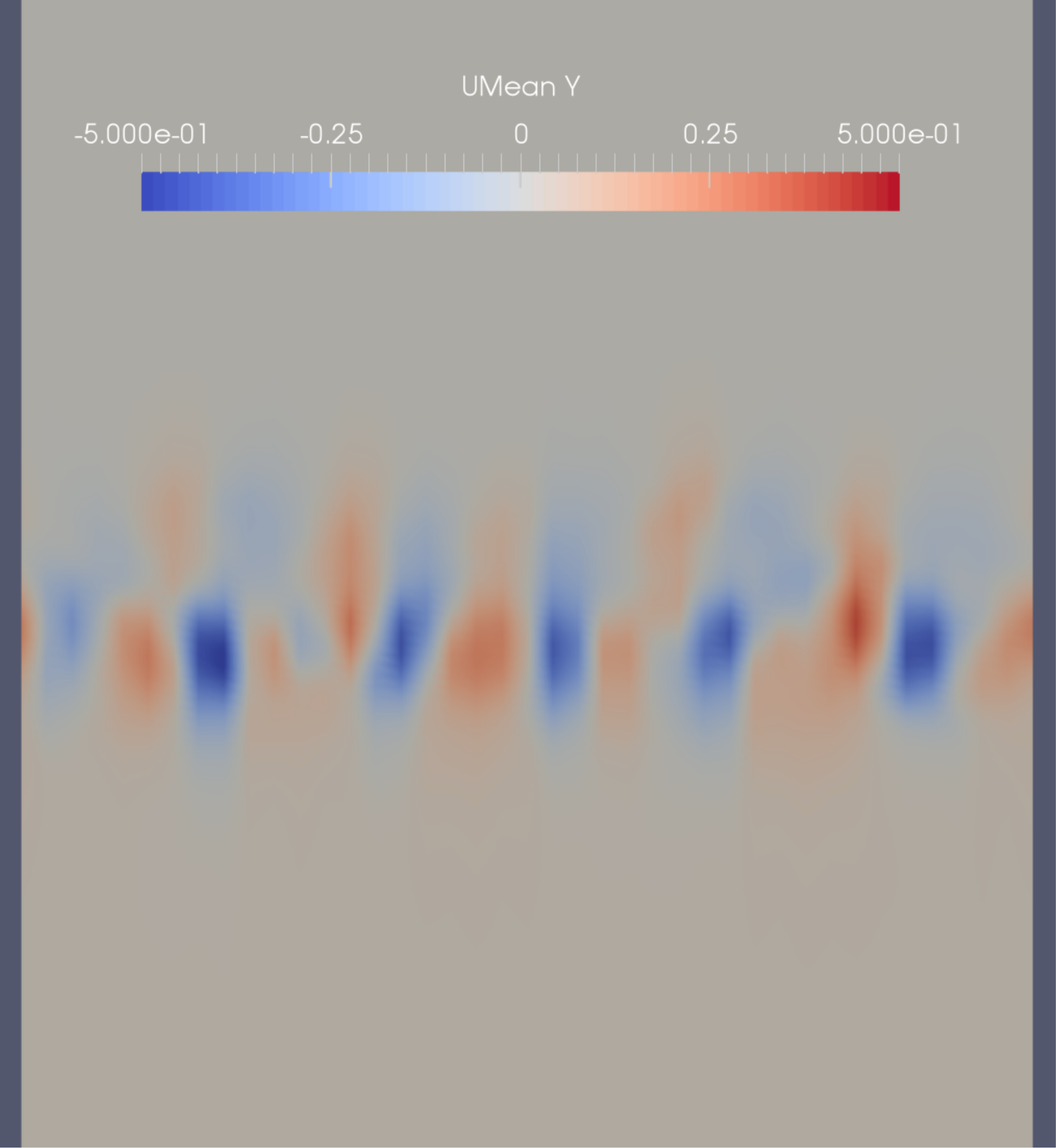} & \includegraphics[width=0.33\linewidth]{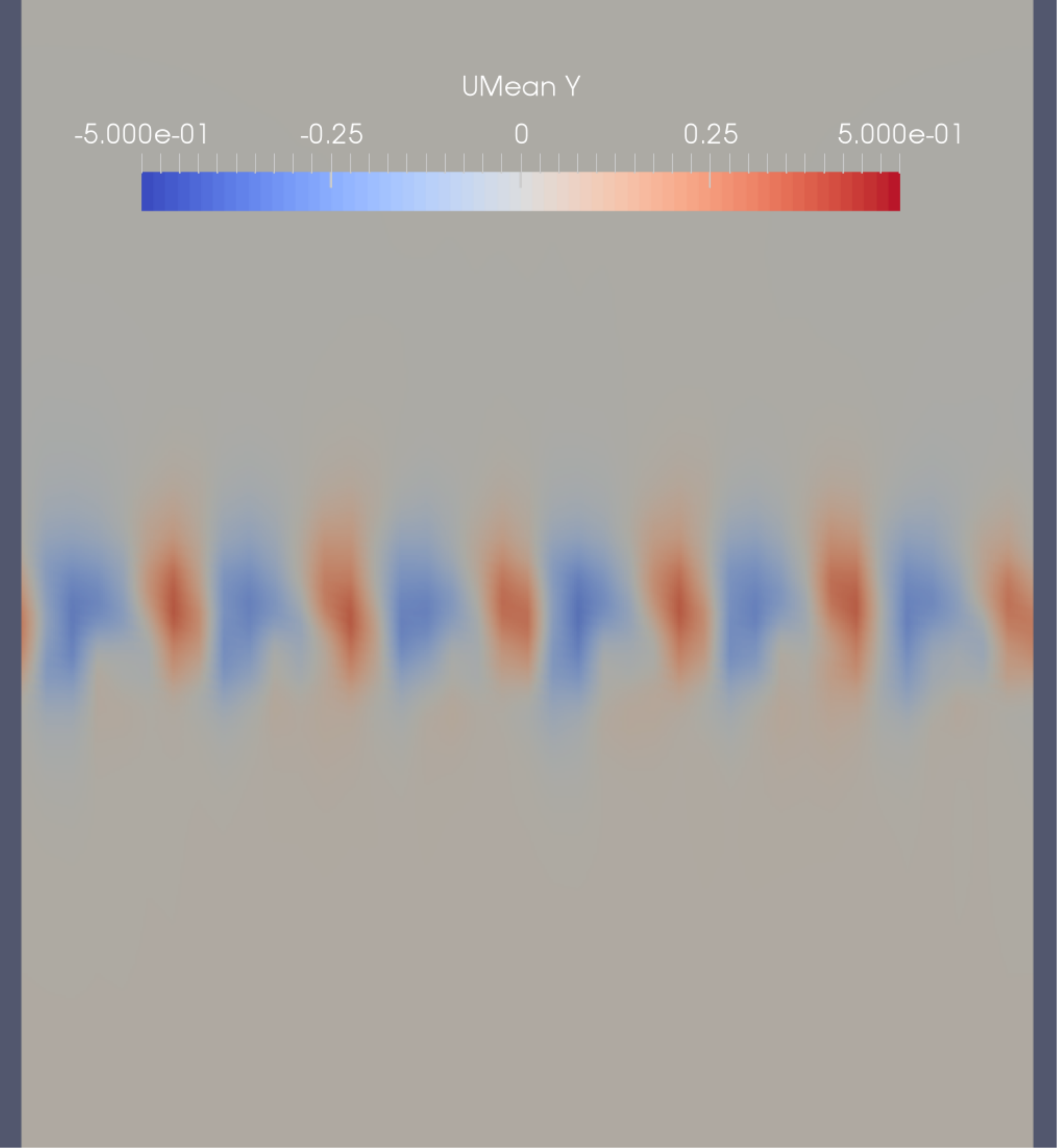} \\
 (a) & (b) & (c)
\end{tabular}
\caption{Isocontours of the time averaged normal velocity $\overline{u_y}$ taken at the streamwise section $x = 16 \Lambda$.
A zoom around the wake region is performed.
Results for (a) the DNS calculation, (b) the LES simulation and (c) the \textit{observer} estimator are shown, respectively.}
\label{fig:SpmixLayVelProfY}
\end{figure}

These conclusions are reinforced by the observation of the isocontours of the time averaged normal velocity $\overline{u_y}$, which are shown in figure \ref{fig:SpmixLayVelProfY}.
Here, data are sampled at the streamwise section $x = 16 \Lambda$, downstream the very last observation window.
The results indicate that the estimator produces results which are qualitatively in agreement with the DNS reference, as the same structure of the flow is conserved even with lower spatial resolution.
On the other hand, LES data appear to be more diffused and the organization of the coherent structures is not the same.

\begin{figure}
 \includegraphics[width=0.8\linewidth]{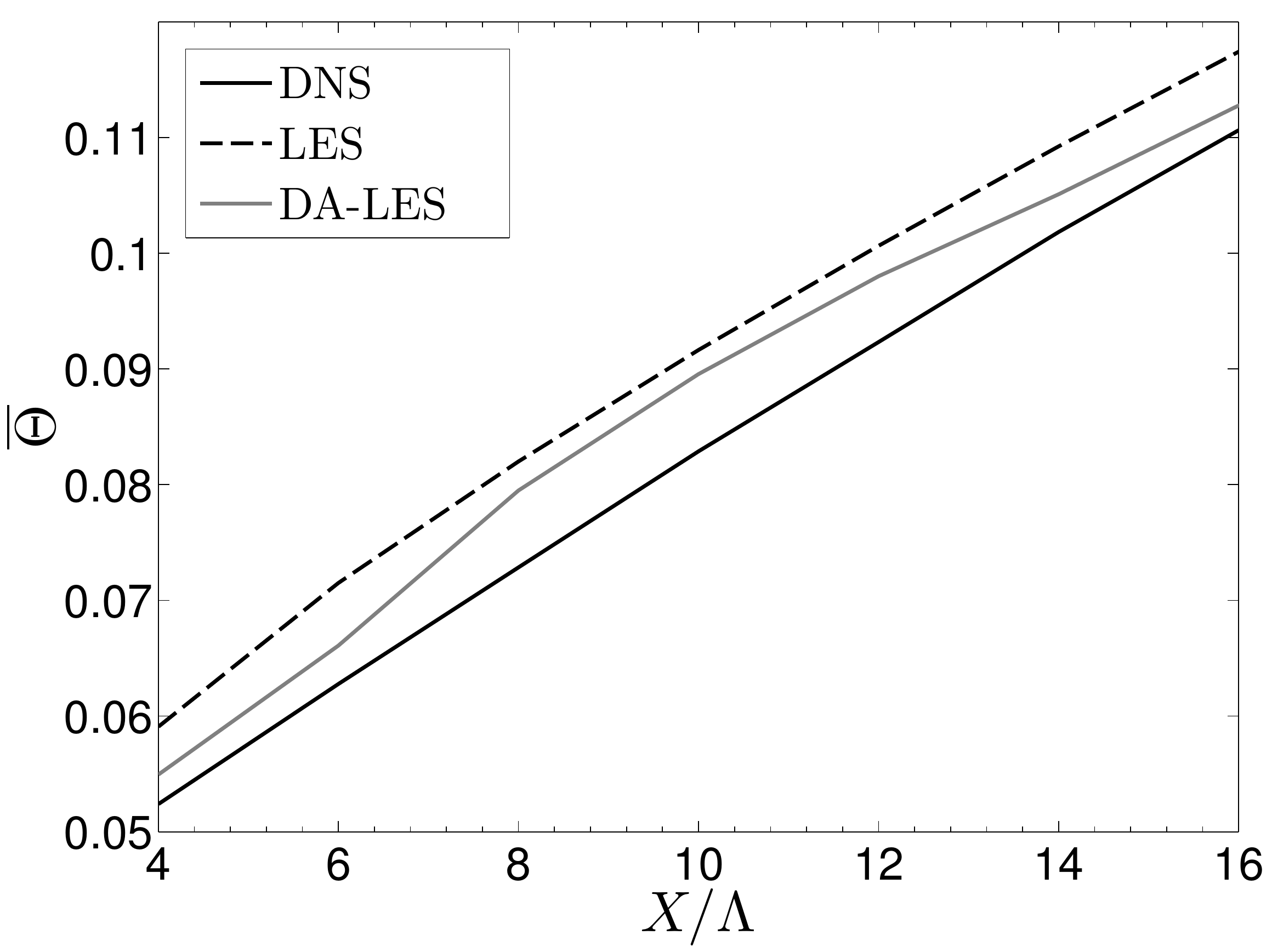}
\caption{Streamwise evolution of the momentum thickness $\overline{\Theta}$.
The results are reported for classical DNS ans LES calculations, as well as for the sequential estimator.}
\label{fig:SpmixLayMomT}
\end{figure}

The streamwise evolution of the momentum thickness $\overline{\Theta}$ is now investigated.
The results, which are reported in figure \ref{fig:SpmixLayMomT}, show that the LES overestimates the evolution of $\overline{\Theta}$.
However, the distance from the DNS data is almost constant, showing that the maximum error produced by the Smagorinsky model comes for the prediction of the laminar state close to the inlet.
On the other hand, the prediction produced by the sequential estimator is roughly in the middle between the LES and DNS results.
This result comes with no operation on the model i.e. the Smagorinsky coefficient has not been modified. However, the augmented prediction could be actively used in a optimization procedure, in order to improve the performance of the LES model.
While these analyses are not subject of investigation of the present research work, future development and applications will be discussed in section \ref{sec:future}.

\subsection{Thick plate}
\label{subsec:ThickPlate}

The case of the flow around a thick plate of height $H_p$ is investigated combining DDES numerical simulation \cite{Shur2008_ijhff} (\textit{model}) and PIV experimental data \cite{Sicot2012_ijhff} (\textit{observation}).
For the Reynolds number $Re=80000$ here investigated, the flow exhibits full turbulent features.
A recirculation bubble is observed at the trailing edge of the plate, whose average length is $\approx 5 H_p$.
This pioneering test case shows a high level of complexity, because of numerous problematic aspects:

\begin{figure}
 \includegraphics[width=0.8\linewidth]{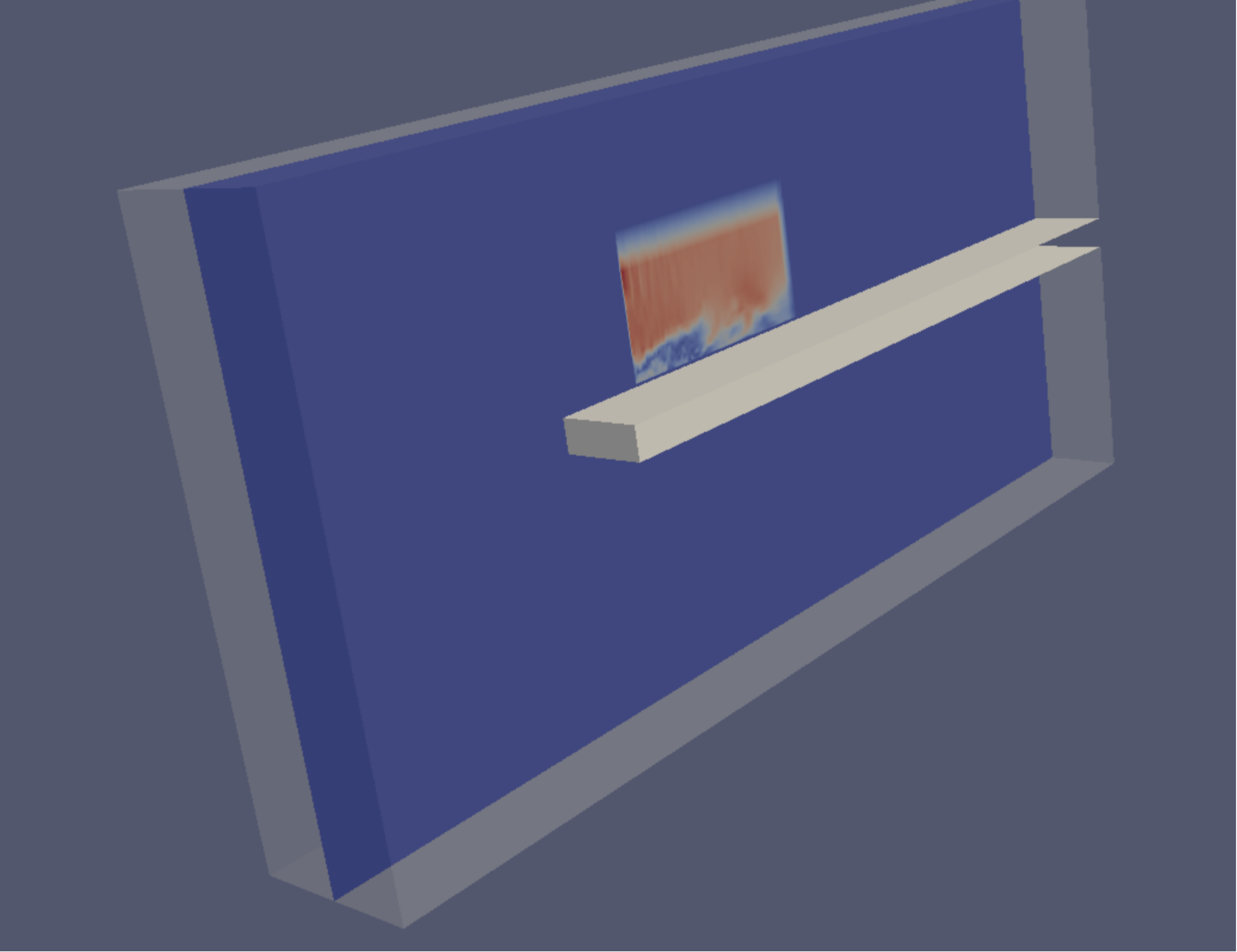}
\caption{Representation of the test case of the flow around a thick plate.
The plate profile is embedded in a three-dimensional physical domain used for DDES, while two-dimensional PIV data are provided for $z = 2.5 H_p$.}
\label{fig:ThickPlatevis}
\end{figure}

\begin{figure}
\begin{tabular}{cc}
 \includegraphics[width=0.48\linewidth]{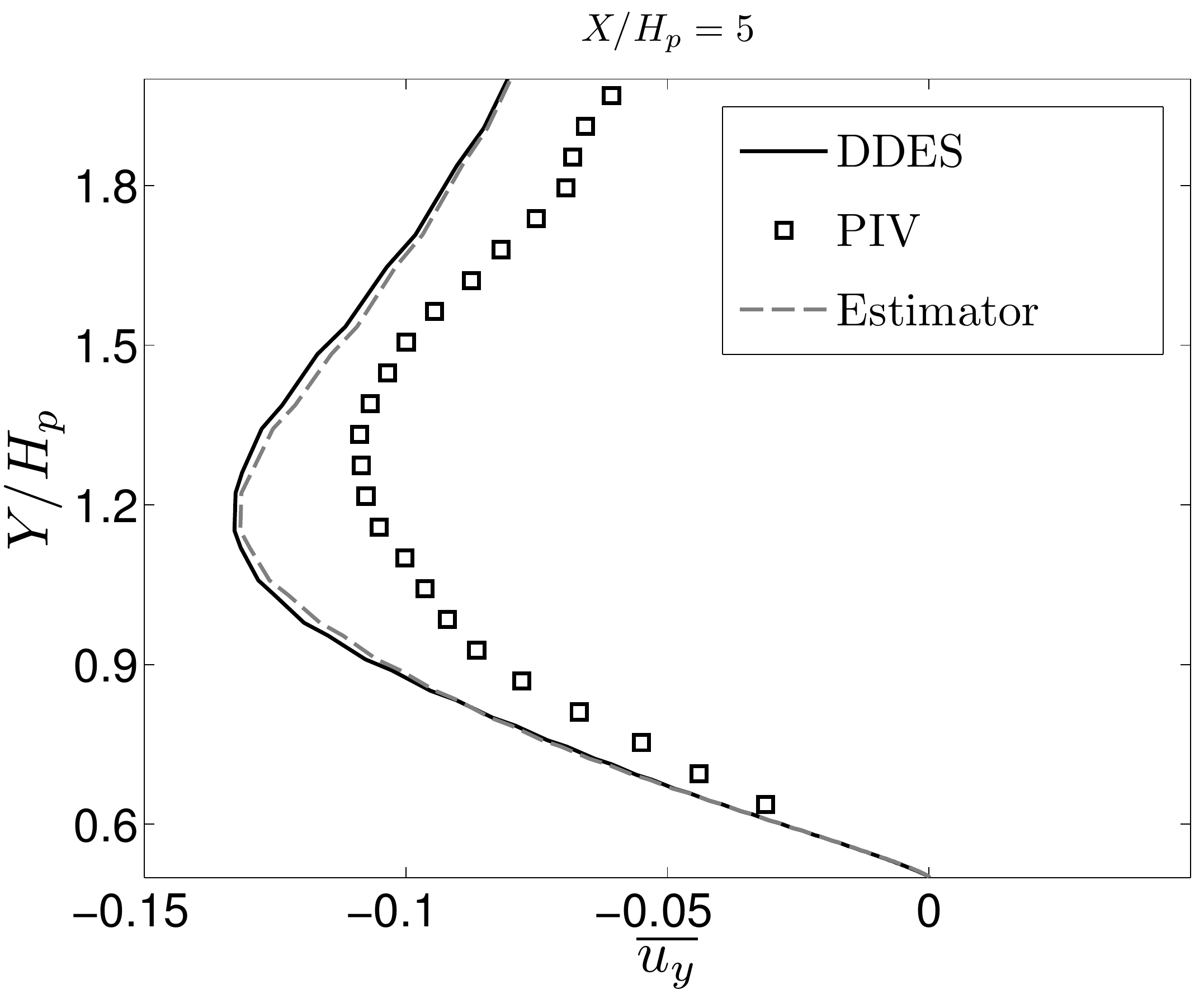} & \includegraphics[width=0.48\linewidth]{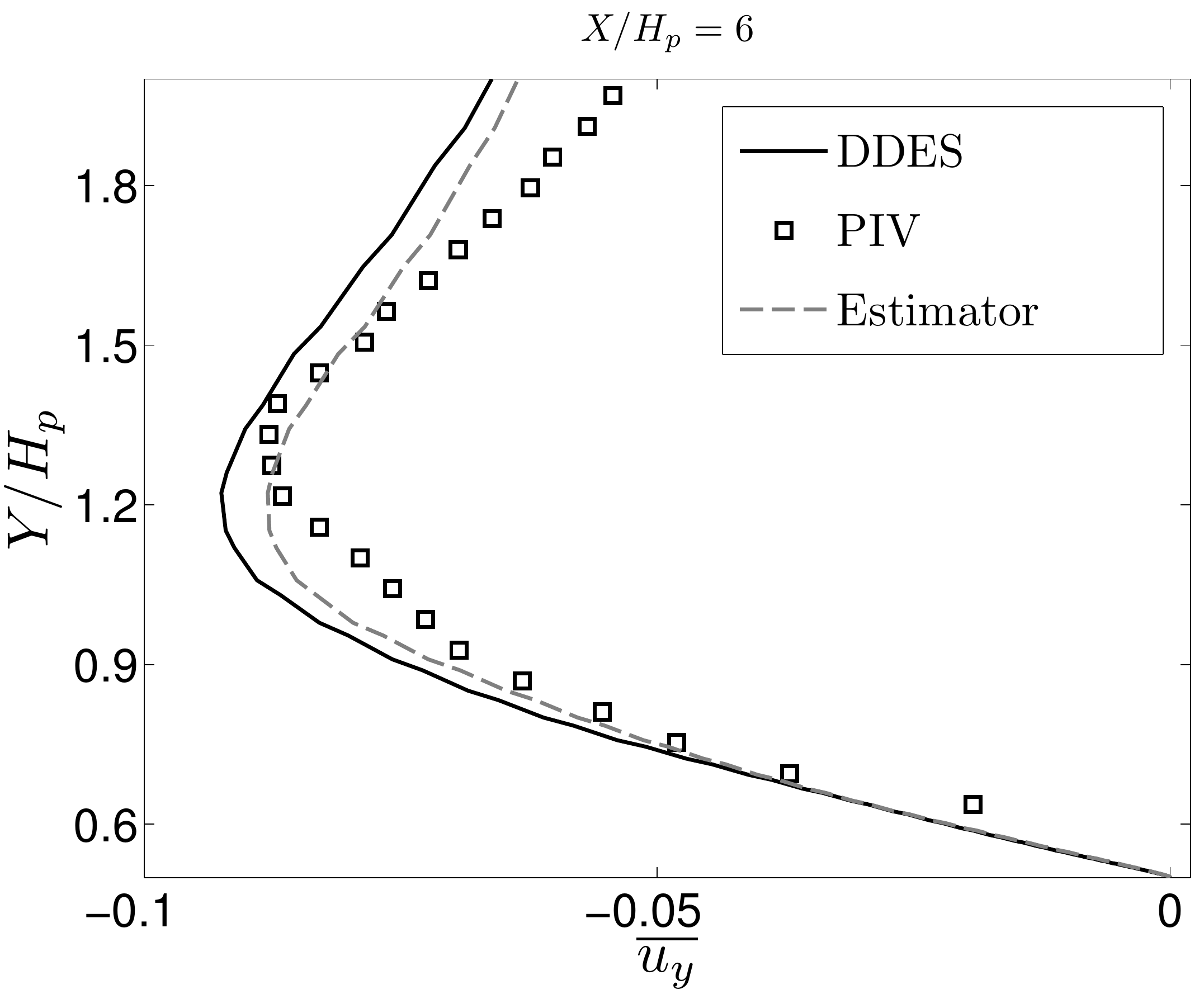} \\
 (a) & (b) \\
 \includegraphics[width=0.48\linewidth]{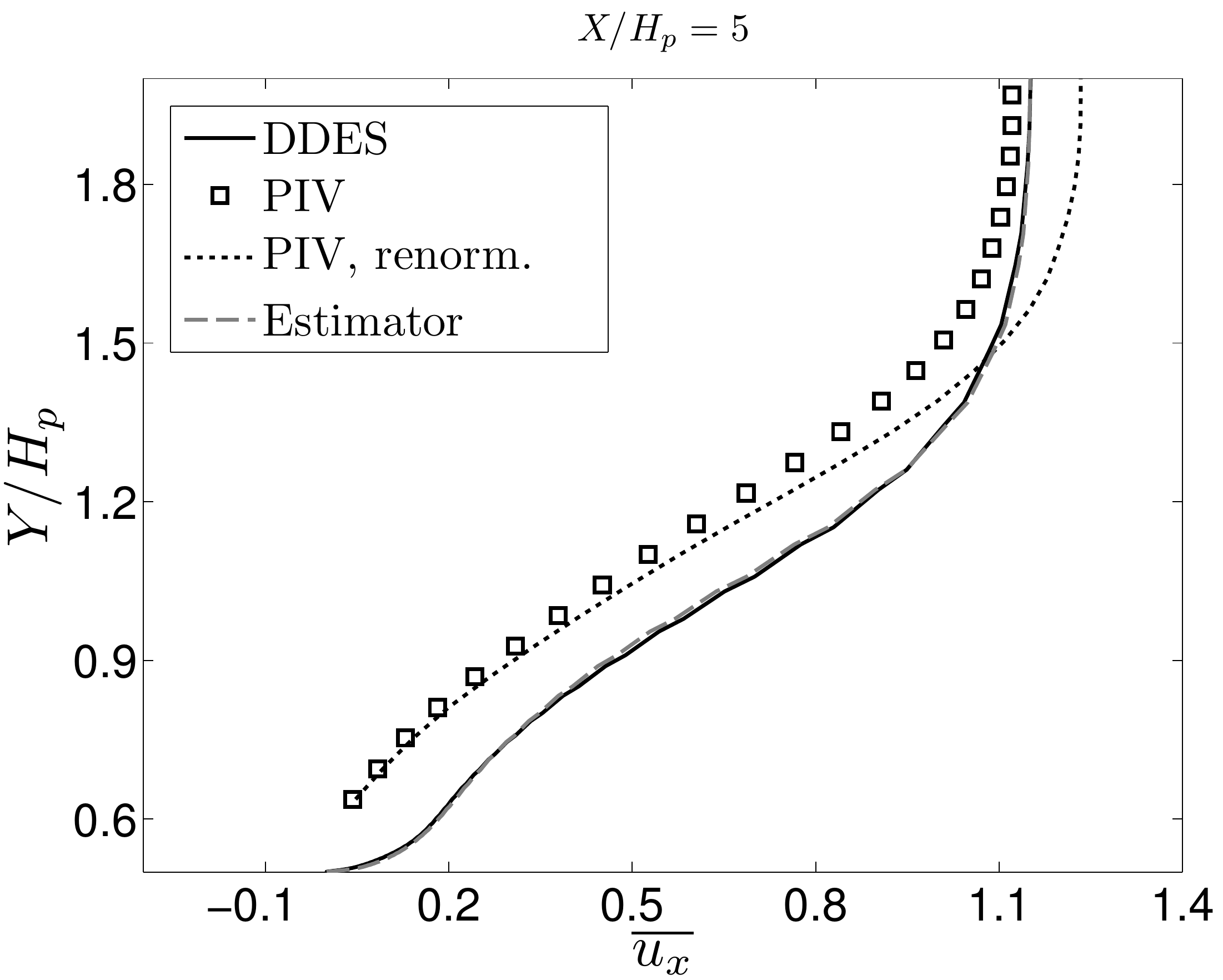} & \includegraphics[width=0.48\linewidth]{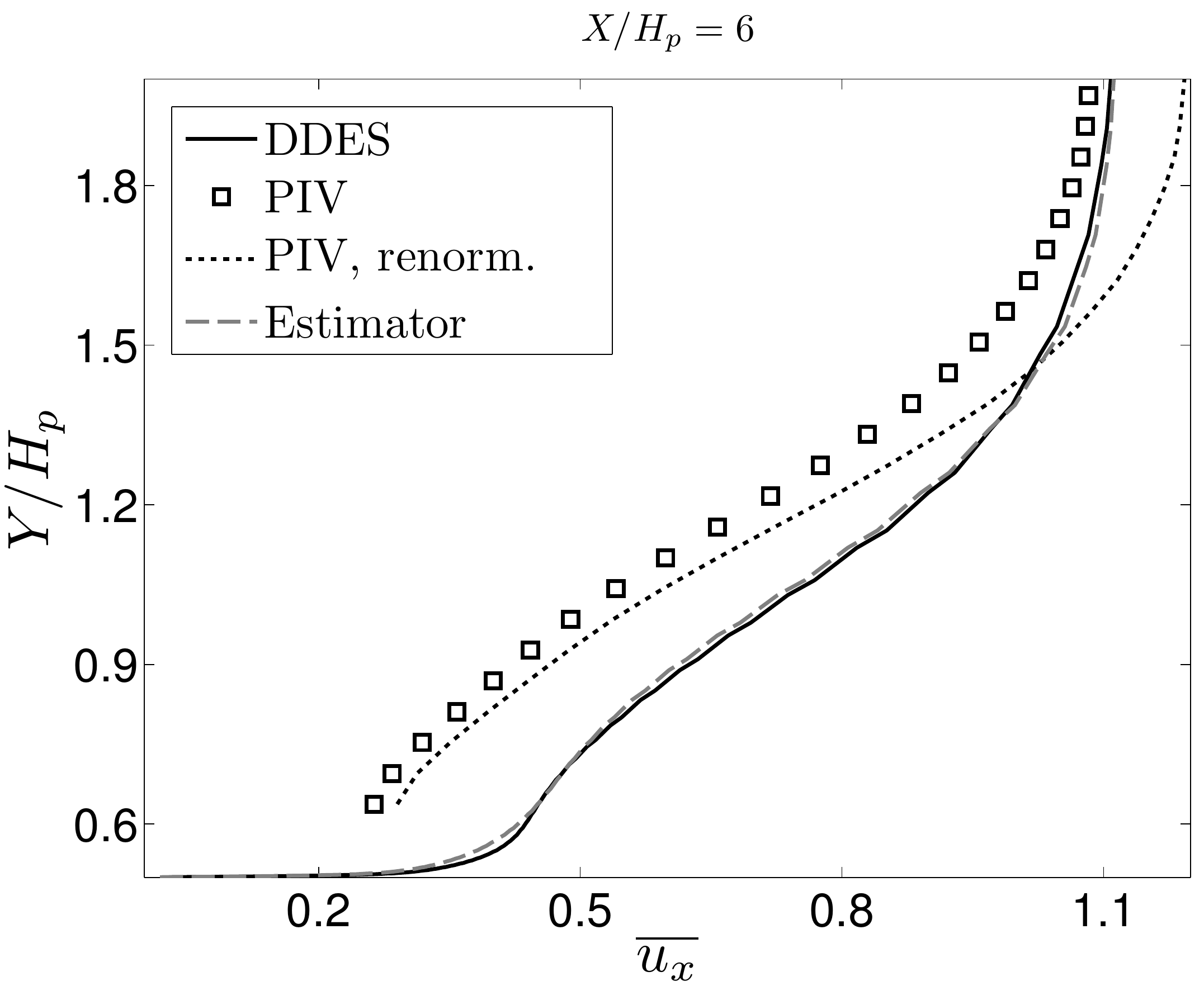} \\
 (c) & (d)
\end{tabular}
\caption{Averaged velocity $\overline{u}$ profiles at the streamwise section (left column) $x = 5 \, H_p$ and (right column) $x = 5 \, H_p$. DDES results and PIV samples are compared with the flow prediction by DA-DDES1 estimator. The (first row) normal velocity $\overline{u_y}$ and the (second row) streamwise velocity $\overline{u_x}$ are reported.
Averages are performed in the spanwise direction and over $T = 100 \, t_A$ times for the numerical simulations.}
\label{fig:ThickPlateAssim0}
\end{figure}

\begin{itemize}
\item{The numerical simulation domain is three-dimensional, while the PIV samples data is two-dimensional, as shown in figure \ref{fig:ThickPlatevis}.
In addition, the observation window is rectangular and it covers a limited space in the physical domain investigated.
Because of these two aspects, the transition between the assimilated state and the pure model prediction could be demanding on the numerical resolution of the pressure field, in order to smooth out this discontinuity.}
\item{PIV data are sampled using an acquisition frequency of $4000 Hz$, which is exactly $1/3$ of the characteristic advection time $t_A = H_p / U_{\infty}$.
On the other hand, the time step imposed in the numerical simulation is around $5 \times 10^{-4} t_A$, which has been chosen in order to grant a maximum Courant number of $Co=0.8$. This implies that the observation is available every $\approx 700$ model time steps only.}
\item{Preliminary comparison of average PIV results and DDES data show a difference of more than $10 \%$ in the mass flow rate, associated with different initial conditions / blockage ratio in the streamwise direction $x$ (see figure \ref{fig:ThickPlateAssim0}).
As a consequence, pure assimilation of the streamwise velocity could be problematic, leading to a badly-constrained optimization problem.}
\end{itemize}

For this reason, a total of three numerical simulations has been performed:
\begin{enumerate}
\item{A classical DDES simulation, which represents the \textit{model}.}
\item{An estimator combining DDES and the normal velocity $u_y$ of the PIV data, which will be referred to as DA-DDES1.
The instantaneous flow field has been assimilated with a constant period of $t_A /3$.
As previously mentioned, this corresponds to approximately $700$ time steps of the numerical simulation.}
  \item{A second estimator referred to as DA-DDES2. In this case, both $u_x$ and $u_y$ are assimilated.
The interest here is to analyze the evolution of the velocity field, when a problematic constraint for the mass flow rate is imposed.}
\end{enumerate}

The physical domain investigated in the numerical simulations is $[-10, \, 25] \times [-8.5, \, 8.5] \times [0, \, 5]$ in $H_p$ units.
The origin of the system is located on the vertical surface of the thick plate.
The physical domain is discretized in $ \approx 7 \times 10^6$ elements.
The recirculation bubble region has been refined in order to capture with sufficient resolution the wall dynamics.
A preliminary simulation has been performed in order dissipate the initial transient and provide a suitable initial condition for the three simulations performed.
The \textit{model} simulation has been run for a total of $T=100 t_A$.
Averages have been performed on the whole time length, which consisted of around $2.7 \times 10^5$ time steps.
For the two estimators simulations, the total time length has been imposed to $T=130 t_A$.
Data Assimilation begins from $t=0$, while averages of the physical quantities have been started from $t=30 t_A$.
This strategy allows for a degree of synchronization between the numerical model and the observation, before averages are calculated.
As in the case of the spatially evolving mixing layer, second order schemes have been employed for the spatial / time derivatives.
The only difference has been the use of a Linear-upwind stabilized transport (LUST) scheme for the divergence of the velocity field, which incorporates a second-order upwind scheme correction in the second-order centered classical scheme.
The PIV data used in the present analysis are distributed over a uniform grid $x-y$ of $141 \times 33$ elements, for a total of $4653$ samples par observation.
Because of the very different structure of the experimental sampling grid with respect to the numerical mesh, PIV data have been interpolated on the center of the mesh elements in the observation window for $z=2.5 \, H_p$, which is represented by the rectangle $x \in [3, \, 10], \, y \in [0.6, \, 3.5]$ in $H_p$ units.
The horizontal surface of the thick plate is located at $y = \pm 0.5$, so that Data Assimilation at the wall is not performed.
In order to simplify the analysis, the PIV interpolation error on the numerical mesh has been considered negligible with respect to the noise of the experimental measurement.

\begin{figure}
 \includegraphics[width=0.8\linewidth]{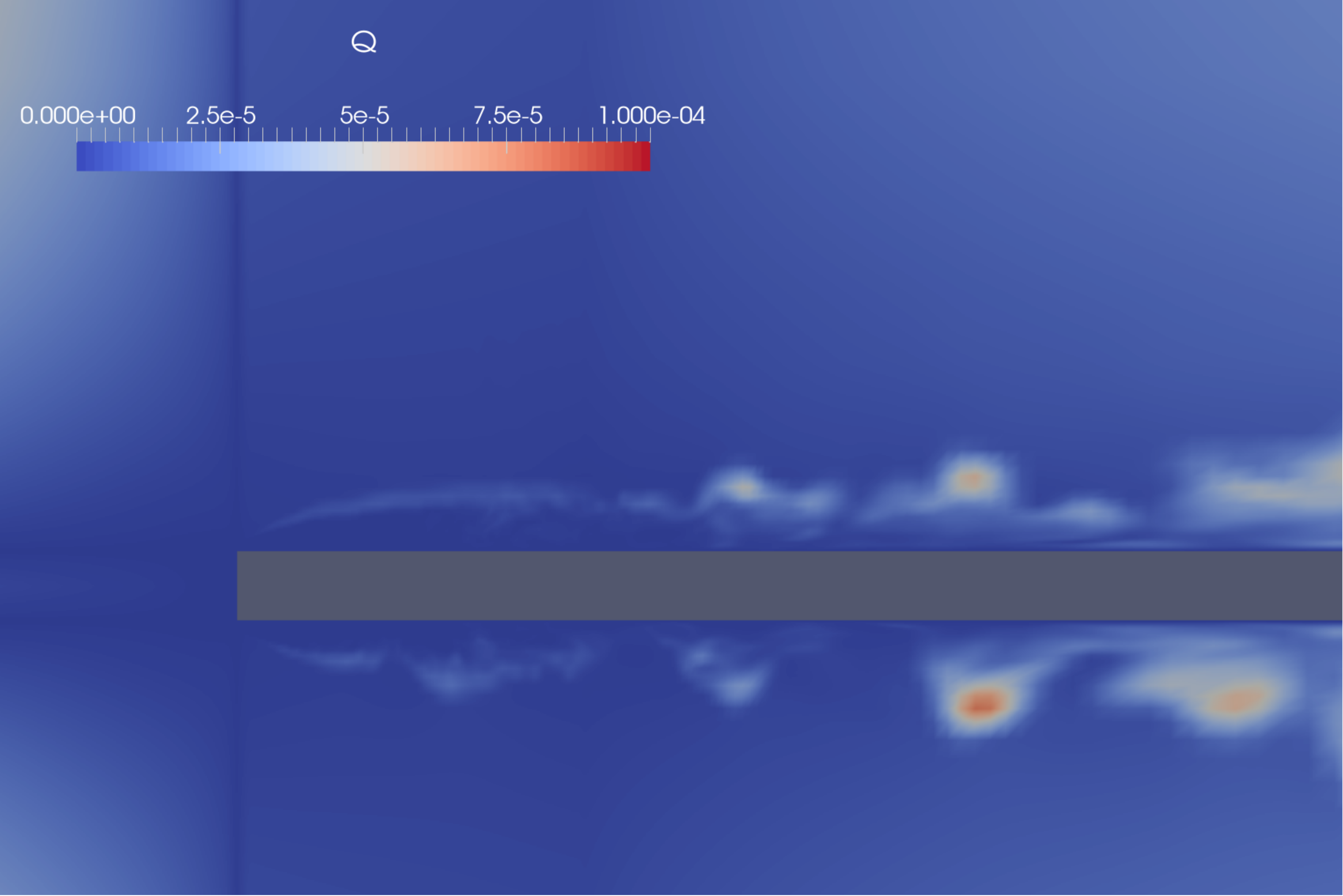}
\caption{Visualization of the instantaneous diagonal elements ($t=0$) for the matrix $Q$ on the assimilation plane $z = 2.5 H_p$.}
\label{fig:ThickPlateQ}
\end{figure}

The matrices setting the level of confidence in the model / observation are now discussed.
The error covariance matrix is set so that $P=0.1 \, I$.
This high level of uncertainty has been chosen in order to speed up the synchronization between the model and the observation.
Similarly to the case of the spatially evolving mixing layer, the diagonal coefficients $Q_i$ of the matrix $Q$ have been locally determined as:

\begin{equation}
\label{eq:Qvalue_FP}
Q_i = C_{TP} \, \left( 1 + \frac{\nu_T}{\nu} \right) \, \sqrt[3]{{\Delta V}^2} \, {\Delta t}  
\end{equation}

\begin{figure}
\begin{tabular}{cc}
 \includegraphics[width=0.48\linewidth]{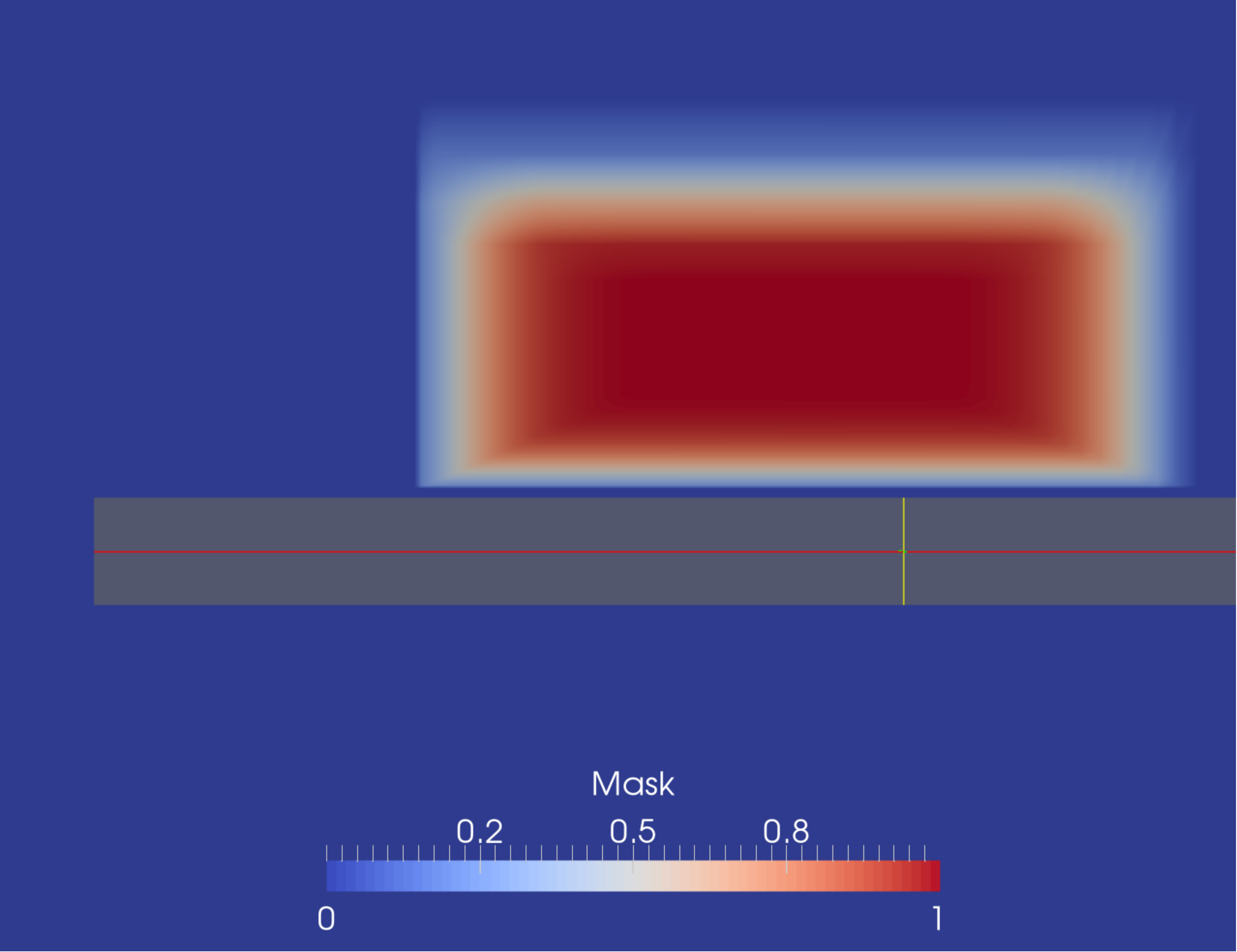} & \includegraphics[width=0.48\linewidth]{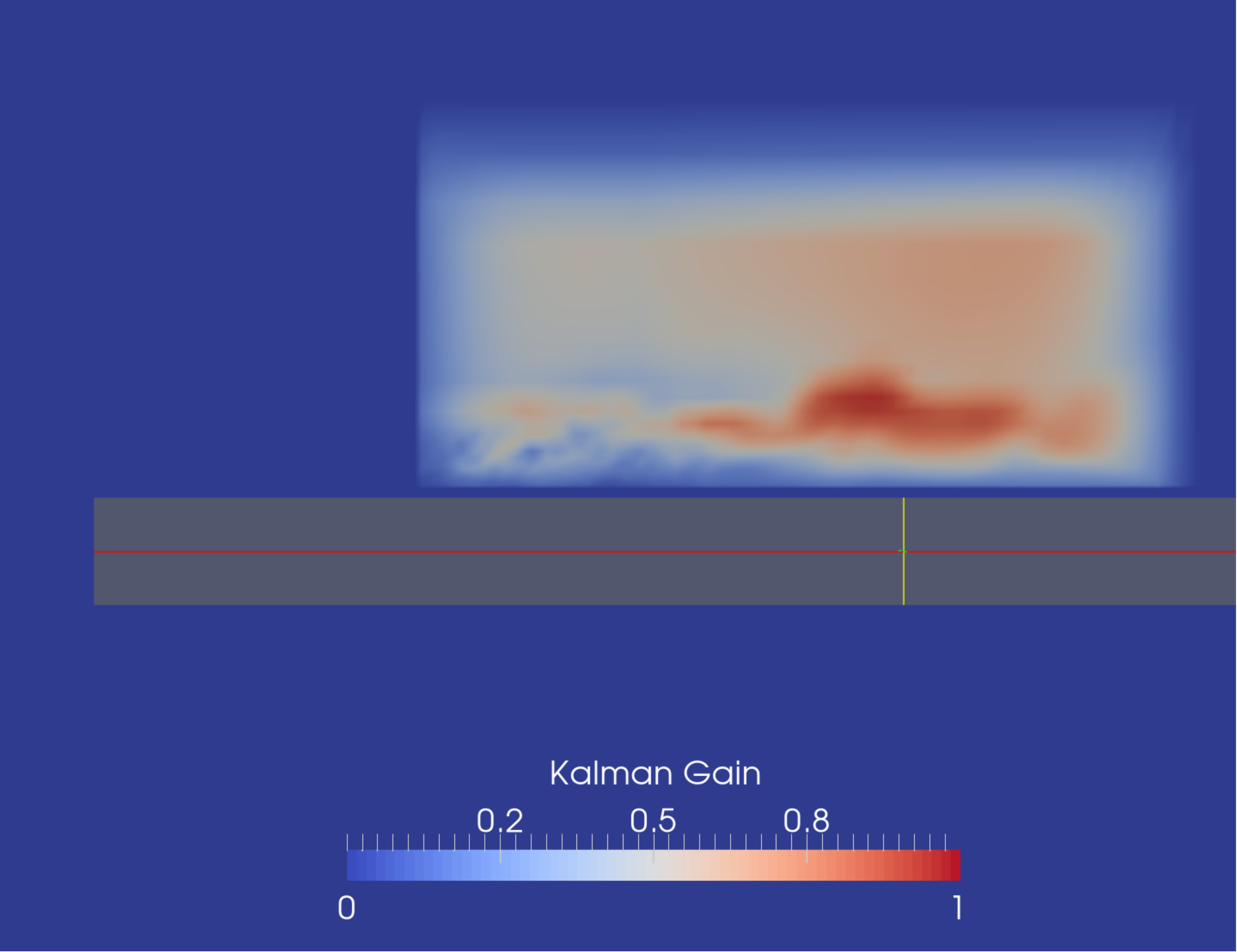} \\
 (a) & (b)
\end{tabular}
\caption{(a) Mask function $M$ used to smooth out the Kalman gain when approaching the border of the observation region.
(b) Instantaneous contours of the diagonal components of the Kalman gain for the streamwise velocity $u_x$ on the assimilation plane, shown for $t=100 t_A$.}
\label{fig:ThickPlateMask}
\end{figure}

A visualization of an instantaneous field for the scalar field $Q$ is shown in figure \ref{fig:ThickPlateQ}.
The value of the constant $C_{TP}$ has been chosen in order to have roughly the same confidence in the model and in the observation during the update phase, considering that approximately $700$ prediction steps are performed between each update.
The precision of the PIV data is estimated to be of the order of $5 \%$, so the covariance matrix $R$ is fixed so that $R= 0.05^2 \, I$.
In addition, in order to prevent problems close to the border of the rectangular assimilation window, a mask function $M$ has been used to smooth out the value of the Kalman gain close to the borders.
In this scenario, a sub-optimal gain in the form $\overline{K} = M \, K$ has been used to improve the stability of the results.
The mask is defined as:

\begin{equation}
\label{eq:mask_FP}
M = \exp{\left(-250\left(\frac{x-x_c}{x_c}\right)^{8} -25\left(\frac{y-y_c}{y_c}\right)^{8}\right)}  
\end{equation}

where $x_c=6.5$ and $y_c=2.05$ are the coordinates of the center of the observation region.
A representation of the mask function within the physical domain is shown in figure \ref{fig:ThickPlateMask} (a), while a visualization of the diagonal component of the resulting Kalman gain for the streamwise velocity is shown in figure \ref{fig:ThickPlateMask} (b). 

First of all, experimental and numerical data provide a very similar estimation of the recirculation bubble length, which is $\approx 5 \, H_p$. The two simulations performed using the estimator are now compared to DDES and PIV data.
The case of Data Assimilation for the velocity $u_y$ only (DA-DDES1) is discussed first.
Average velocity profiles for $\overline{u_x}$ and $\overline{u_y}$ are reported in figure \ref{fig:ThickPlateAssim0} for the streamwise sections $x = 5 \, H_p$ and $x=6 \, H_p$.
These sections are close to the middle of the observation window.
The averages are performed in time and in the spanwise direction for the numerical simulation, while PIV data are clearly averaged over time only.
The analysis of the normal velocity $\overline{u_y}$ in figure \ref{fig:ThickPlateAssim0} (a) and (b) shows that the estimator successfully obtains an hybrid flow prediction for this physical quantity.
The result can be closer to the pure DDES velocity field or to the PIV data depending on the streamwise section investigated.
However, the analysis of spanwise averaged profiles for the numerical simulations indicates that the information is propagated by the Poisson equation in the physical domain, balancing the limited transmission of observed data via the \textit{observer}.
Further confirmation is obtained analyzing the L2 norm of the difference between time averaged numerical prediction for $\overline{u_y}$ and PIV data over the observation window:

\begin{equation}
 N_{FP} = \int_{OW} \frac{\sqrt{(\overline{u_y}-\overline{u_{PIV}})^2}}{\sqrt{(\overline{u_{PIV}})^2 + 10^{-15}}} \, dS
\label{eq:L2Norm_plate}  
\end{equation}

The results, which have been re-normalized over the calculation for DA-DDES1, are reported in table \ref{L2-norms-Mix}.
It appears that the average of the differences from PIV data and the \textit{observer} prediction is approximately half with respect to the same norm for pure DDES (1.473), reinforcing the qualitative result of a flow configuration obtained imposing similar level of confidence in the model and in the observation.
In figure \ref{fig:ThickPlateAssim0} (c) and (d) averages velocity profiles for the streamwise velocity $\overline{u_x}$ are reported.
PIV data are shown as well.
In particular, the dotted line corresponds to a re-normalization of the experimental samples targeting the same average mass flow rate of the numerical simulation.
Even if this quantity is not assimilated, the analysis of figure \ref{fig:ThickPlateAssim0} shows that the prediction provided via the \textit{observer} moves towards the shape of the normalized PIV data.
This observation is fundamental, as it indicates that the whole numerical simulation process tends towards a high-confidence state even if a limited amount of samples is provided.
Thus, assimilation of a physical quantity sampled in a small observation window can actually be beneficial for the prediction of other physical quantities in the whole domain investigated.
This includes quantities that are difficult to measure via experimentation.

 \begin{table}
\centering
\small
\begin{tabular}{l|ccc}
L2 norm error 	& DDES	& DA-DDES1 & DA-DDES2	\\
\hline
$N_{FP}$ & 1.473 & 1 & 1.097
\end{tabular}
\caption{L2 norm of the difference of the numerical profiles when compared to PIV data, for the average normal velocity $\overline{u_{y}}$.
  The norm has been calculated in the observation window for $z=2.5 \, H_p$ and re-normalized over the value calculated for DA-DDES1. \label{L2-norms-Mix}}
\end{table}

\begin{figure}
\begin{tabular}{cc}
 \includegraphics[width=0.48\linewidth]{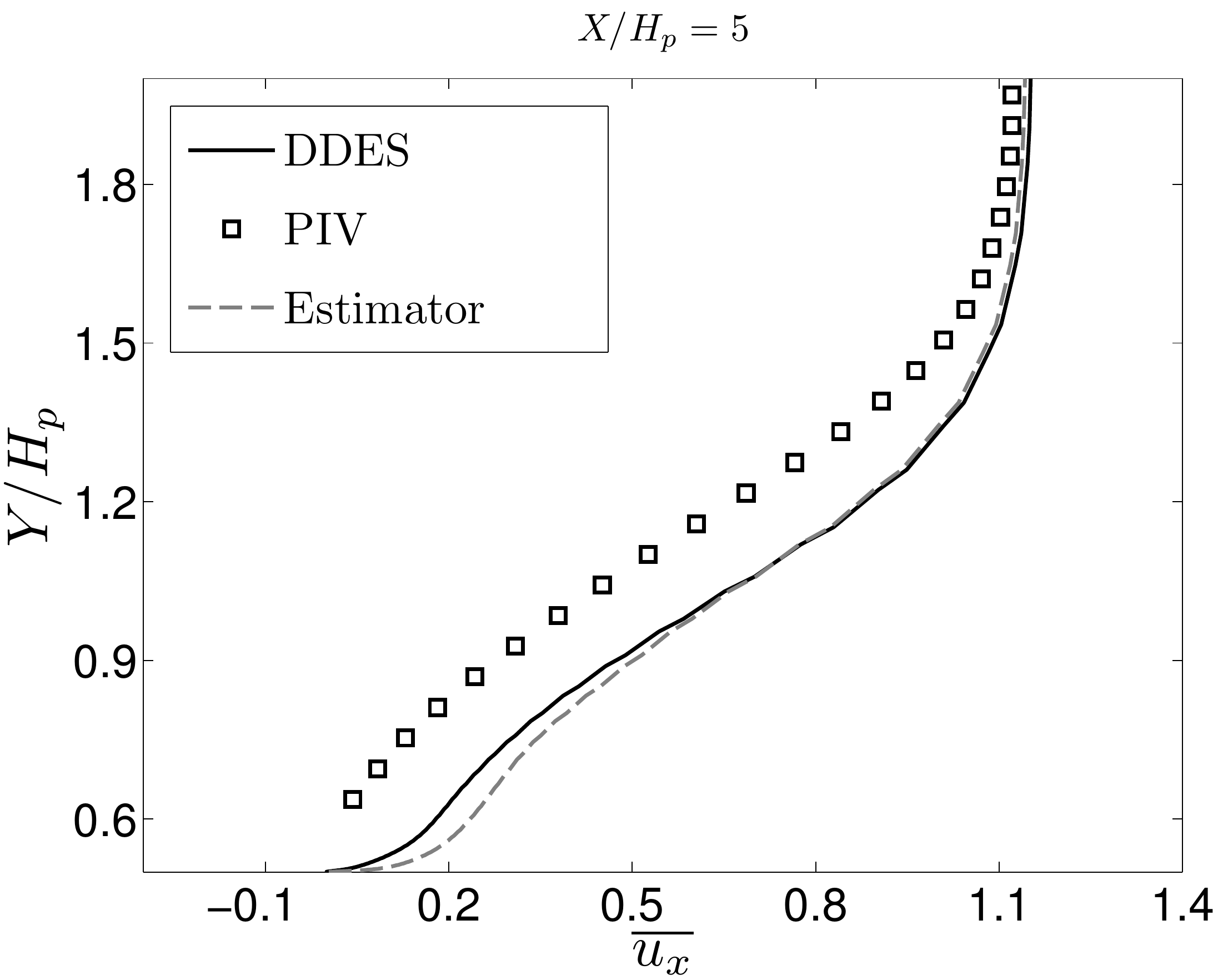} & \includegraphics[width=0.48\linewidth]{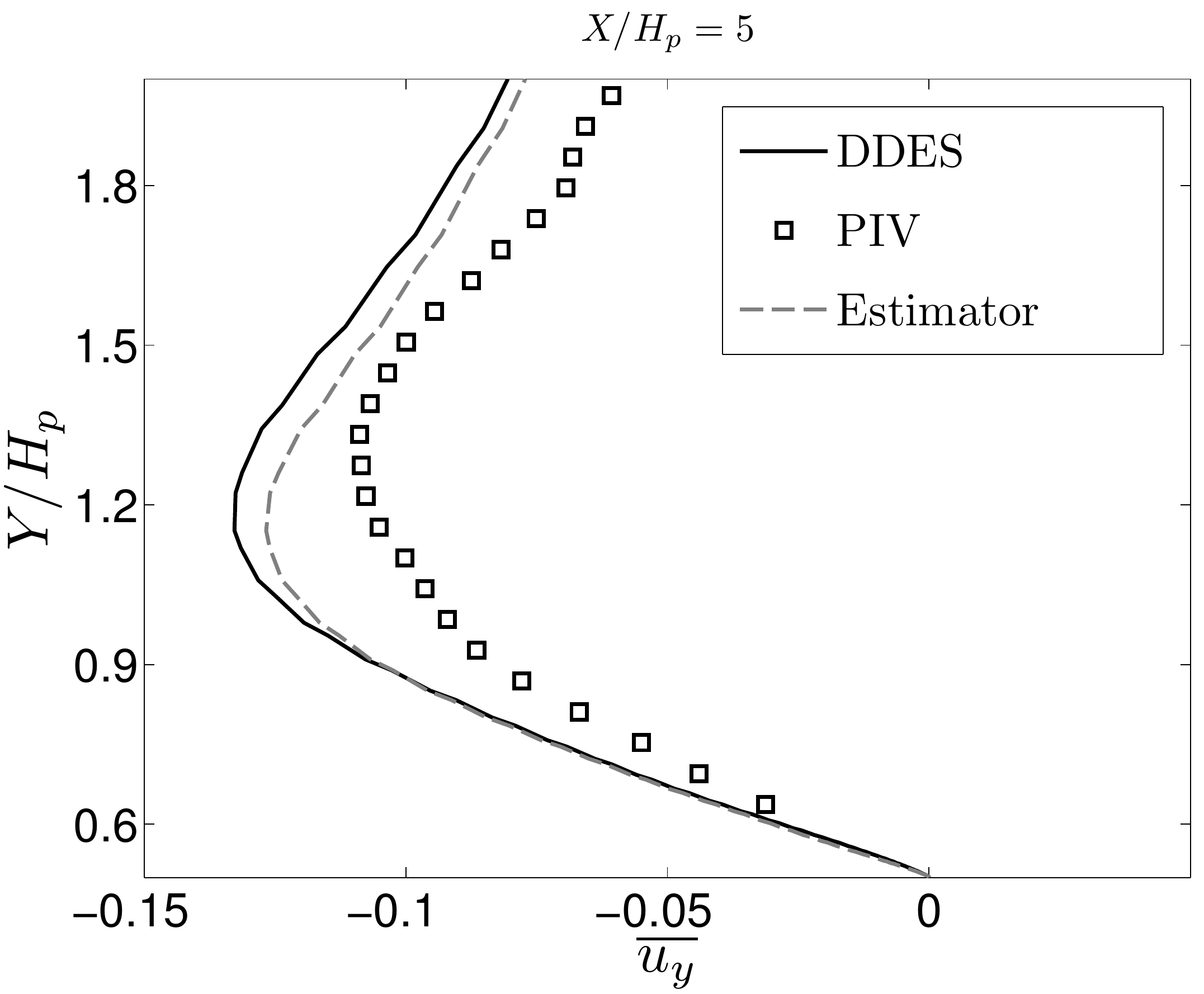} \\
 (a) & (b)
\end{tabular}
\caption{Averaged velocity $\overline{u}$ profiles at the streamwise section $x = 5 \, H_p$.
Averages are performed in the spanwise direction and over $T = 100 \, t_A$ times. DDES results and PIV samples are compared with the flow prediction by DA-DDES2 estimator.
The results are shown for (a) the streamwise velocity $\overline{u_x}$ and (b) the normal velocity $\overline{u_y}$, respectively.}
\label{fig:ThickPlateAssim1}
\end{figure}

The analysis of the velocity profile for the second estimator DA-DDES2 ($u_x$ and $u_y$ assimilated) is now performed.
Average velocity profiles at the streamwise section $x = 5 \, H_p$ are shown in figure \ref{fig:ThickPlateAssim1}.
It is here reminded that, in the case of the numerical simulations, averages are performed both in time and in the spanwise direction $z$.
The estimator provides an hybrid flow prediction from the model / PIV data in the observation region, as shown for the averaged streamwise velocity $\overline{u_x}$ in figure \ref{fig:ThickPlateAssim1} (a).
However, an augmentation of the predicted velocity by the estimator is observable close to the wall.
This effect is the result of the over-constraint condition applied to the problem, as the estimator must (i) provide a weighted prediction based on model results / observation and (ii) to conserve the same mass flow rate of the model.
The difference between the PIV data and the numerical simulations is measured on the observation window by the use of the quadratic function \ref{eq:L2Norm_plate} for both $\overline{u_x}$ and $\overline{u_y}$.
The analysis of the results for the DDES simulation and the DA-DDES2 calculation show that, on average, the estimator is significantly closer to the experimental prediction for $\overline{u_x}$ and $\overline{u_y}$ over the observation window.
However, as the DDES and DA-DDES2 simulations must have the same inlet mass flow, the second must see and acceleration of the flow in the spanwise and normal directions, outside of the observation region.
These results show how easy this numerical problem can be over-constrained, if the assimilation process is not structured carefully. In addition, present results stress as well the idea that suggest that the augmented flow must be actively used to optimize the parametric description of the model, in order to obtain a significant improvement in the overall prediction.
In this case, the augmented prediction should be used to tune the value of the mass inflow rate at the inlet, in order to smooth out the differences between the classical DDES ad the PIV observation.
This topic will be discussed extensively in Section \ref{sec:future}.
The average prediction of the normal velocity is shown in figure \ref{fig:ThickPlateAssim1} (b).
In this case, the DA-DDES2 prediction is successful, as the averaged velocity profile correctly shows intermediate characteristics between the model and the observation.
This is consistent with the rough hypothesis of similar level of confidence in the model and the observation that was initially set, and it is in line with results obtained by the DA-DDES1 estimator.

\section{Future works}
\label{sec:future}

The results presented in section \ref{sec:turbFlows} indicate that the augmented flow prediction performed via \textit{observer} estimator successfully represents a dynamic system which accounts for the level of confidence in the model and in the available observation.
To the Authors' knowledge, these analysis are among the very first successful Data Assimilation applications for the analysis of three-dimensional, unsteady turbulent flows.
In particular, the statistics of several turbulence quantities have been investigated, which represent the main target objective in industrial applications.
These Data Assimilation techniques have the potential to produce a breakthrough in the analysis of turbulent flows, in particular when model and observation are affected by completely different epistemic uncertainties and the level of confidence in the two basic investigative tools is roughly the same.
However, the analysis of the two very different test cases in Section \ref{sec:turbFlows} highlights how results could see groundbreaking improvement accounting for the augmented information:

\begin{enumerate}
\item{To improve $LES$ subgrid scale modeling, in the case of the spatially evolving mixing layer.
The possibility to locally optimize the parameters of the subgrid scale model (i.e. model coefficient) would improve the results of the underlying LES simulation, reducing even more the difference observed when comparing the model and the estimator with DNS results.}
\item{To modify physical parameters of the DDES simulation, such as the mass flow rate, in the case of the the flow around the thick plate.
This would allow the estimator to adjust the mass flow rate of the inlet, in order to avoid the over-constrained problem presented in section \ref{sec:turbFlows} for the estimator DA-DDES2.}  
\end{enumerate}

Accounting for the augmented prediction in order to synergically improve the performance of the model is usually referred to as Linear Quadratic Gaussian (LQG) scheme in the framework of optimal control performed by sequential methods \cite{Bewley2001_pas,Kim2007_arfm,Brunton2015_amr}.
In summary, while present results show that sequential Data Assimilation of turbulent flows is achieved without exponentially increasing computational costs, they also open up to exciting perspectives for the optimization of numerical solvers under a new physical perspective.
This last aspect represents a breakthrough in particular with respect to RANS modeling, where most of the well known models have been extended subject of analysis and tuning in order to improve the overall performances.

\section{Conclusion}
\label{sec:conclusions}

The present research work describes the integration of a Kalman filter based sequential estimator in the PISO algorithm of a segregated solver for incompressible flows.
This technique, which is reminiscent of the proposal for coupled flow solvers by Suzuki \cite{Suzuki2012_jfm}, allows for the derivation of an \textit{augmented} flow state accounting for the level of confidence in the model and in the observation provided.
In addition, the present formulation naturally complies with a zero-divergence condition for the augmented state.
Starting from this proposal, an extended analysis of the error covariance matrix $P$ and the model / observation covariance matrices $Q$ and $R$ has been proposed.
Because of the prohibitive costs associated with a full resolution of these matrices, two reduction strategies have been proposed and assessed.
These strategies dramatically reduce the increase in computational costs of the model, which can be quantified in an augmentation of $10\% - 15\%$ with respect to the classical numerical simulation.
In addition, an extended analysis of the behavior of the elements of the matrix $Q$ has been performed via analysis of simplified numerical cases.
The results have shown that optimized values are linked to the truncation error of the discretization procedure.
Simplified formulae accounting for the precision of the numerical schemes, which are set by the model user, have been elaborated and assessed.

The estimator has been applied to the analysis of a number of flow configurations exhibiting increasing complexity.
The first case analyzed in the two-dimensional case of the flow around a cylinder, $Re=100$.
In order to test the capabilities of the model to synchronize model and observation, data from a previous numerical simulation have been integrated in the estimator in phase opposition.
In addition, the observation window consisted of just $24$ elements at a distance of $\approx 7 \, D$ from the center of the cylinder.
The estimator proved to be very efficient in synchronizing the numerical model with the observed data.
Similar results have been observed within the context of a three-dimensional simulation.
In that case, the estimator produced the synchronization through a three-dimensional transient before returning to a pure two-dimensional evolution.

Finally, the observer has been used to investigated the statistical behavior of two turbulent flow configurations, namely the spatially evolving mixing layer and the flow around a thick plate.
In both cases, the observation has been provided on a limited amount of mesh cells on a two-dimensional window, while the model resolved a full three-dimensional simulation.
The analysis of the results has proven that in both cases the resulting augmented prediction correctly accounts for the level of confidence provided in the model and in the observation.
However, these analysis indicated as well that these techniques, which are at an embryonic stage, can be extended using the \textit{augmented} prediction as a powerful tool for the optimization of the free parameters included in the numerical set-up.

The research work has been developed using computational resources within the framework of the project gen7590 DARI-GENCI.
The Authors would like to thank F. Gava for the precious help provided in early stages of the work and Prof. E. Lamballais for the time he dedicated to fruitful discussion as well as for suggesting the use of the term \textit{augmented} prediction. 

\section{References}

\bibliography{references}

\end{document}